\documentclass[11pt, a4paper]{article} 

\usepackage[ansinew]{inputenc}
\usepackage{amsmath, amssymb, graphics, amsthm}
\usepackage{centernot}
\usepackage{epsfig}
\usepackage{color}
\usepackage{fancyhdr} 
\usepackage{bbold}
\usepackage[normalem]{ulem}

\usepackage{chngcntr}
\counterwithin{figure}{section}

\usepackage[numbers,sort&compress]{natbib}

\newcommand{\bra}[1]{\left\langle #1 \right|}

\newcommand{\ket}[1]{\left| #1 \right\rangle}

\newcommand{\braket}[2]{\left\langle \vphantom {#1 #2} #1 \hphantom{|} \right| \left. \vphantom {#1 #2} #2 \right\rangle}

\newcommand{\braopket}[3]{\left\langle \vphantom {#1 #2 #3} #1 \hphantom{|} \right| #2 \left| \hphantom{|} \vphantom {#1 #2 #3} #3 \right\rangle}

\makeatletter
\@addtoreset{equation}{section}
\makeatother

\newcommand{\be}{\begin{equation}}
\newcommand{\ee}{\end{equation}}
\newcommand{\ba}{\begin{eqnarray}}
\newcommand{\ea}{\end{eqnarray}}

\def\pb#1{\rlap{\lower1.5ex\hbox{$\longleftarrow$}}{#1}}
\def\dpb#1{\rlap{\lower1.5ex\hbox{$\Longleftarrow$}}{#1}}
\def\spb#1{\rlap{\lower1.0ex\hbox{$\leftarrow$}}{#1}}
\def\sdpb#1{\rlap{\lower1.0ex\hbox{$\Leftarrow$}}{#1}}

\title{{\sf An elementary introduction to loop quantum gravity}} 
\author{
{\sf N. Bodendorfer}\thanks{{\sf 
norbert.bodendorfer@fuw.edu.pl}}\\
\\
{\sf  Faculty of Physics, University of Warsaw, Pasteura 5, 02-093, Warsaw, Poland}\\
}
\date{{\small\sf \today}}

\begin{document} 

\maketitle

{\sf

\begin{abstract}
An introduction to loop quantum gravity is given, focussing on the fundamental aspects of the theory, different approaches to the dynamics, as well as possible future directions. It is structured in five lectures, including exercises, and requires only little prior knowledge of quantum mechanics, gauge theory, and general relativity. 
The main aim of these lectures is to provide non-experts with an elementary understanding of loop quantum gravity and to evaluate the state of the art of the field. Technical details are avoided wherever possible. 
\end{abstract}

}

\tableofcontents

\section{Why (loop) quantum gravity}

\subsection{Motivations for studying quantum gravity}

In this section, we are going to gather some motivations for conducting research in quantum gravity. The choice given here represents the personal preferences of the author and includes arguments that he considers to be especially compelling. Clearly, more arguments can be given in favour of studying quantum theories of gravity, and we do not mean to criticise them via our omissions here.

\begin{itemize}
	\item {\bf Geometry is determined by matter, which is quantised} \\
		The classical Einstein equations $G_{\mu \nu} = \frac{8\pi G}{c^4} T_{\mu \nu}$ tell us that geometry, described by the Einstein tensor $G_{\mu \nu} = R_{\mu \nu} - \frac 12 R g_{\mu \nu}$, is determined by the energy-momentum tensor $T_{\mu \nu}$. Quantum field theory on the other hand tells us that matter is quantised, i.e. the energy-momentum tensor becomes an operator $\hat T_{\mu \nu}$ on a Hilbert space. There are essentially two possibilities to reconcile this:
		\begin{enumerate}
		
			\item Also geometry has to be quantised, leading to a ``Quantum Einstein Equation'' of the form $\hat G_{\mu \nu} = \frac{8\pi G}{c^4} \hat T_{\mu \nu}$, or a similar formulation of the quantum theory, e.g. via path integrals, or as an embedding in a more general framework such as string theory.\\
		
			\item	Geometry stays classical and the energy-momentum tensor entering the Einstein equitations is an expectation value in a quantum state depending on the classical geometry. This approach is known as semiclassical gravity and interesting to consider even if one expects the first possibility to be realised in nature. 
		
		\end{enumerate}
		While the second approach seems to be a logical possibility, most researchers consider the first case to be more probable and the second as an approximation to it, including the author.

	\item {\bf Singularities in classical general relativity}\\
	Singularities appear generically in classical general relativity. The most famous ones are the cosmological singularity at the beginning of our universe, the ``big bang'', as well as the singularities at the centre of black holes. Appearance of singularities in physical theories usually signals the breakdown of the theoretical description and the need to go beyond the current framework. In particular, the strong gravitational fields close to such singularities signal that quantum effects might become relevant and alter the classical trajectories. An explicit example of this, curing the singularity, will be discussed in detail in the second lecture.

	\item{\bf Black hole thermodynamics}\\
	Classical black holes exhibit a very intriguing thermodynamic behaviour, and laws resembling the three laws of thermodynamics characterise their behaviour. In particular, one assigns an entropy to black holes which is proportional to their surface area. This observation signals that black holes could have microscopic constituents which are responsible for this entropy. These microscopic constituents are expected to be the degrees of freedom of a suitable quantum theory of gravity, and counting them should result in the black hole's entropy.

	\item {\bf Cutoff for quantum field theory}\\
	In quantum field theory, integrations in the evaluations of Feynman integrals usually lead to infinities, which have to be subtracted by suitable regularisation schemes. On the other hand, one would expect a fundamental theory of nature to be finite. It is conceivable that quantum gravity cures such infinities by providing a suitable physical cutoff, and examples of this have been given for example in string theory and loop quantum gravity. Another popular way to put this problem is to consider a (virtual) photon which exceeds Planck energy: its de Broglie wave length becomes shorter than the Planck length, and one would naively expect a black hole to form. To understand the details of what is happening here, one needs a quantum theory of gravity.

\end{itemize}

\subsection{Possible scenarios for observations}

\begin{itemize}

	\item {\bf Modified dispersion relations / deformed symmetries}\\
		Since quantum gravity is expected to alter spacetime at the Planck scale, it is plausible that quantum theories of gravity might modify the dispersion relations of matter propagating on these quantum spacetimes or deform symmetries such as Lorentz invariance. However, there exist strong bounds from experiments which are sensitive to such effects piling up over a long time or distance, such as observations of particle emission in a supernova \cite{LiberatiTestsOfLorentz}.

	\item {\bf Quantum gravity effects from black holes}\\
		While quantum gravity is believed to resolve the singularities inside a black hole, an observation of this fact is a priori impossible due to the horizons shielding the singularity. However, some scenarios for observations are conceivable: it is first of all possible that quantum gravity effects pile up over time and leak outside of the horizon, as the model in \cite{HaggardBlackHoleFireworks} shows. Associated phenomenology has been discussed in \cite{RovelliPlanckStars, BarrauFastRadioBursts}. On the other hand, it might be possible to observe signatures of evaporating black holes which were formed at colliders \cite{DimopoulosBlackHolesAt}, which however generally requires a lowering of the Planck scale in the TeV range, possibly due to extra dimensions \cite{Arkani-HamedTheHierarchyProblem}. 
		
	\item {\bf Cosmology}\\
	Quantum gravity is expected to provide a resolution for the big bang singularity at the beginning of our universe. In such a scenario, observational effects from the pre-inflationary era due to quantum gravity could still be detectable in the sky. A possible scenario was discussed in, e.g., \cite{AshtekarQuantumGravityExtension}, and more recent work can be found in \cite{AshtekarLoopQuantumCosmologyFrom}.

	\item {\bf Particle spectrum from unification}\\
		A theory of quantum gravity based on unification, such as string theory, can lead to observable effects also due to the prediction of additional matter fields at energies much below the Planck scale. Such scenarios are considered mainly in string theory and often include supersymmetry, see, e.g., \cite{LuestTheLHCString, MaharanaModelsOfParticle}.
	
	\item {\bf Gauge / Gravity}\\
		An indirect way of observing quantum gravity effects is via the gauge / gravity correspondence, which relates quantum field theories and quantum gravity. Via the gauge / gravity dictionary, phenomena happening in quantum gravity, i.e. beyond the classical gravity or classical string theory approximation of the correspondence, then have an analogue in quantum field theory which might be observable, or already observed, see, e.g., \cite{DenefQuantumOscillationsAnd, DenefBlackHoleDeterminants, Caron-HuotHydrodynamicLongTime}. 
	
\end{itemize}

\subsection{Approaches to quantum gravity}

In this section, we will list the largest existing research programmes aimed at finding a quantum theory of gravity, while unfortunately omitting some smaller, yet very interesting, approaches. A much more extensive account is given in \cite{OritiApproachesToQuantum}. For an historical overview, we recommend \cite{RovelliNotesForA}.

\begin{itemize}

	\item {\bf Semiclassical gravity}\\
Semiclassical gravity is a first step towards quantum gravity, where matter fields are treated using full quantum field theory, while the geometry remains classical. However, semiclassical gravity goes beyond quantum field theory on curved spacetimes: the energy-momentum tensor determining the spacetime geometry via Einstein's equations is taken to be the expectation value of the QFT energy-momentum tensor. The state in which this expectation value is evaluated in turn depends on the geometry, and one has to find a self-consistent solution. 
Many of the original problems of semi-classical gravity, see for example \cite{FlanaganDoesBackReaction}, have been addressed recently and the theory can be applied in practise, e.g. as. in \cite{DappiaggiStableCosmologicalModels}.

	\item {\bf Ordinary quantum field theory}\\
The most straight forward approach to quantising gravity itself is to use ordinary perturbative quantum field theory to quantise the deviation of the metric from a given background. While it turned out that general relativity is non-renormalisable in the standard picture \cite{GoroffQuantumGravityAt}, it is possible to use effective field theory techniques in order to have a well-defined notion of perturbative quantum gravity up to some energy scale lower than the Planck scale \cite{DonoghueTheEffectiveField}. Ordinary effective field theory thus can describe a theory of quantum gravity at low energies, whereas it does not aim to understand quantum gravity in extreme situations, such as cosmological or black hole singularities.

	\item {\bf Supergravity}\\
Supergravity has been invented with the hope of providing a unified theory of matter and geometry which is better behaved in the UV than Einstein gravity. As opposed to standard supersymmetric quantum field theories, supergravity exhibits a local supersymmetry relating matter and gravitational degrees of freedom. In the symmetry algebra, this fact is reflected by the generator of local supersymmetries squaring to spacetime-dependent translations, i.e. general coordinate transformations. 

While the local supersymmetry generally improved the UV behaviour of the theories, it turned out that also supergravity theories were non-renormalisable \cite{DeserNonrenormalizabilityOfLast}. The only possible exception seems to be $N=8$ supergravity in four dimensions, which is known to be finite at four loops, but it is unclear what happens beyond \cite{BernUltravioletBehaviorOf}. Nowadays, supergravity is mostly considered within string theory, where $10$-dimensional supergravity appears as a low energy limit. Due to its finiteness properties, string theory can thus be considered as a UV-completion for supergravity. Moreover, $11$-dimensional supergravity is considered as the low-energy limit of M-theory, which is conjectured to have the 5 different string theories as specific limits.

	\item {\bf Asymptotic safety}\\
The underlying idea of asymptotic safety \cite{ReuterQuantumEinsteinGravity} is that while general relativity is perturbatively non-renormalisable, its renormalisation group flow might possess a non-trivial fixed point where the couplings are finite. In order to investigate this possibility, the renormalisation group equations need to be solved. For this, the ``theory space'', i.e. the space of all action functionals respecting the symmetries of the theory, has to be suitably truncated in practise. 
Up to now, much evidence has been gathered that general relativity is asymptotically safe, including matter couplings \cite{DonaMatterMattersIn}, however always in certain truncations, so that the general viability of the asymptotic safety scenario has not been rigorously established so far. Also, one mostly works in the Euclidean. At microscopic scales, one finds a fractal-like effective spacetime and a reduction of the (spectral) dimension from $4$ to $2$ (or $3/2$, which is favoured by holographic arguments \cite{LaihoEvidenceForAsymptotic}, depending on the calculation). Moreover, a derivation of the Higgs mass has been given in the asymptotic safety scenario, correctly predicting it before its actual measurement \cite{ShaposhnikovAsymptoticSafetyOf}.

	\item {\bf Canonical quantisation: Wheeler-de Witt}\\
The oldest approach to full non-perturbative quantum gravity is the Wheeler-de Witt theory \cite{WheelerGeometrodynamics, DeWittQuantumTheoryOfI}, i.e. the canonical quantisation of the Arnowitt-Deser-Misner formulation \cite{ArnowittTheDynamicsOf} of general relativity. In this approach, also known as quantum geometrodynamics, one uses the spatial metric $q_{ab}$ and its conjugate momentum $P^{ab} = \frac{1}{2\kappa} \sqrt{q} \left(K^{ab} - q^{ab} K \right)$, where $K_{ab}$ is the extrinsic curvature, as canonical variables. 

The main problems of the Wheeler-de Witt approach are of mathematical nature: the Hamiltonian constraint operator is extremely difficult to define due to its non-linearity and a Hilbert space to support is is not known. It is therefore strongly desirable to find new canonical variables for general relativity in which the quantisation is more tractable. 
While the so called ``problem of time'', i.e. the absence of a physical background notion of time in general relativity, is present both in the quantum and the classical theory, possible ways to deal with it are known and continuously developed \cite{IshamCanonicalQuantumGravity, GieselScalarMaterialReference}.

	\item {\bf Euclidean quantum gravity}\\
In Euclidean quantum gravity, see \cite{HamberQuantumGravitation} for an overview, a Wick rotation to Euclidean space is performed, in which the gravity path integral is formulated as a path integral over all metrics. Most notably, this approach allows to extract thermodynamic properties of black holes. In practise, the path integral is often approximated by the exponential of the classical on-shell action. 
Its main problematic aspect is that the Wick rotation to Euclidean space is well defined only for a certain limited class of spacetimes, and in particular dynamical phenomena are hard to track. 

	\item {\bf Causal dynamical triangulations}\\
Causal dynamical triangulations (CDT) \cite{AmbjornQuantumGravityVia} is a non-perturbative approach to rigorously define a path integral for general relativity based on a triangulation. It grew out of the Euclidean dynamical triangulations programme, which encountered several difficulties in the 90's, by adding a causality constraint on the triangulations. The path integral is then evaluated using Monte Carlo techniques. The phase diagram of CDT in four dimensions exhibits three phases, one of which is interpreted as a continuum four-dimensional universe. Moreover, the transition between this phase and one other phase is of second order, hinting that one might be able to extract a genuine continuum limit.
More recently, also Euclidean dynamical triangulations has been reconsidered and evidence for a good semiclassical limit has been reported \cite{LaihoLatticeQuantumGravity}.

	\item {\bf String theory}\\
String theory was initially conceived as a theory of the strong interactions, where the particle concept is replaced by one-dimensional strings propagating in some background spacetime \cite{PolchinskiBook1, PolchinskiBook2, MukhiStringTheoryThe}. It was soon realised that the particle spectrum of string theory includes a massless spin 2 excitation, which is identified as the graviton. Moreover, internal consistency requirements demand (in lowest order) the Einstein equations to be satisfied for the background spacetimes. String theory thus is automatically also a candidate for a quantum theory of gravity. 
The main difference of string theory with the other approaches listed here is therefore that the quantisation of gravity is achieved via unification of gravity with the other forces of nature, as opposed to considering the problem of quantum gravity separately. 

The main problem of string theory is that it seems to predict the wrong spacetime dimension: 26 for bosonic strings, 10 for supersymmetric strings, and 11 in the case of M-theory. In order to be compatible with the observed $3+1$ dimensions at the currently accessible energies, one needs to compactify some of the extra dimensions. In this process, a large amount of arbitrariness is introduced and it has remained an open problem to extract predictions from string theory which are independent of the details of the compactification. Also, our knowledge about full non-perturbative string theory is limited, with the main exceptions of D-branes and using AdS/CFT as a definition of string theory.

	\item {\bf Gauge / gravity}\\
The gauge / gravity correspondence \cite{AmmonGaugeGravityDuality}, also known as AdS/CFT, has grown out of string theory \cite{MaldacenaTheLargeN, GubserGaugeTheoryCorrelators, WittenAntiDeSitter}, but was later recognised to be applicable more widely. Its statement is a (in most cases conjectured) duality between a quantum gravity theory on some class of spacetimes, and a gauge theory living on the boundary of the respective spacetime. Once a complete dictionary between gravity and field theory computations is known, one can in principle use the gauge / gravity correspondence as a definition of quantum gravity on that class of spacetimes. 

The main problem of gauge / gravity as a tool to understand quantum gravity is the lack of a complete dictionary between the two theories, in particular for local bulk observables. Also, it is usually very hard to find gauge theory duals of realistic gravity theories already at the classical gravity level (i.e. in an appropriate field theory limit), and many known examples are very special supersymmetric theories.

	\item {\bf Loop quantum gravity}\\
Loop quantum gravity \cite{RovelliQuantumGravity, ThiemannModernCanonicalQuantum, GambiniAFirstCourse} originally started as a canonical quantisation of general relativity, in the spirit of the Wheeler-de Witt approach, however based on connection variables parametrising the phase space of general relativity, e.g. the Ashtekar-Barbero variables \cite{AshtekarNewVariablesFor, BarberoRealAshtekarVariables} based on the gauge group SU$(2)$. The main advantage of these variables is that one can rigorously define a Hilbert space and quantise the Hamiltonian constraint. The application of the main technical and conceptual ideas of loop quantum gravity to quantum cosmology resulted in the subfield of loop quantum cosmology \cite{AshtekarLoopQuantumCosmology, BollietConceptualIssuesIn}, which offers a quantum gravity resolution of the big bang singularity and successfully makes contact to observation. 

The main problem of loop quantum gravity is to obtain general relativity in a suitably defined classical limit. In other words, the fundamental quantum geometry present in loop quantum gravity has to be coarse grained in order to yield a smooth classical spacetime, while the behaviour of matter fields coupled to the theory should be dictated by standard quantum field theory on curved spacetimes in this limit. The situation is thus roughly the opposite of that in string theory. Also, it has not been possible so far to fully constrain the regularisation ambiguities that one encounters in quantising the Hamiltonian constraint. In order to cope with these issues, a path integral approach, known as spin foams \cite{PerezTheSpinFoam}, has been developed, as well as the group field theory approach \cite{BaratinTenQuestionsOn}, which is well suited for dealing with the question of renormalisation.

\end{itemize}

\subsection{General comments on the canonical quantisation programme}

The aim of this lecture series is to provide an introduction to the canonical formulation of loop quantum gravity. As a starting point, it is instructive to consider the canonical quantisation programme quite abstractly and highlight where choices have to be made, and which important steps have to be taken. 

\begin{enumerate}

	\item {\bf Hamiltonian formulation}
	
	As a first step, a Hamiltonian formulation of general relativity has to be constructed. For this, one needs to foliate the spacetime manifold into equal time hypersurfaces, which restricts the allowed spacetimes to be globally hyperbolic.
	This is for example achieved by the Arnowitt-Deser-Misner formulation \cite{ArnowittTheDynamicsOf}, in which the spatial metric $q_{ab}$ and its momentum $P^{ab}$ are the conjugate variables. 
	
	In the process of the Legendre transform, one finds four independent constraints per point on the spatial slice, the Hamiltonian constraint $\mathcal H\approx 0$ and the vector constraints $\mathcal H_a \approx 0$. While the vector constraint simply generates spatial diffeomorphisms, the action of the Hamiltonian constraint is more complicated, as it includes the dynamics of the theory and generates time-like diffeomorphisms only on-shell. Together, these constraints generate the hypersurface deformation algebra, also known as Dirac algebra. 
	The constraints can either (partially) be solved classically, thereby reducing the free phase space coordinates per point to $2+2$, or they can be quantised and subsequently solved. Similarly, one can also enlarge the phase space prior to quantisation and account for the new degrees of freedom by adding constraints.

	\item {\bf Choice of a preferred Poisson-subalgebra}
	
	Before quantisation, we are free to choose a different set of canonical variables to describe the gravitational phase space. 
	An example would be to use SU$(2)$ connection variables in $3+1$ dimensions, the Ashtekar-Barbero variables \cite{AshtekarNewVariablesFor, BarberoRealAshtekarVariables}, which forces us to add an additional Gau{\ss} constraint to the theory. 
	
	Since we cannot quantise all functions on the classical phase space, we have to choose a preferred subalgebra which we later want to represent as operators. This subalgebra should be large enough to separate points on the phase space.
	For example, this subalgebra could include all holonomies that we can construct from a connection parametrising the phase space.

	\item {\bf Hilbert space representation}
		
		We now look for a representation of our preferred subalgebra of phase space functions on a Hilbert space so that the reality conditions of the classical theory are implemented as $\widehat {\mathcal O^*} = \widehat{\mathcal O}^\dagger$ for any $\mathcal O$ in the preferred subalgebra. 
		This Hilbert space is usually denoted as ``kinematical'' if we still have constraints left to impose. 
		
		An important question is to investigate whether the representation is unique under some natural assumptions, since this is generically not the case for infinitely many degrees of freedom. 
	
	\item {\bf Imposing constraints }
	
		If the constraints have not been solved at the classical level, we need to impose them on the kinematical Hilbert space. This process might be gradual, i.e. we could first solve some part of the constraints, leading to a new Hilbert space, and then impose the remaining constraints. It is often the case that solutions to the constraints are not normalisable w.r.t. the kinematical scalar product, so that a new scalar product has to be introduced. This process is for example formalised in the refined algebraic quantisation programme \cite{GiuliniOnTheGenerality, GiuliniGroupAveragingAnd}. 
		
		In this process, attention has to be payed to possible anomalies in the quantum constraint algebra. In particular, the structure functions in the Dirac algebra make this a very difficult task.

	\item {\bf Physical Hilbert space and observables}
	
		Finally, the physical Hilbert space is obtained after all constraints are solved and a suitable scalar product on the space of their solutions has been constructed. It then remains to understand the physical content of the theory, i.e. to extract physics from the gauge invariant ``time-less'' observables. 

\end{enumerate}

\subsection{Arguments for canonical loop quantum gravity}

We will now present a list of arguments for studying loop quantum gravity as a specific approach to quantum gravity. The selection is according to the preferences of the author and thus subject to personal bias. It is neither meant to be exhaustive, nor to imply that these arguments cannot be used in favour of other approaches. 

Before starting, let us summarise in a few sentences the result of the quantisation procedure which starts with a classical reformulation of general relativity in terms of connection variables. The Hilbert space of the theory is spanned by so-called spin network functions, which can be roughly interpreted as spatial lattices labelled with quantum numbers encoding the geometry. Matter fields can be coupled to the theory similarly as in lattice gauge theory. The quantum dynamics evolves those lattices and matter in particular backreacts on them. Another main difference to standard lattice gauge theory is that one can consider superpositions of lattices. 

\begin{itemize}
	\item {\bf No extra structure} \\
		The framework of loop quantum gravity does, for the most part, not introduce any extra structure on top of general relativity and quantum field theory. It follows well established canonical quantisation techniques, based on a specific choice of connection variables and ideas which have been successful in lattice gauge theory. 
						
	\item {\bf Background independence is taken seriously}\\
	No expansion around a certain background spacetime is employed, and background independence is at the core of the quantum theory. While this poses certain technical challenges, it upholds the key lesson of general relativity. In the resulting theory, matter fields are really living on the non-perturbative quantum geometry, as opposed to the perturbative graviton just being another particle. 
	
	\item {\bf Quantum geometry as a UV regulator} \\
	The quantum geometry appearing in loop quantum gravity naturally cuts off UV divergences \cite{ThiemannQSD5} due to discrete spectra of geometric operators and background independence. The Hamiltonian (constraint) of the theory, including standard model matter, is a finite and well defined operator without the need of renormalisation. Loop quantum gravity is thus a candidate for a rigorous definition of quantum field theory.

	\item{\bf Entanglement = geometry} \\
	The quantum geometry described by loop quantum gravity suggests and deep relation between geometry and the entanglement entropy of the gravitational field. Quanta of area always are accompanied by quanta of entanglement \cite{HusainApparentHorizonsBlack, DonnellyEntanglementEntropyIn, BII}.

	\item{\bf Freedom in choice of variables} \\
	In the Hamiltonian approach, there is a large freedom to choose suitable canonical variables before quantisation. In particular, they do not need to be pullbacks of variables in a covariant action principle. By choosing suitable variables, certain problems such as symmetry reductions \cite{BLSI, BIII, BZI, BVI} or the impositions of certain gauge fixings \cite{BSTI, BLSII} can be achieved much more efficiently than with standard variables.

\end{itemize}

\subsection{Criticism of canonical loop quantum gravity} \label{sec:CriticismOfLoops}

In this section, we will gather some criticism which has been expressed towards loop quantum gravity and comment on the current status of the respective issues. Again, the points raised here are the ones most serious in the view of the author, and different lists and opinions could be expressed by others.

\begin{itemize}
	\item {\bf Obtaining general relativity in the appropriate limit} \\
	The kinematics and dynamics of loop quantum gravity are defined in an ultra high energy quantum gravity regime, where the usual notion of continuum spacetime or the idea of fields propagating on a background do not make sense any more. It is thus of utmost importance to understand how general relativity and quantum field theory on curved spacetime emerges in a suitable limit, and what the quantum corrections are. There does not seem to complete agreement on how such a limit should be constructed, and depending on the route chosen, one finds different statements about the status of this endeavour in the literature. The level of complexity of this task can be compared to having a theory of atoms and the aim to compute the properties of solids. 
	
	In order to understand the current status, let us remark that there are two limits which can be taken in order to obtain a spacetime of large scale in loop quantum gravity: large quantum numbers (spins), or many quantum numbers, i.e. very fine spin networks. Much is known about the limit of large spins, where the number of quanta is fixed. Here, one finds strong evidence that the theory reproduces general relativity on large scales both in the canonical \cite{GieselAQG2, GieselAQG3} and in the covariant approach \cite{BarrettLorentzianSpinFoam, HanAsymptoticsOfSpinfoam}. The resulting semiclassical picture however corresponds to Regge-gravity on a given lattice which is specified by the underlying graph on which the (coherent) quantum state labelled by the large spins lives. This limit is usually referred to as the ``semiclassical'' limit in the literature and should not be confused with the following:
	
	On the other hand, one can leave the quantum numbers arbitrary, in particular maximally small, and only increase the number of quanta. This corresponds to a continuum limit and it should be a priori preferred over the large spin limit in the opinion of the author. In particular, the large spin and continuum limit do not need to commute and may in principle lead to different physics, even on macroscopic scales. The problem with this approach is that we know only very little about the dynamics in this sector of the theory, apart from several concrete proposals for its implementation. The dynamics on large scales is then expected to emerge via a coarse graining procedure. More discussion on this point of view can be found in \cite{DittrichTheContinuumLimit, BahrOnBackgroundIndependent}, see also \cite{BahrNumericalEvidenceFor} for recent progress on the related issue of renormalisation. The dynamical emergence of a low spin sector is discussed in \cite{GielenEmergenceOfA} within group field theory. Interesting recent work on constructing quantum states with long range correlations found in quantum field theory can be found in \cite{BianchiSqueezedVacuaIn}. 
	
	As an additional subtlety, we can consider arbitrary superpositions of spin networks, in other words we can have quantum superpositions of ``lattices''. The impact on the dynamics of this feature is so far unclear and may strongly depend on the regularisation details of the Hamiltonian, e.g. whether it is graph preserving or not, and thus superselecting.
	Especially here it becomes clear that while individual basis elements in the Hilbert space can be given a certain interpretation as discrete geometries, generic states could behave quite differently and conclusions based on certain lattice-like truncations of the full theory need to be taken with great care. 
	 Concerning a limit to obtain quantum field theory on curved spacetimes, we point out the pioneering works \cite{SahlmannTowardsTheQFT, SahlmannTowardsTheQFT2, StottmeisterCoherentStatesQuantumI, StottmeisterCoherentStatesQuantumII, StottmeisterCoherentStatesQuantumIII}. 
	
	To conclude, the situation of whether general relativity emerges in the continuum limit is so far unclear, whereas there is strong evidence for a Regge-truncation thereof emerging in the large spin limit. Whether one is satisfied with one or the other limit, or a combination of both, also depends on the following problem.	
			
	\item {\bf Local Lorentz invariance}\\
	It is sometimes suggested that loop quantum gravity is not locally Lorentz invariant in the sense that modified dispersion relations might arise which could be in conflict with observation. Unfortunately, our current understanding of loop quantum gravity does not allow us to answer whether there are Lorentz violations, and how severe they might be. In order to make a meaningful statement, one would essentially have to identify a quantum state corresponding to Minkowski space, which should be thought of as a (possibly infinite) superposition of lattices, put matter fields thereon, and track their dynamics, including back-reaction, on a coarse grained scale where geometry can already be considered smooth. This is a formidable task, and currently out of reach not because of lacking proposals for how to define the involved structures, but mainly due to computational complexity. 	
	
	In order to judge certain statements that one might find in the literature or on the internet, one should keep the following in mind to avoid confusion:
	
	\begin{itemize}
	
		\item {\bf Discrete eigenvalues of geometric operators $\centernot \implies$ Lorentz violations}
		
		One might naively think that discrete eigenvalues of geometric operators violate special relativity: if an observer at rest measures a certain discrete eigenvalue of, say, an area, what does another observer measure who is not at rest? The short answer is that he might in principle measure any value for the area, as long as it is in the (discrete) spectrum of the area operator. However, the expectation value can still transform properly according to special relativity. A well-known example of this is the theory of angular momentum: while the eigenvalues of components of the angular momentum are always (half)-integers, expectation values  transform properly according to the continuous rotation symmetry. This point has been made for example in \cite{RovelliReconcilePlanckScale}, with further discussion in \cite{LivineAboutLorentzInvariance}, where the ideas of \cite{RovelliReconcilePlanckScale} are verified in the context of a simplified toy model related to 3d Euclidean LQG. Similar conclusions are also drawn in other contexts \cite{SnyderQuantizedSpaceTime, DowkerQuantumGravityPhenomenology}.

		\item{\bf Internal gauge groups do not determine isometries of the spacetime}\\
		Different formulations of loop quantum gravity, canonically or covariant, use different internal gauge groups. While the covariant path integral formulations of the Lorentzian theory use either SL$(2,\mathbb C)$ \cite{RovelliLorentzCovarianceOf} or SU$(2)$ \cite{EngleLQGVertexWith} in a gauge-fixed version, the Lorentzian canonical theory in $3+1$ dimensions can be formulated using either SU$(2)$ \cite{BarberoRealAshtekarVariables}, SO$(1,3)$ \cite{AlexandrovHilbertSpaceStructure, CianfraniTowardsLoopQuantum, GeillerALorentzCovariant, BTTI}, or SO$(4)$ \cite{BTTI}. This is because one is coding the spatial metric and its momentum in a connection, whereas the signature of spacetime in the canonical formalism is determined by the Hamiltonian constraint, more precisely a relative sign between two terms. In fact, the structure of spacetime, coded in the hypersurface deformation algebra, is already set at the level of metric variables, and completely insensitive of the additional gauge redundancy that one introduces by passing to connection variables. Also, it does not matter for this whether the connection that one uses can be interpreted as the pullback of some manifestly covariant spacetime connection. 
		While it is a possibility that only a certain choice of variables or internal gauge group leads to a consistent quantum theory in agreement with current bounds on Lorentz violations, such a conclusion cannot be drawn given our current understanding of the theory. 
		
		\item {\bf Expectations for possible Lorentz violations} \\
		In order to parametrise the violation of Lorentz invariance in a model-independent way, one usually constrains the free parameters $c_n$ in a modified dispersion relation such as
		\be
			E^2 = m^2 + p^2 + \sum_{n\geq 3} c_n \frac{p^n}{E_{\text{pl}}^{n-2}} \text{,} \label{eq:Dispersion}
		\ee
		where $E_{\text{pl}} = \sqrt{\hbar c / G}$ is the Planck energy. It is worthwhile to formulate a general expectation at what order one expects the first quantum corrections to appear.
		While an appropriate calculation as outlined above will finally have to decide this question, we can still try to extrapolate from our current understanding of loop quantum gravity at what order effects could occur. 
		At the moment, the best understood setting is homogeneous and isotropic loop quantum cosmology, where we know how to compute the quantum corrections to the classical theory. We obtain a correction of the order $\hbar$, i.e. the effective Friedmann equation 
		\be
			\left( \frac{\dot a}{a} \right)^2 \sim \rho \left(1-\frac{\rho}{\rho_{\text{crit}}} \right), ~~~ \rho_{\text{crit}} \sim \frac{1}{G^2 \hbar} \text{.} \label{eqn:EffectiveFriedmann}
		\ee
		If we furthermore invoke anomaly freedom of the effective constraints for inhomogeneous perturbations in cosmology \cite{CailleteauAnomalyFreeScalar, CailleteauAnomalyFreePerturbations} or in the spherically symmetric setting \cite{BojowaldDiscretenessCorrectionsAnd}, we obtain a universal deformation $\beta \neq 1$ of the propagation equation for gravity and matter as
		\be
			\frac {\partial^2 \phi^2}{\partial t^2} - \frac{\beta}{a^2} \Delta \phi = S[\phi] \label{eq:ModifiedWaveEquation}
		\ee
		where $S[\phi]$ contains source and lower derivative terms and $\beta$ depends on the spatial metric and the extrinsic curvature, see \cite{BarrauAnomalyFreeCosmological, BojowaldSomeImplicationsOf} for an overview. In the simplest case of holonomy corrections, we have $\beta = 1- 2 \rho / \rho_\text{crit}$. While the speed of light is affected in this context, there is no dependence so far on the energy of the particle, as the energy density $\rho$ is that of the background. For the purpose of our estimate, we can however include a qualitative form of backreaction by adding to the background energy density the energy density of the particle under consideration, say a photon of frequency $\omega$ and wave length $\lambda = 2\pi / \omega$. We have $\rho_\text{Photon} \sim E_\text{Photon} / \lambda_\text{Photon}^3 = \hbar \omega^4 / (2 \pi)^3$. Using this most naive approach, we estimate
		\be
			1-\beta \sim \frac{\rho_\text{Photon}}{\rho_\text{crit}} \sim \hbar^2 G^2 \omega^4 \sim \frac{E_\text{Photon}^4}{E_\text{Planck}^4} \text{.}
		\ee
		Therefore, we would expect that a frequency dependent speed of light would not occur before order $n=6$, which seems to be unconstrained so far \cite{LiberatiTestsOfLorentz}, see however also \cite{CollinsLorentzInvarianceAnd, BelenchiaLorentzViolationNaturalness}. It is important to mention again that this estimate is based on a very naive inclusion of backreaction and that the issue of anomaly freedom of the constraint algebra entering the derivation of \eqref{eq:ModifiedWaveEquation} is not well understood in the full theory so far. Therefore, these estimates need to be taken with great care and quantitative predictions should definitely not be drawn from them. However, they show that while it is natural to expect violations of Lorentz invariance from a theory of quantum gravity, these violation can be heavily suppressed because there are two powers of $\hbar$ entering, one from the quantum gravity effect causing the change in propagation speed, and the other from the particle contributing to the energy density determining the magnitude of the effect.

		Experimentally, it turns out that there are very strong constraints on the $n=3$ terms, as well as quite strong constraints on $n=4$ \cite{LiberatiTestsOfLorentz}. The fact that $n=6$ seems to be favoured by our analysis here points to an interesting window for observable effects without apparent conflict to existing observations.

	\end{itemize}

	\item {\bf Testability and ambiguities} \\
	A general problem for theories of quantum gravity is to come up with testable predictions which are independent of free parameters in the theory that can be tuned in order to hide any observable effect below the measurement uncertainty. Furthermore, one would like to have a uniquely defined fundamental theory whose dynamics depend only on a finite number of free parameters. 
	
	In loop quantum cosmology, much progress towards predictions for observable effects have been made \cite{AgulloThePreInflationary, AshtekarLoopQuantumCosmologyFrom, BollietObservationalExclusionOf}, however the choice of parameters in the models, such as the e-foldings during inflation, can so far hide any observable effects. Also, there are different approaches to the dynamics of cosmological perturbations \cite{BollietConceptualIssuesIn}. 
	
	Within full loop quantum gravity (and also loop quantum cosmology), the Barbero-Immirzi parameter $\beta$ \cite{BarberoRealAshtekarVariables, ImmirziQuantumGravityAnd}, a free real parameter entering, e.g., the spectra of geometric operators, constitutes a famous ambiguity. It enters the classical theory in a canonical transformation which is not implementable as a unitary transformation or even an algebra automorphism at the quantum level and thus constitutes a quantisation ambiguity. Since this parameter can enter physical observables, one would like to fix it by an experiment or derive its (only consistent) value by theoretical means. An example for how to fix it by theoretical means is to consider black hole entropy computations and match them to the expected Bekenstein-Hawking entropy. Within the original approach to black hole entropy from loop quantum gravity, this gives a certain value for $\beta$ \cite{AshtekarQuantumGeometryOf}, however this computation neglects a possible running of the gravitational constant, i.e. it identifies the high and low energy Newton constants. More recently, it has been observed that the Bekenstein-Hawking entropy can also be reproduced by an analytic continuation of $\beta$ to the complex self-dual values $\pm i$ \cite{FroddenBlackHoleEntropy}, which interestingly also agrees with a computation of the effective action \cite{BNI}. In fact, the classical theory is easiest when expressed in self-dual variables, so that it might turn out that the value $\beta = \pm i$ could be favoured also in the quantum theory. The current problem is however that the quantum theory is ill-defined for complex $\beta$ and the above mentioned results were obtained via analytic continuation from real beta.

	In addition to quantisation ambiguities resulting from the choice of variables, as above, there are quantisation ambiguities in the regularisation of the Hamiltonian constraint, and thus the dynamics of the theory. These go somewhat beyond factor ordering, as the techniques used in regularising the Hamiltonian also involve classical Poisson bracket identities which are used to construct otherwise ill-defined operators \cite{ThiemannQSD1}. The requirement of anomaly freedom of the quantum constraint algebra already removed many of those ambiguities in Thiemann's original construction \cite{ThiemannQSD1}. The precise notion of anomaly freedom used in \cite{ThiemannQSD1} has been criticised on the ground that it corresponds to a certain ``on-shell'' notion \cite{NicolaiLoopQuantumGravity}, which is however the one that makes sense in the context of \cite{ThiemannQSD1}. More recent work based on a slightly changed quantisation seems to make significant progress towards the goal of implementing a satisfactory ``off-shell'' version of the quantum constraint algebra, at least in simplified toy models, \cite{LaddhaTheDiffeomorphismConstraint, TomlinTowardsAnAnomaly, HendersonConstraintAlgebraInI, LaddhaHamiltonianConstraintIn}. The possible regularisations then seem to be very constrained, although no unique prescription has emerged so far. 
	
	To conclude, it is so far not possible to extract definite predictions from full loop quantum gravity due to the existence of ambiguities in its construction, whose influence on the dynamics is not well understood. However, there is promising recent work on the removal of such ambiguities, which could eventually lead to clear predictions which would allow to falsify the theory.

	\item {\bf Quantising hydrodynamics} \\
	It is possible that the dynamics of classical general relativity arise in a thermodynamic limit from some more fundamental theory. This possibility was in particular stressed by Jacobson in \cite{JacobsonThermodynamicsofSpacetime}, where the Einstein equations are derived as an equation of state under some natural assumptions. If such a scenario is realised in nature, then it is questionable whether one should directly quantise general relativity to obtain a fundamental theory of quantum gravity, as e.g. done in canonical LQG, or whether one should simply try to find a fundamental theory that reduces to general relativity in a suitable limit. 
	
	This criticism is certainly warranted to a certain degree. However, taking it at face value, it makes the task of ``guessing'' the correct fundamental theory from scratch very difficult. Instead, one can approach the issue of the quantum dynamics in loop quantum gravity from a naturalness point of view, as e.g. done in group field theory or spinfoam models. Here, one takes only the kinematical setting emerging from the canonical quantisation, selects the quantum dynamics by appealing to simplicity, and checks whether they yield the Einstein equations in a suitable limit. In fact, the kinematical structure of loop quantum gravity is consistent with the basic ingredients of Jacobson's derivation \cite{ChirocoSpacetimeThermodynamicsWithout}. It should also be stressed that even if the scenario of \cite{JacobsonThermodynamicsofSpacetime} is realised, then one could still learn from quantising gravity: in an analogy to solid state physics, instead of obtaining the ``atoms of space'', one would obtain ``phonons of space'', which still could provide valuable insights.
	
	To conclude, research in loop quantum gravity makes the assumption that we can learn about the microscopic degrees of freedom of quantum gravity by quantising diffeomorphism invariant theories in a background independent way. 
	While this strategy may turn out to be misguided, it may also be very hard to find the correct microscopic description without any such hints.

\end{itemize}

\subsection{Exercises}

Read about what other people have to say about quantum gravity and make up your own mind.

\section{Elements of loop quantum gravity through cosmology} \label{sec:LQC}

In this section, we will introduce some of the main aspects of loop quantum gravity at the example of a simple cosmological model, following \cite{AshtekarRobustnessOfKey}. We consider a homogeneous and isotropic spacetime in the presence of a massless and minimally coupled scalar field. Already with this simple example, many important features of loop quantum gravity are visible and can be grasped much easier than in the context of full general relativity, which involves many more technicalities. We set $c = \hbar = 12 \pi G = 1$ ($12 \pi G = 1$ instead of the usual $8 \pi G = 1$ gives the biggest simplification here). 

The introductory lectures in \cite{AshtekarAnIntroductionTo} give a far more complete account of the subject of this section and we refer the interested reader there for more details. In this section, we mainly follow \cite{AshtekarRobustnessOfKey} (with a few changes incorporating e.g. a more natural choice of physical scalar product and Dirac observables based on \cite{KaminksiQuantumConstraintsDirac}, see also \cite{CraigDynamicalEigenfunctionsAnd}, and \cite{BojowaldLargeScaleEffective, MielczarekExactSolutionsFor} for other exactly soluble models). Seminal papers on loop quantum cosmology include \cite{BojowaldAbsenceOfSingularity, AshtekarMathematicalStructureOf, AshtekarQuantumNatureOf}. More recent developments of the subject are summarised in \cite{AshtekarLoopQuantumCosmology, BollietConceptualIssuesIn}. Since this is only a short presentation of a specific result, we cannot discuss all the motivation and insights that were necessary to arrive at the final picture sketched here, and we refer the interested reader to the cited literature.  

As a word of caution, we need to mention that the qualitative picture obtained in this section may not survive a more detailed analysis when inhomogeneities are taken into account. In particular, invoking anomaly freedom of the algebra of effective constraints, one arrives at a scenario where the signature of spacetime is changing before the bounce that replaces the big bang happens, see \cite{BojowaldSomeImplicationsOf} for an overview.  These issues are however subject to current research and no final conclusions have been drawn.

\subsection{Preliminaries}

We will leave a more detailed look at full general relativity for the next lecture. For now, let just recall that the Einstein equations describe how matter influences the geometry of spacetime. Spacetime is described by a Lorentzian $4$-metric $g_{\mu \nu}$, with $\mu, \nu = 0,1,2,3$. In the case of a homogeneous and isotropic spacetime, we may write the line element associated to the metric as
\be
	ds^2 = - N^2 dt^2 + a(t)^2 \left(dx^2 + dy^2 + dz^2\right) \text{,}
\ee
where $N$ is the lapse function. Choosing $N$ is a matter of gauge,  and we will set it to $1$ in the following section on the classical dynamics, meaning that $t$ measures proper time. For quantisation, the most convenient choice is $N = a^3$, as it renders the theory exactly solvable. 
Let us pass now to the Hamiltonian description of this system. Again, we will leave details for the next lecture and simply postulate now that the resulting Hamiltonian for homogeneous and isotropic cosmology minimally coupled to a massless scalar field $\phi$ has the form 
\be
	 H = N \left(\frac{P_\phi^2}{2 v} - \frac{b^2 v}{2} \right)
\ee	
where $b= -3 \dot a / a$, $v = a^3 V_0$, $P_\phi = - a^3 \dot \phi V_0$, and $V_0$ is the fiducial\footnote{$V_0$ comes about since the Hamiltonian is actually an integral of the $3$-dimensional spatial slice. Here, all densities (in the differential geometric sense) are multiplied by $V_0$, i.e. integrated, whereas scalars such as $\phi$ are simply independent of the spatial coordinate. Since $V_0$ needs to be finite, we could work with a fiducial cell in the case of a non-compact universe, but will not look at those details here.} (coordinate) volume of the universe. The interpretation of $v$ is thus to measure the volume of the universe. The non-vanishing Poisson brackets are
\be
	\{ v,b\} = 1 ~~ \text{and} ~~ \{ \phi,P_\phi \} = 1 \text{.} \label{eq:CosPB}
\ee

A peculiar feature of general relativity is that its Hamiltonian is constrained to vanish; we write $H \approx 0$ following \cite{DiracLecturesOnQuantum}. This ``weak equality'' means that we can use $H = 0$ only at the end of computations, in particular after evaluating Poisson brackets. In turn, $H$ generates gauge transformations. In our case, this is the freedom of choosing a time coordinate. Thus, it is not a gauge invariant statement to ask for the volume $v$ of the universe at a certain time $t$. Rather, we should ask for the value of $v$ when some other event takes place, e.g. when the scalar field $\phi$ takes a certain value. We will come back to this issue soon. 
In the next lecture, this gauge invariance will be generalised to also include the freedom to choose spatial coordinates, which is (largely) fixed in the current context by our choice of metric.

\subsection{Classical dynamics}

Let us derive and solve the equations of motion for $N=1$. Hamilton's equations give 
\begin{align}
	\dot v &= \{ v, H \} = - bv  &\dot b &= \{ b, H \} =  \frac{P_\phi^2}{2 v^2} + \frac{1}{2} b^2  \label{eq:Hamvb} \\
	\dot \phi &= \{ \phi, H \} = \frac{P_\phi}{v} &\dot P_\phi &= \{ P_\phi, H \} = 0 \text{.}  \label{eq:Hamphi} 
\end{align}
Furthermore, we can now use
\be
	H = 0 ~~~ \Leftrightarrow ~~~ P_\phi^2 =  b^2 v^2 \text{.} \label{eq:HgN}
\ee
First, we note that $P_\phi$ is a constant of motion. We insert \eqref{eq:HgN} into \eqref{eq:Hamvb} to obtain $\dot b = b^2$, giving $b = \frac{-1}{t-t_0}$. Inserting this into the first equation of \eqref{eq:Hamvb} gives $\dot v = \frac{v}{t-t_0}$, leading to
\be
	v = \pm |P_\phi| (t-t_0) \text{.} \label{eqn:Solv}
\ee
The sign of $v$ is not determined by the equations of motion and it corresponds to an arbitrary choice of orientation of the manifold. In the classical theory, it would be most natural to restrict to positive $v$, but in the quantum theory we will need impose a symmetry condition on the wave function relating positive and negative $v$. 
Already here we see that we reach $v = 0$ within finite proper time. $v=0$, i.e. $t=t_0$, can be identified as the big bang singularity, or a big crunch in the time reversed picture. 

Next, let us solve the equations of motion for the scalar field. Inserting \eqref{eqn:Solv} into \eqref{eq:Hamphi}, we find
\be
	\dot \phi = \pm  \text{Sign}(P_\phi) \frac{1}{t-t_0} ~~~ \Leftrightarrow ~~~ \phi - \tilde \phi_0 = \pm \text{Sign}(P_\phi )   \,\text{Log} \left(  t-t_0 \right) \text{.}
\ee
We can thus express $v$ as a function of $\phi$ and $P_\phi$ as
\be
	v = \pm \exp \left( \pm  \text{Sign}(P_\phi) (\phi - \phi_0) \right) \text{.}
\ee
with some new constant $\phi_0$. Next to $P_\phi$, this also suggests another function commuting with the constraint, a so called Dirac observable, 
\be
	\left. v \right|_{\phi = \tilde \phi} :=    v \, \exp \left(  \mp \text{Sign}(P_\phi) (\phi- \tilde \phi)  \right) \label{eq:ObsNonP} \text{.}
\ee
$\left. v \right|_{\phi = \tilde \phi}$ is simply the value that the volume $v$ of the universe takes at scalar field time $\tilde \phi$. The two independent Dirac observables are thus \eqref{eq:ObsNonP} as well as $P_\phi$, leaving us with $1+1$ phase space degrees of freedom. Physically, this means that at some point $\tilde \phi$ in scalar field time, we can fix $v$ and $P_\phi$. 

As we have seen, the evolution predicted by general relativity leads to a singularity in this simple model. While it was initially believed that such a phenomenon was due to the high level of symmetry involved, it was later shown that singularities occur generically \cite{SenovillaSingularityTheoremsIn}. It is believed that a quantum theory of gravity could provide a natural resolution of singularities by quantum effects. In fact, the energy density close to the singularity surpasses the Planck density and quantum gravitational effects should become relevant. In the following, we will see how the initial singularity can be resolved by quantum effects within loop quantum cosmology, however persists in the metric based Wheeler-de Witt approach.

\subsection{Wheeler-de Witt quantisation}

We continue to follow \cite{AshtekarRobustnessOfKey}, up to some minor changes in the choice of Dirac observables and scalar products.
The Poisson brackets \eqref{eq:CosPB} along with the quantisation rule $\{ \cdot, \cdot \} \rightarrow - i [\cdot, \cdot ]$ suggest that we look for self-adjoint operators satisfying
\be
	[ \hat v, \hat b ] =  i ~~ \text{and} ~~ [\hat  \phi, \hat P_\phi ] =  i \text{.} \label{eq:CosCom} 
\ee
Using wave functions $\chi(b)$, this can be done for example as
\begin{align}
	&\hat b \chi(b, \phi)  = b \chi(b, \phi) &&& \hat \phi  \chi(b, \phi) & = \phi \chi(b, \phi) \\
	&\hat v \chi(b, \phi)  =  i \frac{\partial}{\partial b}  \chi(b, \phi) &&& \hat P_\phi  \chi(b, \phi) & = -  i \frac{\partial}{\partial \phi}  \chi(b, \phi) \text{.} \label{eq:OpsWdWChi2}
\end{align}
on the Hilbert space $\mathcal L^2(\mathbb R^2, db\, d\phi)$.
With the necessary hindsight, we rescale our wave functions as $\Psi(b) = \sqrt{|b|} \chi(b)$. The previous operators become
\begin{align}
	&\hat b \Psi(b, \phi)  = b \Psi(b, \phi) &&& \hat \phi  \Psi(b, \phi) & = \phi \Psi(b, \phi) \\
	&\hat v \Psi(b, \phi)  = \sqrt{|b|} i \frac{\partial}{\partial b} \frac{1}{\sqrt{|b|}} \Psi(b, \phi) &&& \hat P_\phi  \Psi(b, \phi) & = -  i \frac{\partial}{\partial \phi}  \Psi(b, \phi) \text{.} \label{eq:OpsWdWPsi2}
\end{align}
on the Hilbert space $\mathcal L^2(\mathbb R^2, db / |b| \, d\phi)$.
Next, we quantise the Hamiltonian constraint $H = 0$, which we do in the form \eqref{eq:HgN}, i.e. for $N = a^3$ with the symmetric ordering $\sqrt{|b|} v |b| v \sqrt{|b|}$ (on $\mathcal L^2(\mathbb R^2, db\, d\phi)$), resulting in 
\be
	\partial_\phi^2 \Psi(b, \phi) =  (|b| \partial_b)^2 \Psi(b, \phi) \text{.} \label{eq:HConstQuant}
\ee
By definition, physical states need to satisfy \eqref{eq:HConstQuant}. In addition, we have to incorporate another symmetry of our model, the invariance under large gauge transformations which invert the spatial orientation (i.e. parity transformations). Physical states then have to transform in an irreducible representation and wave functionals therefore have to be either even or odd functions in $b$. One can therefore restrict to functions of positive $b$ only (or negative). 
Then, it is convenient to perform the variable transformation
\be
	y =  \log b ~~~ \Leftrightarrow ~~~ b = e^{ y} \text{.} \label{eq:VarTrafoWDW}
\ee
The Hamiltonian constraint \eqref{eq:HConstQuant} now simplifies to
\be
	\partial_\phi^2 \Psi(y, \phi) = \partial_y^2 \Psi(y, \phi) =: - \Theta \Psi(y, \phi) \text{,} \label{eq:HamQuantSimpl}
\ee
which is nothing else than the $1+1$-dimensional Klein-Gordon equation, with $\phi$ interpreted as time and $y$ as the spatial coordiate. 
Admissible physical states are positive frequency solutions to \eqref{eq:HamQuantSimpl} and thus have to satisfy 
\be
	- i \partial_\phi \Psi(y, \phi) = \sqrt{\Theta} \Psi(y, \phi) \text{.}  \label{eq:HamQuantSimplPos}
\ee
We can construct them from an initial state $\Psi(y,\phi_0) =  \int_{- \infty}^{ \infty} dk \tilde \Psi(k) e^{-i k y}$ as
\begin{align}
	\Psi(y, \phi) &=   \int_{- \infty}^{ \infty} dk \tilde \Psi(k) e^{-iky + i |k|(\phi - \phi_0)} \nonumber \\
			&=   \int_{- \infty}^{ 0} dk \tilde \Psi(k) e^{-ik(\phi + y)} e^{  i k  \phi_0} +    \int_0^{ \infty} dk \tilde \Psi(k) e^{ik(\phi - y)} e^{ - i k  \phi_0} \nonumber \\
			&=: \Psi_L(y_+) + \Psi_R(y_-) \text{,} \label{eq:SolKG}
\end{align}
where $y_{\pm} = \phi \pm y$ and left and right moving states where labelled by $L$ and $R$ respectively.

As is often the case, the solutions \eqref{eq:SolKG} to the quantum constraint \eqref{eq:HamQuantSimplPos} are not normalisable w.r.t. the kinematical scalar product. We therefore need to introduce a new scalar product on the physical Hilbert space. The framework of refined algebraic quantisation \cite{HiguchiQuantumLinearizationInstabilities, GiuliniOnTheGenerality} offers a systematic way of doing so, however is somewhat technical and thus not ideally suited for an introductory treatment. An exhaustive discussion in the context of loop quantum cosmology, including the construction of Dirac observables, can be found in \cite{KaminksiQuantumConstraintsDirac}.
Here, we will choose a simple shortcut leading to the same results:

Instead of applying the whole machinery of refined algebraic quantisation, we can simply interpret \eqref{eq:HamQuantSimplPos} as a Schr\"odinger equation with Hamiltonian $ \sqrt{\Theta}$ in the time variable $\phi$. The Hilbert space is then simply given by $\mathcal L^2(\mathbb R, dy)$, and the evolution of wave functions is governed by \eqref{eq:HamQuantSimplPos}. The solutions constructed in \eqref{eq:SolKG} remain valid and are in particular normalisable w.r.t. $\mathcal L^2(\mathbb R, dy)$ for appropriate $\tilde \Psi(k)$. 

We are now in a position to address the question of whether the big bang or big crunch singularities get resolved. For this, we note that the relevant parameter is 
\be
	\rho := P_\phi^2 / 2 v^2 \text{,}
\ee
which can be interpreted as the energy density of the scalar field, or equivalently, due to $H=0$, as the spacetime curvature. Since $P_\phi$ is a constant of motion, we can check for a singularity in the evolution of a given state $\Psi$ by looking at the expectation value of the total volume of the universe as a function of $\phi$.

Based on our solutions solutions \eqref{eq:SolKG}, we can now compute this expectation value in the Schr\"odinger picture, that is for evolving states and a fixed time volume operator. 
Let us first consider a purely left-moving state. We find
\begin{align}
	\braopket{\Psi_L(\phi)}{  \hat v}{ \Psi_L(\phi)}& =  \int_{- \infty}^{\infty} dy \, \overline{\Psi_L(y+\phi)} e^{-y/2} i \partial_y e^{-y/2} \Psi_L(y+\phi)  \nonumber \\
	&=  \int_{- \infty}^{\infty} d(y+\phi) \, \overline{\Psi_L(y+\phi)} e^{-(y+\phi)/2} i \partial_{y+\phi} e^{-(y+\phi)/2}  \Psi_L(y+\phi)\, e^\phi  \nonumber \\
	&=  \int_{- \infty}^{\infty} d y' \, \overline{\Psi_L(y')} e^{-(y')/2} i \partial_{y'} e^{-(y')/2}  \Psi_L(y')\, e^\phi  \nonumber \\
	&=: V_0 \, e^{\phi} \text{.} \label{eq:CompVofPhiWdW}
\end{align}
We note that $V_0$ is a constant that can be computed from $\Psi_L$.
It follows that the volume becomes zero as $\phi \rightarrow  -\infty$, resulting in the blowup of the energy density $\rho$ and thus a singularity. 
For right-moving states, we find a contracting universe as $\phi$ increases, i.e. the same result up to the substitution $e^\phi \rightarrow e^{-\phi}$.
Some more comments about the singularity resolution, including superpositions of left and right moving states are given in section \eqref{sec:Super2}.

\subsection{Loop quantum cosmology} \label{sec:LQCBouncebv}

We will now repeat the same steps as above using a slightly different quantisation inspired by loop quantum gravity. Let us consider wave functions $\tilde \chi(v)$ (suppressing for now the $\phi$-dependence) which have support\footnote{In loop quantum cosmology, one usually takes $v \in \mathbb R$, which corresponds to using square integrable functions on the Bohr compactification of the real line as wave functions. We will avoid these technicalities here as the Bohr compactification is not needed when using our current variables.} only at $v \in \mathbb Z$. Such a choice can be thought of to be inspired by the discrete eigenvalues of the geometric operators in loop quantum gravity, which in this case would correspond to the integers labelling the irreducible representations of U$(1)$. 
Since $\tilde \chi(v)$ is discontinuous as a function of $v$, there can be no operator corresponding to $b$, which would be a derivative w.r.t. $v$. However, $e^{i n b}$ for $n \in \mathbb Z$ can be given a well-defined meaning as
\be
	\widehat{e^{i n b}} \, \tilde \chi(v) = \tilde \chi(v+n) \text{.}
\ee 
Both $\hat v$ and $\widehat{e^{i n b}} $ are self-adjoint w.r.t. the kinematical scalar product
\be
	\braket{\tilde \chi}{\tilde \chi'} =  \pi \sum_{v \in \mathbb Z} \, \overline{ \tilde \chi(v) }\tilde \chi'(v). 
\ee

On this Hilbert space, we can approximate $b \rightarrow \sin b = \frac{e^{i  b} - e^{-i  b}}{2i}$. This approximation is good whenever $b \ll 1$. Due to the Hamiltonian constraint $H \approx 0$, this corresponds to $\rho \ll 1$, i.e. to a matter energy density much smaller than the Planck density. Since we would anyway expect for quantum gravity effects to become relevant at the Planck scale, the approximation $b \rightarrow \sin b$ is perfectly acceptable as a means to construct a quantum theory of gravity which gives standard cosmology for $\rho \ll 1$, but features UV-modifications with the potential to cure the singularity encountered before.

In order to be mimic the computation in the Wheeler-de Witt framework, we perform a Fourier transform to go to the $b$-representation:
\begin{align}
	\chi(b) = \sum_{v \in  \mathbb Z} e^{i v b} \, \tilde \chi(v), ~~~~ \tilde \chi(v) = \frac{1}{2\pi} \int_{-\pi}^\pi db \, e^{-ivb} \chi(b) \label{eq:FourierTransformLQC}
\end{align}
$\chi(b)$ is $2 \pi$-periodic and we restrict us to the support $(-\pi, \pi)$. Invariance under parity again leads us to consider only symmetric $\Psi(b)$ and we choose to work only with functions on the interval $(0, \pi)$. Again, it is convenient to rescale the wave functions as $\Psi(b) = \sqrt{\sin b}  \, \chi(b)$. Then, the inner product reads
\be
	\braket{\Psi}{\Psi'}= \int_{0}^{\pi} \frac{db}{\sin b} \, \overline{ \Psi(b) }\Psi'(b)
\ee
The quantised Hamiltonian constraint, in the same ordering as chosen in the Wheeler-de Witt case, now reads 
\be
	\partial_\phi^2 \Psi(b, \phi) =  (\sin(b) \partial_b)^2 \Psi(b, \phi) \text{.} \label{eq:HConstQuantLQC}
\ee

Next, we map the interval $(0, \pi)$ to $(-\infty, \infty)$ via the variable transformation
\be
	x = \log \left( \tan (b/2) \right)  ~~~ \Leftrightarrow ~~~ b = 2 \tan^{-1}(e^x) \text{.} \label{eq:VarTrafoLQC}
\ee
It follows that
\be
	\partial_x = \sin(b) \partial_b, ~~~~\partial_b = \cosh(x) \partial_x, ~~~~ dx = \frac{1}{\sin b} db, ~~~~ db = \frac{1}{\cosh x} dx \text{.}
\ee
The Hamiltonian constraint thus again takes the form
\be
	\partial_\phi^2 \Psi(x, \phi) = \partial_x^2 \Psi(x, \phi) =: - \Theta \Psi(x, \phi) \text{,} \label{eq:HamQuantSimplLQC}
\ee
however with a different interpretation of $x$ as opposed to $y$ in \eqref{eq:HamQuantSimpl}.

We can now go through the same procedure as above: we choose positive frequency solutions to \eqref{eq:HamQuantSimplLQC}, interpreted as a Schr\"odinger equation for the time $\phi$. We again compute the expectation value of the volume as a function of $\phi$ by repeating the computation \eqref{eq:CompVofPhiWdW} with the substitution $e^{-y} \rightarrow \cosh(x)$, coming from the difference in the variable transformations \eqref{eq:VarTrafoWDW} and \eqref{eq:VarTrafoLQC}, and thus ultimately from the difference in the choice of algebra, i.e. the non-existence of $\hat b$ in LQC and the resulting substitution $b \rightarrow \sin b$:
\begin{align}
	\braopket{\Psi_L(\phi)}{  \hat v}{ \Psi_L(\phi)}& =  \int_{- \infty}^{\infty} dx \, \overline{\Psi_L(x+\phi)} \sqrt{\cosh(x)} i \partial_x \sqrt{\cosh(x)} \Psi_L(x+\phi)  \nonumber \\
	&= \int_{- \infty}^{\infty} dx \, \overline{\Psi_L(x+\phi)} \left( \cosh(x) i \partial_x + \frac{i}{2}\sinh(x)  \right) \Psi_L(x+\phi)  \nonumber \\
	&= \text{Re} \int_{- \infty}^{\infty} dx \, \overline{\Psi_L(x+\phi)} \frac{1}{2} \left( e^{x+\phi} e^{-\phi} + e^{-x-\phi} e^{\phi}   \right) i \partial_x   \Psi_L(x+\phi)  \nonumber \\
	&=: V_+ \, e^{\phi} + V_- \, e^{-\phi} \nonumber \\ 
	&=: V_\text{min} \braket{\Psi_L}{ \Psi_L} \cosh(\phi-\phi_\text{bounce})\text{,} \label{eq:CompVofPhiLQC}
\end{align}
with
\be
	V_\pm =\text{Re} \int_{- \infty}^{\infty} dx \, \overline{\Psi_L(x)} \frac{1}{2} e^{\mp x}  i \partial_x   \Psi_L(x), ~~~~~ V_{\text{min}}  = \frac{2 \sqrt{V_+ V_-}}{\braket{\Psi_L}{\Psi_L}}, ~~~~~ \phi_{\text{bounce}} =  \frac{1}{2} \log \frac{V_-}{V_+} \text{.}
\ee

Thus, we find, as opposed to the Wheeler-de Witt theory, that the expectation value of the spatial volume has a lower bound $V_{\text{min}}$. Moreover, the evolution is symmetric around the bounce point.
The singularity that persisted in the Wheeler-de Witt approach at the level of expectation values of the volume operator is thus resolved if one chooses a quantisation strategy inspired by loop quantum gravity (see however section \ref{sec:Super2}). For right-moving states, we would obtain the same behaviour.

\subsection{Kinematical scalar products and ordering}

For clarity, we recall the kinematical scalar product written in terms of different variables and wave functions, suppressing the $\phi$-dependence for simplicity. 
For defining and solving the Wheeler-de Witt equation, we used the kinematical scalar product
\be
	\braket{\Psi}{\Psi'}_{\text{kin, LQC}} = \int_{-\infty}^{\infty} dx \, \overline{ \Psi(x)} \Psi'(x), ~~~~~ \braket{\Psi}{\Psi'}_{\text{kin, WdW}} = \int_{-\infty}^{\infty} dy \, \overline{ \Psi(y)} \Psi'(y)
\ee
in the gravitational sector. Transforming back to wave functions of $b$, we obtain
\be
	\braket{\Psi}{\Psi'}_{\text{kin, LQC}} = \int_{0}^{\pi} \frac{db}{\sin b} \, \overline{ \Psi(b)} \Psi'(b), ~~~~~ \braket{\Psi}{\Psi'}_{\text{kin, WdW}} = \int_{0}^{\infty} \frac{db}{b} \, \overline{ \Psi(b) }\Psi'(b) \text{.}
\ee
It is clear that $\Theta$ as given in \eqref{eq:HConstQuant} and \eqref{eq:HConstQuantLQC} is self-adjoint in those scalar products. In order to have standard scalar products also in the $v$ representation, we define $\chi(b) = \Psi(b) / \sqrt{\sin b}$ in the LQC case and $\chi(b) = \Psi(b) / \sqrt{b}$ in the WdW case. After a Fourier transform similar to \eqref{eq:FourierTransformLQC}, it follows that 
\be
	\braket{\Psi}{\Psi'}_{\text{kin, LQC}} =  \pi \sum_{v \in \mathbb Z} \, \overline{ \tilde \chi(v)} \tilde \chi'(v), ~~~~~ \braket{\Psi}{\Psi'}_{\text{kin, WdW}} =   \pi \int_{-\infty}^{\infty} dv\, \overline{ \tilde \chi(v) }\tilde \chi'(v) \text{.} \label{eq:ScalarProductsChi}
\ee
The ordering that we chose for the operator $\Theta$ when using the wave functions $\tilde \chi(v)$ with the scalar products \eqref{eq:ScalarProductsChi} is thus 
\be
	- \Theta_{\chi, \text{LQC}} = \sqrt{\sin b} ~ v ~ \sin b ~ v ~ \sqrt{\sin b}, ~~~~~-  \Theta_{\chi, \text{WdW}} = \sqrt{ b} ~ v ~  b ~ v ~ \sqrt{ b}
\ee
This is equivalent to using the wave functions $\Psi(b)$ with ordering $\sin b\, \partial_b \sin b \, \partial_b$ or $\Psi(x)$ with the (trivial) ordering $\partial_x \partial_x$.

In the Schr\"odinger picture used above, the scalar product is simply the kinematical scalar product stripped of the $\phi$-integration. This scalar product can also be arrived at as the physical scalar product by refined algebraic quantisation\footnote{Here, some subtleties arise due to the precise form of the constraint that one inputs in expressions like $\ket{\Psi}_{\text{phys}} = \delta(H) \ket{\Psi}_{\text{kin}}$. In particular, factors of $\{ \phi, H\}$ will rescale the physical states due to the properties of the Dirac-$\delta$. These rescalings are however accounted for in a proper definition of the corresponding Dirac observables, e.g. the volume of the universe at a given scalar field time, agreeing with the Schr\"odinger picture, see \cite{KaminksiQuantumConstraintsDirac}.}.

\subsection{Superselection and superpositions} \label{sec:Super2}

Our treatment here so far differs slightly from \cite{AshtekarRobustnessOfKey} in that we did not take into account superselection induced by the action of the Hamiltonian constraint. For instance, in the $v$-representation, the gravitational part (in the given ordering) acts as 
\be
	(\widehat{\sin b}) \hat v (\widehat{\sin b}) \hat v \tilde \Psi(v) = - \frac 14 v \left( (v+1) \tilde \Psi(v+2) + (v-1) \tilde \Psi(v-2) - 2v \tilde \Psi(v) \right) 
\ee
Therefore, wave functions with support on $2 \mathbb Z$ and $2 \mathbb Z + 1$ are superselected. Restriction to one of these sectors imposes a constraint on the wave functions $ \Psi(x)$ as follows. From the Fourier transform \eqref{eq:FourierTransformLQC}, we see that a wave function in the $v$-representation (with $\tilde \Psi(v) = \tilde \Psi(-v)$) with support on $2 \mathbb Z$ becomes a wave function in the $b$-representation which is symmetric around $b = \pi / 2$. On the other hand, support on $2 \mathbb Z+1$ results in a function which is antisymmetric around $b = \pi / 2$. This then translates to (anti)symmetry for $\Psi(x)$ around $x=0$.

We chose not to include this discussion in the main derivation for the following reason. In the loop quantum cosmology computation, considering symmetric $\Psi(x)$ does not change the result. However, in case of the Wheeler-de Witt quantisation, considering (anti)symmetric $\Psi(y)$, which are also superselected by the analogous operator $\partial_y^2$,  corresponds to taking superpositions of expanding and contracting branches. One might naively think that from such a superposition one would obtain a non-singular universe also in the Wheeler-de Witt case. This is however misleading, as one needs to invoke a quantum mechanical formalism applicable to the universe as a whole, as well as to sequences of events, such as the consistent histories approach. In doing this, one finds that the probability for a singularity to occur in the Wheeler-de Witt case is $1$ \cite{CraigConsistentProbabilitiesInWheeler}, while it is $0$ in the case of loop quantum cosmology \cite{CraigConsistentProbabilitiesInLoop}. These more rigorous computations thus back the conclusions that we have drawn here. See however also \cite{FalcianoWheelerDeWittQuantization} for a treatment using the Bohmian approach to quantum mechanics, where different conclusions are reached. A recent review is given in \cite{CraigTheConsistentHistories}.

\subsection{Outlook on full theory}

In order to establish a connection to the full theory, we need to slightly change our variables\footnote{Loop quantum cosmology can also be formulated in these variables, and in fact was originally due to their resemblance to full LQG.}. Instead of using $v$ and $b$, we use $a^2$ and $b a$. Then, $a^2$ can be integrated over a surface, while $b a$ corresponds geometrically a one-form and wants to be integrated over a line. This corresponds in the full theory to computing holonomies from the SU(2)-connection $A_a^i$, the analogue of $b a$, and fluxes form the densitised triad $E^a_i$, which corresponds to $a^2$. 

The substitution $b \rightarrow \sin b$ in the full theory corresponds to the fact that there exists no operator corresponding to the connection $A_a^i$ and we thus have to approximate all expressions involving it via holonomies constructed from $A_a^i$. 

One can be more explicit in constructing embeddings of loop quantum cosmology into the full theory if one quantises using the diagonal metric gauge and using a set of variables which resembles those of loop quantum cosmology more closely \cite{BIII, BVI}.

\subsection{Exercises}

\begin{enumerate}

	\item {\bf Warm up:}\\ 
	Convince yourself that other choices of time, e.g. $N = a^3$, do not affect the equations of motion upon using $H=0$ after evaluating Poisson brackets. \\

	\item {\bf Maximum energy density:}\\ 
	The volume at the bounce point can be chosen arbitrarily. Show however that the energy density at the bounce point is bounded from above by a critical energy density $\rho_\text{crit}$. Show that the effective Friedmann equation \eqref{eqn:EffectiveFriedmann} captures this effect and reproduces \eqref{eq:CompVofPhiLQC}.\\

	\item {\bf Barbero-Immirzi parameter:}\\ 
	Instead of using $b$ and $v$ as canonical variable, we could also use $\beta b$ and $v / \beta$ for $\beta \in \mathbb R \backslash \{0\}$. Since these new variables are still canonically conjugate, we can base the same quantisation also on them. What is the consequence for the spectrum of the volume operator and the maximal energy density of the universe? \\
	The direct analog of $\beta$ in full general relativity is known as the Barbero-Immirzi parameter and we will encounter it in the next lectures. 
	
	\item {\bf Inverse triad corrections:}\\ 
	The substitution $b \rightarrow \sin b$ is known as a holonomy correction in loop quantum cosmology. More complicated models also consider so called inverse triad corrections, which for example arise for the choice $N=1$ in the quantum theory. Then, one has to define an operator corresponding to $1/v$. Since zero is in the spectrum of $v$, this cannot simply be done via the spectral theorem. \\
	Consider the $v$-representation with wave functions $\chi(v)$ 
	and scalar product \eqref{eq:ScalarProductsChi}. Show that classically $\mp 2i\{\sqrt{|v|}, e^{\pm ib} \} e^{\mp ib} = \text{Sign}(v)/ \sqrt{|v|}$. This expression can be directly promoted to a quantum operator via the quantisation map $ \{\cdot , \cdot \} \rightarrow -i [\hat \cdot,\hat \cdot]$. Show that it results in a self-adjoint operator for both choices $\pm$ and consider the ordering suggested by $-i\{\sqrt{|v|}, e^{+ ib} \} e^{- ib}  + i\{\sqrt{|v|}, e^{- ib} \} e^{+ ib}$. Compute the action of this operator on a wave function $\chi(v)$ and show that it acts as $ (\sqrt{|v+1|} - \sqrt{|v-1|}) \chi(v)$. Taylor expand for large $v$ and show that the leading term is $\text{Sign}(v)/ \sqrt{|v|}$. Show that the so defined operator has an upper bound. Square it to obtain an operator corresponding to $1/|v|$. Modify this prescription to construct an operator corresponding to $|v|^{\epsilon}$ with $\epsilon \in (-1,0)$.
	Manipulations of this kind are known as Thiemann's tricks, going back to \cite{ThiemannQSD1}, where they are essential in regularising the Hamiltonian constraint.

\end{enumerate}

\section{General relativity in the connection formulation and quantum kinematics}

In this section, we will derive the classical foundations of loop quantum gravity and perform a kinematical quantisation. We will gloss over many technical details and focus on the core ideas and concepts. For an advanced in depth discussion, we refer the reader to \cite{ThiemannModernCanonicalQuantum}. For an introductory textbook, see \cite{GambiniAFirstCourse}, as well as \cite{RovelliQuantumGravity} for an intermediate level. Other introductory papers with varying degree of details include \cite{ThiemannLecturesOnLoop, PerezIntroductionToLoop, AshtekarAnIntroductionTo, MercuriIntroductionToLoop, GieselFromClassicalTo, BilsonLQGForThe}. Original literature includes \cite{RovelliLoopSpaceRepresentation, AshtekarRepresentationsOfThe, AshtekarRepresentationTheoryOf, AshtekarDifferentialGeometryOn, AshtekarProjectiveTechniquesAnd, MarolfOnTheSupport, AshtekarQuantizationOfDiffeomorphism}. We set $8 \pi G = c = 1$. 

\subsection{Canonical general relativity}

A more in depth introduction to loop quantum gravity would start at this point by introducing the Arnowitt-Deser-Misner (ADM) formulation \cite{ArnowittTheDynamicsOf} of general relativity, along with the Dirac procedure to treat constrained systems \cite{DiracLecturesOnQuantum}. While a thorough knowledge of both subjects is certainly important once doing research in the subject, most details are not really relevant in a first introduction. We will thus take a different approach here and simply state how the ADM formulation looks like and how exactly it works. 

\begin{figure}[h]
	\centering	\includegraphics[trim = 19mm 185mm 15mm 11mm, clip, scale=0.55]{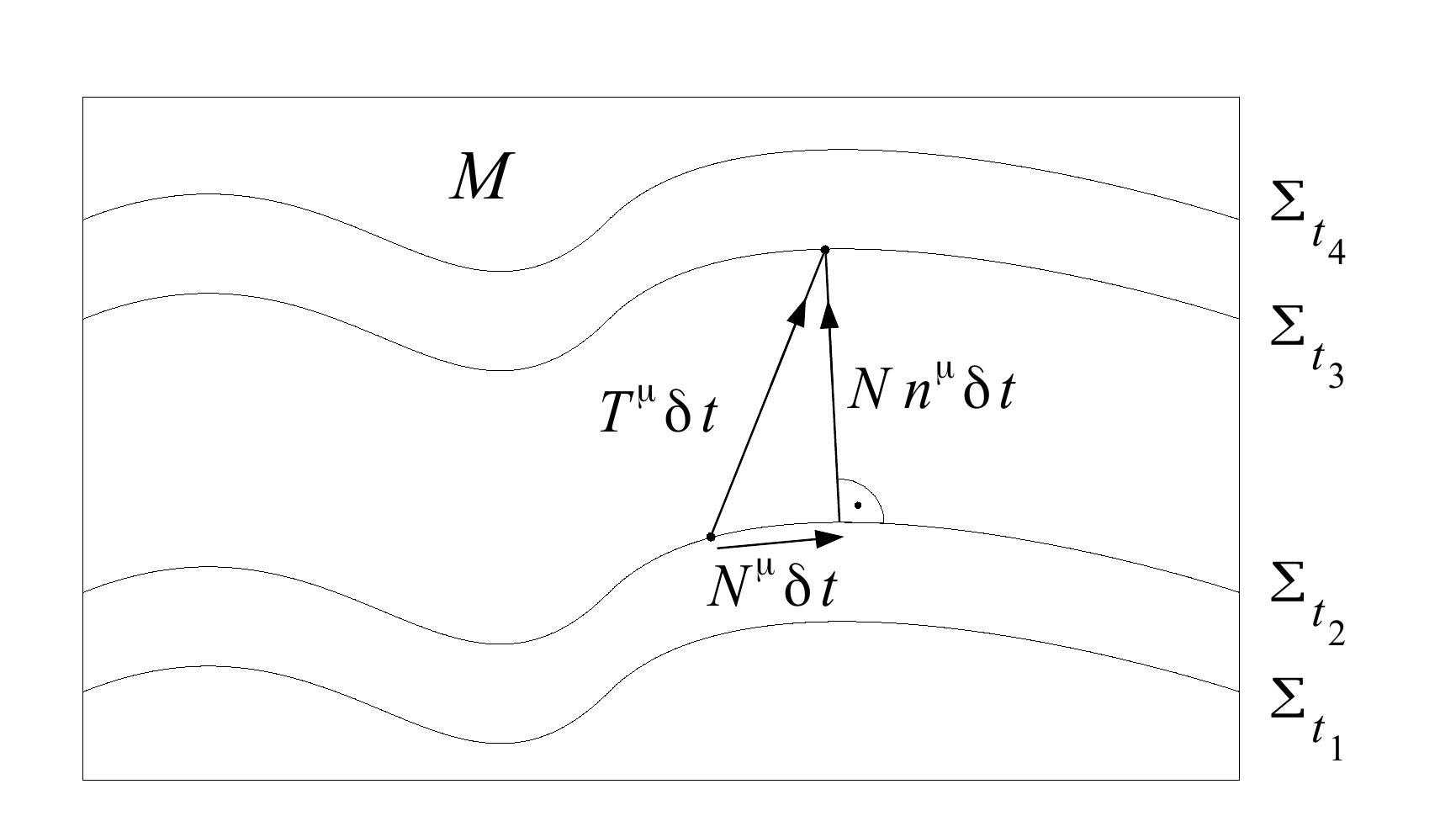}
		\caption{An infinitesimal time evolution step between two neighbouring Cauchy surfaces is shown. The evolution is along the vector field $T^\mu$ and splits into a lapse component orthogonal to $\Sigma$ and a shift component tangential to $\Sigma$.}
		 \label{fig:Split}
	\end{figure}

As a first step, the four-dimensional spacetime manifold $M$ is foliated into three-dimensional ``equal time'' spatial slices $\Sigma_t$, which is equivalent to imposing global hyperbolicity (well-definedness of initial value problems) on the class of spacetimes we consider. 
For simplicity, we will consider $\Sigma$ to be compact, so that boundary terms can be neglected.
The split also induces a split in the spacetime metric tensor $g_{\mu \nu}$ as
\be
	g_{\mu \nu} =  \left( \begin{array}{cc}
-N^2 + N^a N_a & N_a   \\
N_a & q_{ab}  \end{array} \right) 
~~~~~~~
g^{\mu \nu} =  \left( \begin{array}{cc}
-1/N^2 & N^a / N^2  \\
N^a / N^2 & q^{ab} - N^a N^b / N^2  \end{array} \right) \text{,}
\ee
where $q_{ab}$ is the Euclidean three-metric induced on $\Sigma$, $N$ is called the lapse function, and $N^a$ is called the shift vector. $a,b, \ldots$ denote spatial tensor indices on $\Sigma$. $N$ and $N^a$ result form a decomposition of the time evolution vector field $T^\mu = N n^\mu + N^\mu$, with $n_\mu N^\mu = 0$ and $n^\mu$ being the unit normal on $\Sigma$, i.e. $n^\mu n_\mu = -1$, see figure \ref{fig:Split}. 

In addition, we need a conjugate variable to the spatial metric $q_{ab}$. For this, we define the extrinsic curvature 
\be
	K_{\mu \nu} = \frac{1}{2} \mathcal L_n g_{\mu \nu} \text{,}
\ee
where $\mathcal L_n$ denotes the Lie derivative w.r.t. the vector field $n^\mu$ (see the exercises). It turns out that $n^\mu K_{\mu \nu}= n^\nu K_{\mu \nu} = 0$, so we can write the extrinsic curvature as $K_{ab}$, i.e. an object on $\Sigma$. From $K_{ab}$, we construct 
\be
	P^{ab} = \frac{\sqrt{\det{q}}}{2} \left(K^{ab} - q^{ab} K \right) \text{,}
\ee
which fully captures the information in $K_{ab}$. The non-vanishing Poisson brackets turn out to be
\be
	\left\{q_{ab}(x), P^{cd}(y) \right\} = \delta_{(a}^c \delta_{b)}^d \delta^{(3)}(x,y) \text{.} \label{eq:ADMBrackets}
\ee

The Hamiltonian of the ADM formulation is given by
\be
	H = \mathcal H[N] + \mathcal H_a[ N^a] 
\ee
with 
\begin{align}
	\mathcal H[N]&= \int_\Sigma d^3x \, N \left( \frac{2}{\sqrt{q}} \left( P^{ab} P_{ab} - \frac{1}{2} P^2 \right) - \frac{\sqrt q}{2} \stackrel{(3)} R\right)\\
	\mathcal H_a[N^a] &= -2 \int_\Sigma d^3x \, N^a \nabla_b P^b {}_a \text{.}
\end{align}
$\mathcal H$ and $\mathcal H_a$ are denoted as the Hamiltonian constraint and spatial diffeomorphism constraint. Consistency of the dynamics forces both of them to vanish, we write
\be
	\mathcal H \approx 0, ~~~~ \mathcal H_a \approx 0 \text{.}
\ee
Here, $\approx$ denotes a so called weak equality, i.e. an equality that can be used only after Poisson brackets have been computed. This means that the Hamiltonian $H$ is weakly zero, however Poisson brackets involving it are generically non-zero. $\mathcal H$ and $\mathcal H_a$ form an algebra, the so called hypersurface deformation, or Dirac algebra
\begin{align}
	\left\{\mathcal H[M], \mathcal H[N] \right\}&=  \mathcal H_a \left[q^{ab} \left( M \partial_b N - N \partial_b M \right) \right] \\
	\left\{\mathcal H[M], \mathcal H_a[N^a] \right\}&= - \mathcal H \left[ \mathcal L_N M \right] \label{eq:PBHHa} \\
	\left\{\mathcal H_a[M^a], \mathcal H_a[N^a] \right\}&= - \mathcal H_a \left[ \mathcal L_N M^a \right] \text{.} \label{eq:PBHaHa}
\end{align}
It is important to note that this algebra features structure functions, so that it is not a Lie algebra. In particular, this leads to problems when trying to find operators which satisfy this algebra (see exercises in section \ref{sec:DynamicsAndOutlook}). 

$\mathcal H$ and $\mathcal H_a$ should be regarded as generators of gauge symmetries.
In particular, only those phase space functions which Poisson-commute with both $\mathcal H$ and $\mathcal H_a$ have some invariant physical meaning, e.g., are independent of the choice of coordinates. We will call such functions Dirac observables $\mathcal O$. 

To understand the symmetries generated by $\mathcal H_a$, we compute
\be
	\left\{ q_{ab}, \mathcal H_a[N^a] \right\} = \mathcal L_{\vec N} q_{ab}, ~~~ \left\{ P^{ab}, \mathcal H_a[N^a] \right\} = \mathcal L_{\vec N} P^{ab} \text{.} \label{eq:LieDerFromHa}
\ee
We see that $\mathcal H_a[N^a]$ generates infinitesimal spatial diffeomorphisms along the vector field $\vec N$. The action of $\mathcal H$ is more involved and it is harder to interpret it properly, as it also encodes the dynamics of the theory. It can be shown that if the Einstein equations are satisfied, then $H$ generates diffeomorphisms orthogonal to $\Sigma$, see for example \cite{ThiemannModernCanonicalQuantum}. On the other hand, without the equations of motion holding, the symmetry group of canonical general relativity is distinct from the group of four-diffeomorphisms, see section 1.4 in \cite{ThiemannModernCanonicalQuantum} for a discussion.

The lapse and shift functions appear in the ADM formulation only as arbitrary Lagrange multipliers in the Hamiltonian. They correspond to a choice of gauge and become relevant when we want to reconstruct the complete spacetime from the canonical data on a single Cauchy surface $\Sigma$. The choice of $N$ and $N^a$ determines the relative positions of neighbouring Cauchy surfaces as shown in figure \ref{fig:Split}. In other words, it tells us where in the spacetime we end up after an infinitesimal time evolution generated by the Hamiltonian $H$. 

In total, the constraints thus delete $4+4$ phase space degrees of freedom, four by the equations $\mathcal H \approx 0$ and $\mathcal H_a \approx 0$, and four additional degrees of freedom by selecting observables for which $\{ \mathcal O, \mathcal H\} = 0 = \{ \mathcal O, \mathcal H_a\}$. We are thus left with 2+2 phase space degrees of freedom per point.

\subsection{Connection variables}

In a next step, we will need to change our variables. First, we introduce an additional local SU$(2)$ gauge symmetry\footnote{In principle, any internal gauge group could be used, as long as an equivalence to general relativity is ensured, e.g. by imposing additional constraints. In 3+1 dimensions, this can also be done by using the groups SO$(1,3)$ \cite{AlexandrovSU(2)LoopQuantum, CianfraniTowardsLoopQuantum, GeillerALorentzCovariant, BTTII} and SO$(4)$ \cite{BTTI}. For non-compact gauge groups however, many of the techniques used to construct the Hilbert space are not available \cite{AshtekarRepresentationsOfThe, AshtekarRepresentationTheoryOf, AshtekarDifferentialGeometryOn, AshtekarProjectiveTechniquesAnd, MarolfOnTheSupport, AshtekarQuantizationOfDiffeomorphism}.} in our framework by coordinatising our phase space by $E^{a}_i$ and $K_a^i$, $i = 1,2,3$, which are related to the ADM variables as
\be
	q q^{ab} = E^{ai} E^{b}_i, ~~~~~ \sqrt{q} K_{a} {}^b = K_{ai}  E^{bi} \text{.} \label{eq:ADMVarsFromKE}
\ee
We can thus interpret $E^{a}_i$ as a densitised tetrad $\sqrt{q} e^{a}_i$, with $q^{ab} = e^{a}_i e^{bi}$, and write the extrinsic curvature as $K_{ab} = K_{ai} e^i_b$, using the co-tetrad $e_b^i$ (we restrict to positive orientation here to avoid additional sign factors). Internal indices $i,j$ are trivially raised and lowered by the Kronecker $\delta_{ij}$.
The new non-vanishing Poisson brackets are 
\be
	\left\{K_a^i(x), E^{b}_j (y) \right\} = \delta^{(3)}(x,y) \delta_a^b \delta^i_j \text{.} \label{eq:KEBracket}
\ee
Since we now have $3+3$ more phase space degrees of freedom, we need to introduce an additional constraint, the Gau{\ss} law 
\be
	G_{ij}[\Lambda^{ij}] =  \int_\Sigma d^3x \, \Lambda^{ij}  K_{a[i}  E^{a}_{j]} \approx 0 \text{.} \label{eq:GaussLaw}
\ee
It generates internal SU$(2)$ gauge transformations 
\be
	\left\{ K_{a}^{i}, G_{kl}[\Lambda^{kl}]  \right\} = \Lambda^{i} {}_j K_{a}^j, ~~~~~\left\{ E_{i}^{a}, G_{kl}[\Lambda^{kl}]  \right\} = \Lambda_{i} {}^j E^{a}_j \label{eq:TransEKGauss}
\ee
under which observables have to be invariant. In particular, this applies to the combinations \eqref{eq:ADMVarsFromKE}, and thus the ADM variables. In order to link this new formulation to the ADM formulation, we now have to show that the ADM Poisson brackets \eqref{eq:ADMBrackets} are reproduced by \eqref{eq:KEBracket}, i.e. we need to show that 
\be
	\left\{q_{ab}[E, K](x), P^{cd}[E,K](y) \right\}_{\{K,E\}} = \delta_{(a}^c \delta_{b)}^d \delta^{(3)}(x,y) + G_{ij}[...] \text{.} \label{eq:ADMPBFromEK} 
\ee
i.e. up to terms proportional to the Gau{\ss} law. This can be done, although it involves a little algebra. We can now simply express the Hamiltonian and spatial diffeomorphism constraints in terms of our new variables (see exercises). Due to \eqref{eq:ADMPBFromEK}, they will generate the same dynamics up to SU$(2)$ gauge transformations. Since observables are by definition also invariant under SU$(2)$ gauge transformations, the dynamics generated by our extended formulation is in fact identical to that of the ADM formulation. 

Although we now have an internal SU$(2)$ gauge freedom, \eqref{eq:TransEKGauss} tells us that none of our variables transform as a connection. There is however a natural connection that one can build, the spin connection $\Gamma_a^{i}$, defined as $\nabla_a e_b^i := \partial_a e_b^i - \Gamma_{ab}^c e_c^i + \epsilon^{ijk} \Gamma_a^j e_b^k = 0$. We can thus choose the new canonical variables
\be
	A_{a}^i = \Gamma_a^i + \beta K_a^i, ~~~~ \tilde E^a_i = \frac{1}{\beta} E^{a}_i \text{,} 
\ee
where $\beta$ is a free real parameter known as the Barbero-Immirzi parameter. It constitutes a 1-parameter ambiguity in the construction of our connection variables and will obtain a physical meaning only later at the quantum level. First, let us check that $A_a^i$ transforms indeed as a connection. For this, it can be shown (see exercises), that
\be
	G_{ij}[ - \epsilon^{ijk} \Lambda_K ]  =   \int_\Sigma d^3x\, \Lambda_K D_a \tilde E^{a}_i :=  \int_\Sigma d^3x\, \Lambda_K \left( \partial_a \tilde E^{a}_i + \epsilon^{ijk} A_a^j \tilde E^a_k \right) := G_k [\Lambda^k] \text{,} \label{eq:GaussAE}
\ee
where $D_a$ acts only on internal indices, and consequently
\be
	\left\{ A_{a}^{i}, G_{k}[\Lambda^{k}]  \right\} = -D_a \Lambda^{k} = -\partial_a \Lambda^{k} - \epsilon^{ijk} A_a^j \Lambda^{k} , ~~~~~\left\{ \tilde E_{i}^{a}, G_{k}[\Lambda^{k}]  \right\} = \epsilon^{ijk} \Lambda^{j} \tilde E^{a}_k \label{eq:TransAEGauss} \text{.}
\ee
$A_a^i$ now transforms as a connection, so we need to compute now the new Poisson brackets. It turns out that the transformation we just performed is canonical, i.e. that
\be
	\left\{A_a^i(x), \tilde E^{b}_j (y) \right\} = \delta^{(3)}(x,y) \delta_a^b \delta^i_j  \label{eq:AEBracket}
\ee
are the only non-vanishing Poisson brackets (see exercises). 
For notational simplicity, we will drop in the following the twiddle over $\tilde E^a_i$ and comment on this in case of possible confusion. 
We can thus again express the ADM constraints in terms of our new variables (up to terms proportional to $G_{ij}$) and arrive at
\begin{align}
	\mathcal H[N]&= \int_\Sigma d^3x \, N \left(\beta^2 \frac{E^{ai} E^{bj}}{2 \sqrt{q}} \epsilon^{ijk} F_{ab}^k - \frac{\left(1+\beta^2\right)}{\sqrt q} K_{[a}^i K_{b]}^{j} E^{ai} E^{bj} \right) \label{eq:HamConstrAE}\\f
	\mathcal H_a[N^a] &= \int_\Sigma d^3x \, E^{ai} \mathcal L_{\vec N} A_{ai} \text{,} \label{eq:LieDerFromHaAE}
\end{align}
where $F_{ab}^i = 2 \partial_{[a} A_{b]}^i + \epsilon^{ijk} A_a^j A_b^k$ and $K_a^i = (A_a^i - \Gamma_a^i) / \beta$ as well as $\sqrt{q} = \sqrt{ E^a_i E^b_j E^c_k \epsilon_{abc} \epsilon^{ijk} /6}$. In particular, we see that a great simplification is achieved when $\beta = \pm i$. Unfortunately, it is not known how to formulate the quantum theory in the case of non-real $\beta$ due to complicated reality conditions.

\subsection{Holonomies and fluxes} \label{sec:HolFlux}

Due to the usual obstructions, we cannot quantise all functions on phase space, but we need to pick a certain subset. This subset should be point-separating, i.e. to allow us to reconstruct other phase space functions with arbitrary precision. The choice of such a preferred subset in loop quantum gravity is closely related to the choice of variables in lattice gauge theory: we choose holonomies, i.e. parallel transports of our connection, and fluxes, i.e. smearings of the conjugate variable $E^a_i$ over surfaces. The main difference to lattice gauge theory is however that we do not consider only a given set of holonomies and fluxes specified by a choice of lattice, but all possible holonomies and fluxes obtained by choosing arbitrary curves and surfaces. 

More precisely, we define holonomies $h^j_c(A)$ along a curve $c : [0,1] \rightarrow \Sigma$ in a certain representation $j$ of SU$(2)$ as the solution to the equation
\be
	\frac{d}{d t} h^j_{c}(t) = h^j_{c}(t) A(c(t))    \label{eq:HolDef}
\ee
evaluated at $t=1$, where $A(c(t)) = A_a^i(c(t)) \dot c^a (t) \tau^{(j)}_i$ and $\tau^{(j)}_i$ are the 3 ($i = 1,2,3$) generators of SU$(2)$ in the representation $j$ (chosen such that $\tau^\dagger  = -\tau$). The solution can be written as a path ordered exponential as (see exercises)
\be
	 h_c(A) = \mathcal{P} \exp \left( \int_c A \right) = \mathbb{1} + \sum_{n=1}^\infty \int_0^1 dt_1 \, \int_{t_1}^1 dt_2 \ldots \int_{t_{n-1}}^1 dt_n A(c(t_1)) \ldots A(c(t_n)) \text{.} \label{eq:HolPathOrder}
\ee
Clearly, by taking the limit of an infinitesimally short path, we can reconstruct the connection. 

Next, we construct fluxes by integrating $E^a_i$, contracted with a smearing function $n^i$, over a surface $S$ as
\be
	E_n(S) := \int_S E^a_i \, n^i d S_a = \int_S E^a_i \, n^i \epsilon_{abc} dx^b \wedge dx^c \text{.}
\ee
 Again, by taking the limit of infinitesimally small surfaces and suitable $n^i$, we can reconstruct $E^a_i$. 
 
In order to construct the quantum theory, we need to know the Poisson brackets of holonomies and fluxes. Computing these requires a regularisation procedure and a discussion of the intersection properties of the involved surfaces and curves. We will not do this in detail here, but only restrict to the most important Poisson bracket in the simplest non-trivial intersection. A complete discussion can be found in \cite{ThiemannModernCanonicalQuantum}.

We consider the Poisson bracket of a holonomy with a flux and choose our surface $S$ such that it intersects the curve $c$ exactly once in the point $c(s)$ and from below w.r.t. the orientation of $S$. In this case, we obtain
\be
	\left\{h^j_c(A), E_n(S) \right\} = h^j_{c_1} (A) \left( \tau^{(j)}_i n^i \right) h^j_{c_2} (A) \label{eq:HolFluxBracket}
\ee 
where $c_1 = c |_{t \in [0, s]}$ and $c_2 = c |_{t \in [s, 1]}$. In other cases of interest, one considers curves which terminate on the surface as well as all possible choices of orientations. An interesting subtlety arises when one considers the Poisson bracket of two fluxes, however we will not detail this in these lectures. For a pedagogical treatment, see e.g. \cite{PerezIntroductionToLoop, GieselFromClassicalTo}.

\subsection{Quantisation} \label{sec:Quantisation}

\subsubsection{Hilbert space and elementary operators}

We are now in a  position to quantise our theory. For this, we need to find a Hilbert space on which we can define operators corresponding to holonomies and fluxes. As elements in our Hilbert space, we choose complex valued wave functions $\Psi$ which depend on a finite number of holonomies. Such functions are called cylindrical functions in loop quantum gravity and we write them as 
\be
	\Psi(A) = \Psi(h^{j_1}_{c_1}(A), \ldots, h^{j_n}_{c_n}(A)), ~~~ n \in \mathbb N_0 \text{.} 
\ee
We will also use the bra-ket-notation $\ket{\Psi}$ to underline that $\Psi(A)$ is a Hilbert space element. 
Operators corresponding to holonomies simply act by multiplication in this Hilbert space:
\be
	\hat {h}^j_c \ket{\Psi}  = h^j_c(A) \Psi(h^{j_1}_{c_1}(A), \ldots, h^{j_n}_{c_n}(A))
\ee
The collection of curves $c_i$ on which a cylindrical function depends constitutes a graph $\gamma$, which we can also consider as a coloured graph, where each edge (curve) $c_i$ is labelled by a spin $j_i$. Since we are free to choose $j=0$, we can in particular have trivial dependencies if we want. Such possible trivial dependencies lead to the notion of cylindrical consistency, which roughly says that the theory should be invariant under adding trivial ($j=0$) edges, as well as orientation flips. We will not spell out the details here, but only remark for later that given two graphs $\gamma_1$ and $\gamma_2$, we can always find a graph $\gamma_3$ which is finer than both $\gamma_1$ and $\gamma_2$, i.e. all edges in $\gamma_1$ or $\gamma_2$ are included in $\gamma_3$. In particular, we can express two cylindrical functions $\Psi_1$ and $\Psi_2$ as functions on $\gamma_3$ by adding trivial dependencies.

Fluxes act as derivative operators according to the classical relations such as \eqref{eq:HolFluxBracket}. Again, in the simplest single-intersection case leading to \eqref{eq:HolFluxBracket}, where we take  $\Psi = \Psi(h_c^j(A))$, we have
\be
	\hat {E}_n(S) \ket{\Psi}  = -i \hbar \int_S n^i  \frac{\delta}{\delta A_a^i} d S_a  \ket{\Psi} = -i \hbar \frac{ \partial \Psi(h^{j}_{c}) }{\partial (h^{j}_{c})_\alpha {}^\beta } \left(  h^j_{c_1}  \left( \tau^{(j)}_i n^i \right) h^j_{c_2} \right)_{\alpha} {}^\beta \text{.} \label{eq:DefQuantFlux}
\ee
Clearly, the operators we just defined satisfy the commutation relation \eqref{eq:HolFluxBracket}. It can be checked that also the other commutation relations not detailed here are satisfied, but this goes beyond the scope of these introductory lectures. 

We still need to equip our Hilbert space with a scalar product. Since the holonomies on which our wave functions depend are elements of SU$(2)$, the Haar measure is a natural choice for constructing a scalar product. We define the kinematical scalar product
\begin{align}
	\braket{\Psi}{\Phi}_{\text{kin}} &= \int d \mu_{\text{AL}} \overline{\Psi(A)} \Phi(A) \\
						&= \int d \mu_{\text{AL}} \overline{\Psi(h^{j_1}_{c_1}(A), \ldots, h^{j_n}_{c_n}(A))}  \Phi(h^{j_1}_{c_1}(A), \ldots, h^{j_n}_{c_n}(A))\\
						&:=  \int_{\text{SU}(2)^n} dg_1  \ldots dg_n \, \overline{\Psi(g_1, \ldots, g_n)} \Phi(g_1, \ldots, g_n) \text{,} \label{eq:ScalarPKin}
\end{align}
where AL labels to the Ashtekar-Lewandowski measure $\mu_{AL}$. It is immediate to verify that the two cylindrical functions are orthogonal if they depend non-trivially on holonomies defined on different edges. Also, it can be checked that the classical reality conditions $h_c^j \in \text{SU}(2)$ and $E^a_i = \bar E^a_i$ are implemented. 

In a last step, we can complete the Hilbert space spanned by the cylindrical functions w.r.t. the scalar product. This leads to the so called space of distributional connections, which we take as the quantum configuration space. The details of this completion do are not relevant for this introductory course, and we refer to \cite{ThiemannModernCanonicalQuantum} for details. Interestingly, it turns out that the representation of the holonomy-flux algebra that we just constructed is unique under mild technical assumptions such as spatial diffeomorphism invariance and irreducibility, known as the LOST theorem \cite{LewandowskiUniquenessOfDiffeomorphism}.

\subsubsection{Gau{\ss} law and spin networks}

The quantisation that we performed so far is called kinematical, since the dynamics of the theory, encoded in the constraints, in particular the Hamiltonian constraint, have not been taken into account so far. Let us start by imposing the Gau{\ss} law. For this, we would need to promote \eqref{eq:GaussAE} to an operator on our Hilbert space and compute its kernel. This can be done in full generality \cite{ThiemannModernCanonicalQuantum}. In order to avoid any technical overhead, we will choose a shortcut in this paper and simply compute the action of the classical Gau{\ss} law on cylindrical functions seen as phase space functions and demand invariance. This procedure will lead to the same result as quantising \eqref{eq:GaussAE}.  

Holonomies have the useful property that they are transforming under gauge transformations only at their endpoints, 
\be
	h^j_c(A) \rightarrow U\left(c(0)\right)  h^j_c(A) \,U^{-1}\left(c(1)\right) \label{eq:HolGaussFin}
\ee
or written infinitesimally (see exercises) 
\be
	\left\{ h^j_c(A), G_k[\Lambda^k] \right\} = - h^j_c(A) \Lambda^k\left(c(1)\right) \tau^{(j)}_k + \Lambda^k\left(c(0)\right) \tau^{(j)}_k h^j_c(A)  \label{eq:HolGaussInf} \text{.}
\ee

In order to construct a gauge invariant state, we therefore have to choose the cylindrical functions such that the transformations at the endpoints of holonomies cancel each other. This means that we need to look for tensors which are invariant w.r.t. the action of SU$(2)$ and contract all holonomies ending or starting at a given vertex with such a tensor, in a way that no free indices remain. 

The simplest example is known as a Wilson loop: we take a single curve with $c(0) = c(1)$, a loop, and simply trance over the holonomy. This corresponds to contracting its group indices with the Kronecker delta $\delta^\alpha {}_\beta$, which is an invariant tensor of SU$(2)$. Next, for a three-valent vertex, the invariant tensors turn out to be the Clebsch-Gordan coefficients familiar from the quantum mechanics of angular momentum, or the related Wigner 3J-symbols, which enjoy a higher symmetry. All higher invariant tensors can be build from contracting 3J-symbols in a suitable way.   

The gauge invariant (w.r.t. the Gau{\ss} law) part of the Hilbert space is thus spanned by the above type of a cylindrical function, called {\it spin network}. Briefly, we state that a spin network is defined by a graph $\gamma$ whose edges $c_i$ are coloured with SU$(2)$ spins $j_i$, and whose vertices $v_k$ are coloured with invariant tensors $\iota_k$ in the tensor product of the edge representations incident at $v_k$. For a spin network, we introduce the notation $T_{\gamma, \vec j, \vec \iota}$, which contains all relevant information. 
We limit the invariant tensors $\vec \iota$ within this notation to an orthonormal subset so that 
\be
	\braket{T_{\gamma, \vec j, \vec \iota}}{T_{\gamma, \vec j', \vec \iota'}}_{\text{kin}} = \delta_{\vec j, \vec j'}  \delta_{\vec \iota, \vec \iota'} \text{.}
\ee
The spin networks $T_{\gamma, \vec j, \vec \iota}$ then form an orthonormal basis of the kinematical Hilbert space restricted to the graph $\gamma$. 

\subsubsection{Spatial diffeomorphisms} 

Spatial diffeomorphisms are imposed in loop quantum gravity as finite transformations, since their generator does not exist as an operator on the Hilbert space (see exercises). The process of imposing finite diffeomorphisms is however relatively straight forward. In particular, the kinematical scalar product \eqref{eq:ScalarPKin} turns out to be invariant under spatial diffeomorphisms. 

On a holonomy, a finite spatial diffeomorphism $\phi$ acts as
\be
	\phi \left( h_c^j(A) \right) =  h_c^j\left(\phi^*A \right) = h_{\phi(c)}^j \left(A \right) \text{.} \label{eq:FiniteDiffeoAction}
\ee
We remark that for a 1-parameter family $\phi^{\vec N}_t$ diffeomorphism along a vector field $\vec N$ with $\phi_0^{\vec N} = \mathbb{1}$, we have $\phi^{\vec N}_t {}^* A = A + t \mathcal L_{\vec N} A + O(t^2)$, so that the Lie derivatives in \eqref{eq:LieDerFromHaAE} correspond to the infinitesimal transformations of \eqref{eq:FiniteDiffeoAction}. 
Finite spatial diffeomorphisms thus simply act by moving the graphs $\gamma$ on which spin networks are defined. 

Solving the spatial diffeomorphism constraint then amounts to demanding invariance under finite spatial diffeomorphisms. For this, one can employ a group averaging procedure \cite{GiuliniGroupAveragingAnd}, which roughly states that one should start with a given spin network and average it over the action of the spatial diffeomorphism group. Due to the special properties of the scalar product \eqref{eq:ScalarPKin}, such a procedure works even despite the infinite volume of the diffeomorphism group. However, solutions to the spatial diffeomorphism constraint are not elements of the kinematical Hilbert space, but have to be defined in its dual.

Given a spin network $T_{\gamma, \vec j, \vec \iota}$, we define the dual state 
\be
	 \bra{\eta[T_{\gamma, \vec j, \vec \iota}]} := \frac{1}{d_{\gamma, \vec j, \vec \iota}}~~ \sum_{\phi \in \text{Diff}_{\text{}}(\Sigma) / \text{Diff}_\gamma} ~~~ \bra{T_{\phi(\gamma), \phi(\vec j), \phi(\vec \iota) }} , \label{eq:RiggingMapDiffeo}
\ee
where $\text{Diff}_\gamma$ is the set of diffeomorphisms preserving $\gamma$ and $d_{\gamma, \vec j, \vec \iota}$ are symmetry factors depending on the spin network. $ \eta[T_{\gamma, \vec j, \vec \iota}] $ has the property that 
\be
	\braket{ \eta[T_{\gamma, \vec j, \vec \iota}] }{T_{\gamma', \vec j', \vec \iota' }}_{\text{kin}} =\braket{ \eta[T_{\gamma, \vec j, \vec \iota}] }{T_{\phi(\gamma'), \phi(\vec j'), \phi(\vec \iota') }}_{\text{kin}} 
\ee
for any diffeomorphism $\phi$. In particular, 
\be
	\braket{ \eta[T_{\gamma, \vec j, \vec \iota}] }{T_{\gamma', \vec j', \vec \iota' }}_{\text{kin}}  = \delta_{\vec j, \vec j'} \delta_{\vec \iota, \vec \iota'}
\ee
if $\gamma$ and $\gamma'$ are related by a diffeomorphism (this equation has to be understood in the sense of a common refinement as discussed above).
If we would not have divided by $\gamma$-preserving diffeomorphisms, we would have an uncountably infinite number of terms corresponding to the same spin network on the right hand side. Likewise, the symmetry factor $d_{\gamma, \vec j, \vec \iota}$ corrects for graph symmetries which map the coloured graph onto itself. 

It is possible now to define a new scalar product $\braket{\cdot}{\cdot}_{\text{diff}}$ between diffeomorphism invariant states, however we will not discuss this further here. Details about this construction can be found in \cite{AshtekarQuantizationOfDiffeomorphism, ThiemannModernCanonicalQuantum}.

\subsection{Outlook}

We are now done with defining the Hilbert space and implementing the ``kinematical'' constraints, i.e. those which do not contain the dynamics of our theory. In the next section, we will discuss the physical properties of (diffeomorphism invariant) spin network states by constructing geometric operators which act on them. A discussion of the dynamics, obtained by quantising the Hamiltonian constraint as well as other means, will take place in section \ref{sec:DynamicsAndOutlook}.

\subsection{Exercises}

\begin{enumerate}

	\item {\bf Lie derivative} \\The following is independent of the number of dimensions, we choose $a,b$ to label tensor indices, e.g. on a spatial slice. The Lie derivative  $\mathcal L$ with respect to a vector field $N^a \equiv \vec N$, an infinitesimal version of a diffeomorphism along $\vec N$, is defined as follows: On geometric scalars $s$, it acts as $n^\mu \partial_\mu s$.  On vector fields $v^a$, it acts as $\mathcal L_{\vec N} v^a = N^b \partial_b v^a - v^b \partial_b N^a$. The action on other tensor fields follows by the Leibnitz rule, e.g., by defining $w_{ab} = v_a v_b$, and demanding that, e.g., $v_a v^a$ transforms as a scalar. Show that 
	\begin{enumerate}
	
		\item $\mathcal L_{\vec N} v_a = N^b \partial_b v_a + v_b \partial_a N^b$
		
		\item $\mathcal L_{\vec N} q_{ab} = N^c \partial_c q_{ab} + q_{ac}\partial_{b} N^c + q_{bc}\partial_{a} N^c  = N^c \partial_c q_{ab} + 2 q_{c(a}\partial_{b)} N^c = 2 \nabla_{(a} N_{b)}$\\ 		where $\nabla_a N_b = \partial_a N_b - \Gamma_{ab}^c N_c$ and $\Gamma_{ab}^c := \frac{1}{2} q^{cd} \left( \partial_a q_{bd} + \partial_b q_{ad} - \partial_d q_{ab} \right)$ are the Christoffel symbols and $T_{(ab)}  :=( T_{ab}+T_{ba})/2$ denotes symmetrisation with weight 1. 
		
		\item $\mathcal L_{\vec N} \sqrt{q}^{n} = N^c \partial_c \sqrt{q}^{n} + n \, \sqrt{q}^{n} \partial_c N^c$, ~~ with $q = \det q_{ab}$
		
		\item $\mathcal L_{\vec N} \sqrt{q} = \partial_a \left( \sqrt{q} N^a \right)$. This means that in geometrically well-defined integrals (where the integrand has density weight $n=1$), we can partially integrate the Lie derivative. 
		
	\end{enumerate}
	
	\item {\bf Spatial diffeomorphism constraint}\\
	Show that up to a boundary term, $\mathcal H_a[N^a] = \int_\Sigma d^3x \, P^{ab} \mathcal L_{\vec N} q_{ab}$. Using the results of the previous exercise, compute the action of $\mathcal H_a[N^a]$ on $q_{ab}$ and $P^{ab}$. Proof equation \eqref{eq:PBHHa} and \eqref{eq:PBHaHa} by using the Leibnitz rule of the Lie derivative. Convince yourself that the total volume of compact $\Sigma$ Poisson commutes with the spatial diffeomorphism constraint. Find other observables with the same property. \\
	Show that the spatial diffeomorphism constraint in $E, K$ variables, again up to a boundary term, reads $\mathcal H_a[N^a] = \int_\Sigma d^3x \, E^{ai} \mathcal L_{\vec N} K_{ai}$, where the Lie derivative does not act on SU$(2)$ indices $i,j$.

	\item {\bf ADM Poisson brackets from $\{K, E\}=\delta$}\\
	Proof \eqref{eq:ADMPBFromEK}. For this, express the ADM variables $q_{ab}$, $P^{ab}$ in terms of $K_{a}^i$, $E^a_i$, and evaluate their Poisson brackets using \eqref{eq:KEBracket}.
	
	\item {\bf Gau{\ss} law}\\
	Show that the Gau{\ss} law forms a closed subalgebra. Proof \eqref{eq:GaussAE}. For this, show that the additional terms needed to make $D_a$ act also on tensor indices via the Christoffel symbols cancel. 
	
	\item {\bf Canonical transformation (hard!)}\\
	Show that \eqref{eq:AEBracket} is the only non-vanishing Poisson bracket. For this, it is useful to first show that $\Gamma_a^i$ has a generating potential, i.e. that $\Gamma_a^i = \delta F / \delta E_i^a$ with $F = \int_\Sigma d^3x E^{a}_i \Gamma_a^i$. The explicit expression for $\Gamma_a^i$ can be derived from $\partial_a e_b^i - \Gamma_{ab}^c e_c^i + \epsilon^{ijk} \Gamma_a^j e_b^k = 0$.
	
	\item {\bf Hamiltonian constraint}\\
	Show that the Hamiltonian constraint can be written as \eqref{eq:HamConstrAE} up to terms proportional to the Gau{\ss} law. For this, it is useful to define $\Gamma_{aij} = -\epsilon_{ijk} \Gamma_a^k$, for which the defining equation reads $\partial_a e_b^i - \Gamma_{ab}^c e_c^i + \Gamma_{a} {}^i {}_k e_b^k = 0$. 
	Then, we define the field strengths $R(\Gamma)_{abij} = 2 \partial_{[a} \Gamma_{b]ij} + \left[\Gamma_a, \Gamma_b \right]_{ij}$ and $R_{abi} = 2 \partial_{[a} \Gamma_{b]}^i + \epsilon^{ijk} \Gamma_a^j \Gamma_b^k$. What is the relation between $R_{abi}$ and $R_{abij}$?
	Show that $R_{abij} e_c^i e_d^j = R_{abcd}$, where $R_{abcd}$ is the Riemann tensor defined with the convention $[\nabla_a, \nabla_b]  u_c = R_{abc} {}^d u_d$. Show that in components $R_{ab} {}^c {}_{d} = 2 \partial_{[a} \Gamma_{b]d}^c + 2 \Gamma_{d[c}^f \Gamma_{a]f}^c$.
		
	\item {\bf Holonomies and fluxes}\\
	Show that \eqref{eq:HolPathOrder} is a solution of \eqref{eq:HolDef}. Convince yourself that with this definition, a parallel transport happens ``from left to right'', i.e. $v^\alpha(c(0)) \mapsto v^\alpha(c(0)) h^j_c(A)_{\alpha} {}^\beta$. Show that $h^j_{c_1 \circ c_2}(A) = h^j_{c_1}(A) h^j_{c_2}(A)$ and $h^j_{c^{-1}}(A) = h^j_c(A)^{-1}$ by using ${}^\dagger = {}^{-1}$ for SU(2). Convince yourself about \eqref{eq:HolFluxBracket}.

	\item {\bf Scalar product}\\
	Show that the scalar product is discontinuous in the following sense. Take two copies of an arbitrary cylindrical function. Add another non-trivial holonomy defined on a curve $c$ to one of the cylindrical functions. Compute the scalar product. What happens in the limit when $c$ is shrunk to a point? 
	Similarly, consider a definition of an operator corresponding to $A_a^i$ as a derivative w.r.t. the curve length of a holonomy, evaluated at zero curve length. 
	By Stone's theorem, it follows that there cannot exist an operator corresponding to $A_a^i$ (in unexponentiated form) in the quantum theory. 
	
	\item {\bf Solving the Gau{\ss} law}\\
	Show \eqref{eq:HolGaussInf} by computing the Poisson bracket. Show that this is the infinitesimal generator of \eqref{eq:HolGaussFin}. Verify explicitly that a Wilson loop is gauge invariant. Show by considering infinitesimal gauge transformations that in the tensor product of three spin $1$ representations, the totally anti-symmetric Levi-Civita symbol $\epsilon^{ijk}$ is an invariant tensor. 
	
	\item {\bf Ashtekar-Lewandowski vacuum} \\
	Consider the constant cylindrical function $\Psi = 1$. This state is known at the Ashtekar-Lewandowski vacuum. Show that it corresponds to a maximally degenerate spatial geometry by evaluating the vacuum expectation value of flux operators.

	\item {\bf The virtue of $\beta = \pm i$}\\
	Set $\beta = \pm i$. Compute the Poisson bracket of two Hamiltonian constraints as given in \eqref{eq:LieDerFromHaAE}. For this, first show that $\delta F_{ab}^i = 2 D_{[a} \delta A_{b]}^i$. Attempt the same computation for $\beta \neq \pm i$ or in metric variables and compare the level of difficulty. \\However, note that the reality conditions in this case become $A_a^i + \bar A_a^i = 2 \Gamma_a^i(E)$, so that the non-polynomiality is hidden, but not overcome.
	
\end{enumerate}

\section{Geometric operators, matter, and quantum geometry}

In this section, we will discuss the construction of geometric operators and compute their action on spin network states. The strategy to derive those operators is to regularise them in terms of holonomies and fluxes and to compute their action in the limit of infinitely small holonomies and fluxes, where the exact classical expression is approached. 
We will again skip many technical details and only consider the most instructive cases. For a rigorous treatment, see \cite{ThiemannModernCanonicalQuantum}. For original literature, see \cite{SmolinRecentDevelopmentsIn, AshtekarQuantumTheoryOf1, RovelliDiscretenessOfArea, AshtekarDifferentialGeometryOn, AshtekarQuantumTheoryOf2}. 

Next, we are going to discuss how to couple matter to this theory. The emerging picture is very similar to lattice gauge theory, with the additional twist that the lattices themselves are now dynamical objects describing the gravitational sector of the theory. For original literature, see \cite{ThiemannKinematicalHilbertSpaces, ThiemannQSD5, AshtekarPolymerAndFock}, as well as \cite{AshtekarGravitonsAndLoops, VaradarajanPhotonsFromQuantized, VaradarajanGravitonsFromA, AshtekarPolymerAndFock} for a comparison to standard Fock representations. A textbook treatment is given in \cite{ThiemannModernCanonicalQuantum}.

\subsection{Area operator}

The seminal paper on the area operator in \cite{SmolinRecentDevelopmentsIn}, whereas the construction was made rigorous in \cite{AshtekarQuantumTheoryOf1}. In order to construct an operator corresponding to the area $A(S)$ of a given surface $S$ embedded in $\Sigma$, we need to write the classical quantity $A(S)$ in terms of holonomies and fluxes. Since $A(S)$ is only a function of the spatial metric, fluxes are already enough to define this operator, which drastically simplifies its action on spin network states as we will see. 

Classically, we have 
\be
	A(S) = \int_U d^2u \, \sqrt{ \det (X^*q) (u)}
\ee
where $X : U \rightarrow S$ is an embedding of the coordinate chart $U$ into $S$, and $X^*q$ is the induced metric on $S$. We partition $U$ into disjoint subsets $U_i$ as $U = \cup_i U_i$, so that 
\be
	A(S) = \sum_i \int_{U_i} d^2u \, \sqrt{ \det (X^*q) (u)} \text{.}
\ee
In the limit of small $U_i$, we have, noting the $\beta$-dependence of $E^a_i$, 
\be
	 \int_{U_i} d^2u \, \sqrt{ \det (X^*q) (u)} =  \int_{X(U_i)}  \sqrt{q q^{ab} d s_a d s_b} \approx  \beta \sqrt{E^k(U_i) E_k(U_i)} \label{eq:Induced2dAreaApprox}
\ee
where $E^k(U_i) = \int_{U_i}  E^a_k dU_a$ (see exercises). Therefore, 
\be
	A(S) = \lim_{U_i \rightarrow 0} \beta \sum_i  \sqrt{E^k(U_i) E_k(U_i)} \text{.} \label{eq:ClassRegArea} 
\ee

\begin{figure}[h]
	\centering\includegraphics[trim = 02mm 140mm 50mm 70mm, clip, scale=0.6]{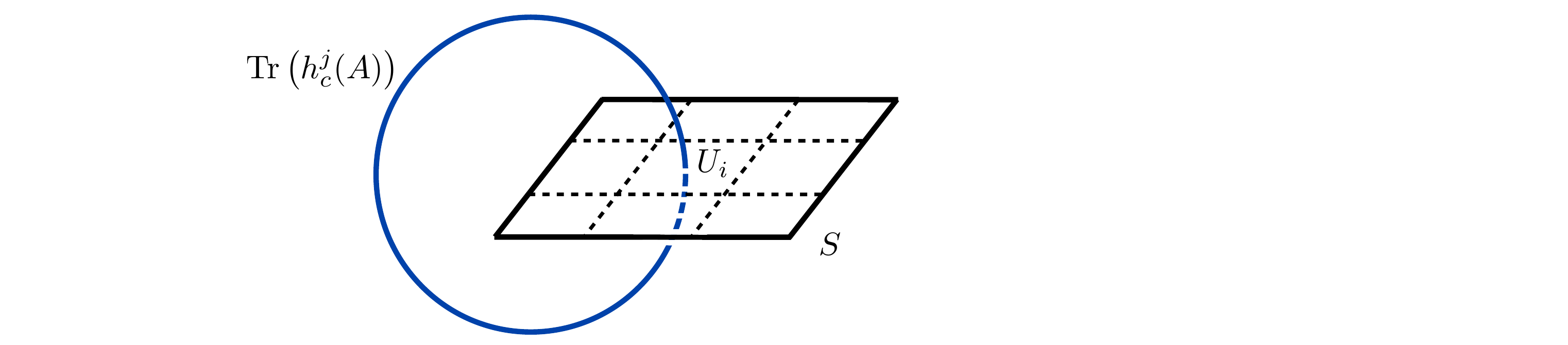} 
		\caption{The setup for the simplest non-trivial action of the area operator $\hat A(S)$ is depicted. The surface $S$ is partitioned into small subsurfaces $U_i$. The operator acts on a single Wilson loop which interests $S$ exactly once.}
		\label{fig:AreaOperatorAction}
	\end{figure}

We can promote \eqref{eq:ClassRegArea} to on operator by substituting \eqref{eq:DefQuantFlux} for the classical flux. 
In order to deal with the square root, we first compute the action of $\hat E^k(U_i) \hat E_k(U_i)$ on a Wilson loop $\Psi = \text{Tr} \left(h_c^j(A)\right)$ in the case where the curve $c$ intersects $U_i$ exactly once, see figure \ref{fig:AreaOperatorAction}.
We obtain
\begin{align}
	 \hat E^k(U_i) \hat E_k(U_i) \ket{\text{Tr} \left(h_c^j(A)\right)} &= - i \hbar \hat E^k(U_i) \ket {\text{Tr} \left(  h^j_{c_1}  \left( \tau^{(j)}_k  \right) h^j_{c_2}  \right)} \\
	 & = - \hbar^2 \ket{ \text{Tr} \left(  h^j_{c_1}  \left( \tau^{(j)}_k  (\tau^{(j)}) ^k   \right) h^j_{c_2} \right)} \\
	 & = \hbar^2 j (j+1) \ket{ \text{Tr}  \left(h_c^j(A)\right) } \text{.}
\end{align}

Due to the diagonal action, we can define the square root of this operator via the spectral theorem ($\hat E^k(U_i)$ can also be shown to be self-adjoint), so that
\be
	\sqrt{ \hat E^k(U_i) \hat E_k(U_i) } \ket{\text{Tr} \left(h_c^j(A)\right)} = \hbar \sqrt{ j (j+1) } \ket{\text{Tr}  \left(h_c^j(A)\right) } \
\ee

We are now in a position to construct the area operator. Again, we stick to the simplest possible case, where the area $S$ intersects a Wilson loop $\text{Tr} \left(h_c^j(A)\right)$ exactly once, and not in a vertex (where the formulas become more complicated). Since contributions to \eqref{eq:ClassRegArea} vanish whereever $U_i$ does not intersect the holonomy, we simply get
\be
	\hat A (S) \ket{\text{Tr} \left(h_c^j(A)\right)} = \hbar \beta \sqrt{j(j+1)} \ket{\text{Tr} \left(h_c^j(A)\right)} \text{.} \label{eq:AreaOpSingleHol}
\ee

It is now straight forward to generalise this result to more complicated (non-vertex) intersections: for every (non-vertex) intersection with an edge of a spin network coloured with a spin $j$, the eigenvalue of the area operator increases by $\hbar \beta \sqrt{j(j+1)}$. In case of an intersection at a vertex, the resulting formula is more complicated and given in \cite{ThiemannModernCanonicalQuantum}.

Due to our suppression the gravitational constant ($8 \pi G = 1$), we actually have $\hbar  = 8 \pi  \hbar G= 8 \pi \l_p^2$, so that the smallest eigenvalue of the area operator, called area gap, is obtained for a single $j=1/2$ intersection as $4 \sqrt{3}  \pi \, \beta \, l_p^2$. We see that $\beta$ sets the scale of the quantum geometry w.r.t. the Planck scale.

\subsection{Volume operator}

The volume operator $\hat V(R)$ of a region $R$ \cite{SmolinRecentDevelopmentsIn, RovelliDiscretenessOfArea, AshtekarQuantumTheoryOf2} is constructed in a similar way than the area operator by regularising the classical expression 
\be
	V(R) = \int_R d^3x \sqrt{\det q_{ab}} \text{.}
\ee
The process of regularisation turns out to be more involved than for the area operator and in particular the final expression is a lot more complicated. Here, we will only derive an important property of the volume operator which will be key for the interpretation of spin networks as providing states corresponding to quantum geometries. 

\begin{figure}[h]
	\centering\includegraphics[trim = 02mm 10mm 5mm 10mm, clip, scale=0.3]{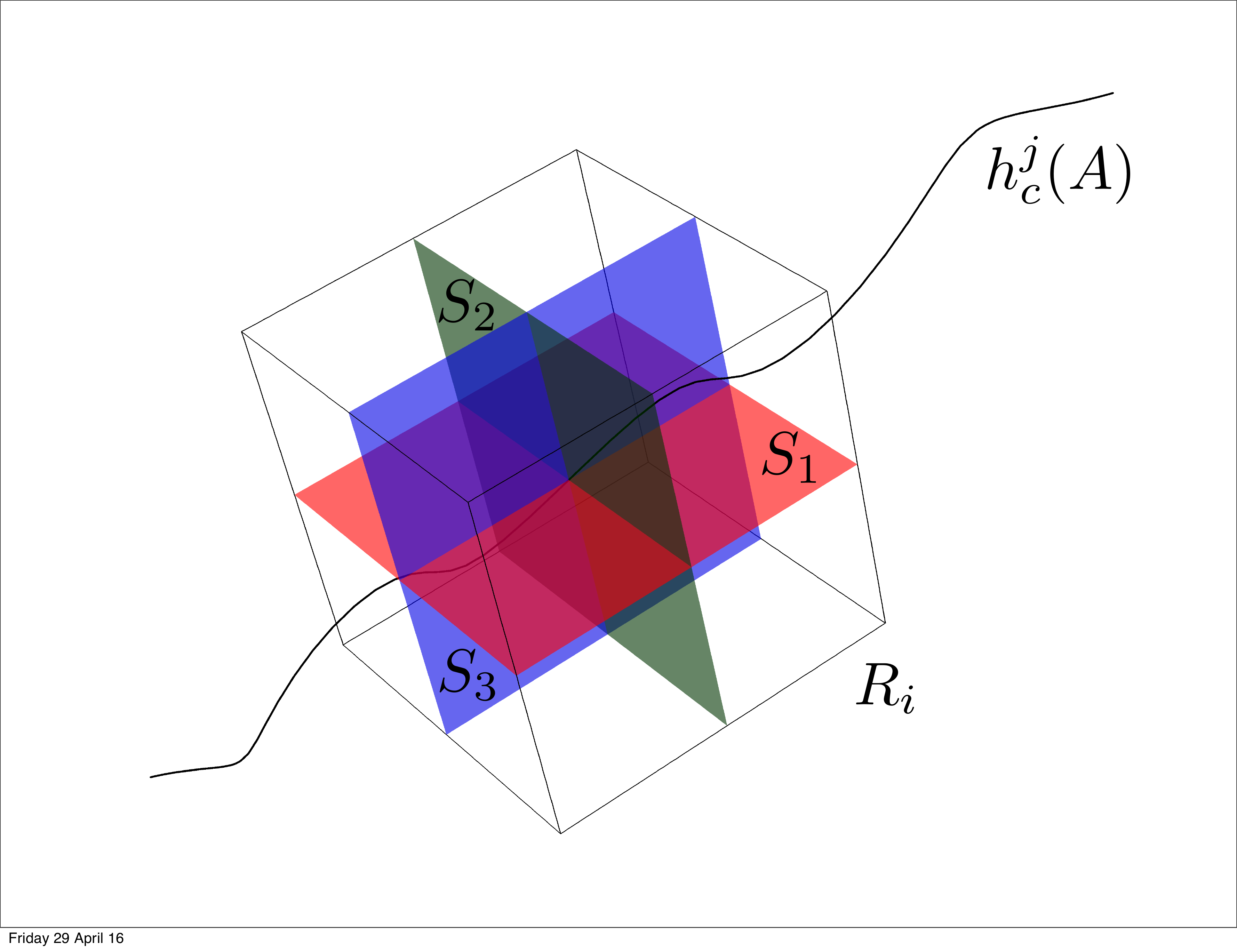} 
		\caption{The construction underlying the volume operator $\hat V(R)$ acting on a single holonomy is sketched. The region $R$ is partitioned into small cubes, of which only $R_i$ is shown. The three fluxes in the volume operator restricted to $R_i$ are evaluated on the surfaces $S_1$, $S_2$, and $S_3$, whose intersection is the only intersection of the curve $c$ with any of the surfaces. }
		\label{fig:VolumeHolonomy}
	\end{figure}

In terms of densitised triads, the classical volume reads
\be
	V(R) = \beta^{3/2} \int_R d^3x \sqrt{| E^a_i E^b_j E^c_k \epsilon_{abc} \epsilon^{ijk} /6 |} \text{.}
\ee
To promote this expression to an operator, we would again partition $R$ into small regions $R_i$, so that in the limit of infinitesimally small regions, we can replace $ E^a_i E^b_j E^c_k \epsilon_{abc} \epsilon^{ijk} $ by the product of three fluxes: 
\be
	\int_{R_i} d^3x \sqrt{| E^a_i E^b_j E^c_k \epsilon_{abc} \epsilon^{ijk} /6 |} \approx \sqrt{| E_i(S_1) E_j(S_2) E_k(S_3) \epsilon^{ijk} |}
\ee
Then, we again evaluate the action of the term under the square root when acting on a spin network. Let us first consider the case of a (part of a) holonomy contained in $R_i$ depicted in figure \ref{fig:VolumeHolonomy}. Let us evaluate the action of two rightmost fluxes. We get
\be
	\hat E_i(S_1) \hat E_k(S_2) \ket{h_c^j(A)} = - \hbar^2 \frac{1}{2} \ket{  h^j_{c_1}  \left( \tau^{(j)}_i   \tau^{(j)}_k +  \tau^{(j)}_k   \tau^{(j)}_i \right) h^j_{c_2}  } \text{.} \label{eq:VolTwoFluxAction}
\ee
The symmetrisation on the right hand side comes from the precise regularisation of the Poisson bracket such as \eqref{eq:HolFluxBracket}, on which we did not comment on here. Roughly speaking, one needs to extend surfaces to thin 3-dimensional regions and curves to 3-dimensional tubes. As this has should be done symmetrically, one obtains symmetrised expressions such as in \eqref{eq:VolTwoFluxAction}. 

Due to the contraction of \eqref{eq:VolTwoFluxAction} with $\epsilon^{ijk}$ in the volume, the action of the volume operator vanishes on a holonomy. It follows that a non-vanishing action of the volume operator requires that the spin network that we act upon contains at least one three-valent\footnote{It turns out that the volume operator also vanishes on gauge (Gau{\ss}) invariant three-valent vertices, but not necessarily on gauge-variant ones.} vertex in the region $R$. Then, the three fluxes can act on three different holonomies, and thus result in the generators $\tau^{(j)}_i$ being contracted with three different indices of the invariant map. Such constructions do not vanish in general, but yield another invariant map. 

We will not display here the complete expression for the volume operator, which is in fact quite lengthy and not very illuminating, see e.g. \cite{ThiemannModernCanonicalQuantum}. It is however clear from the previous discussion that the volume operator preserves spin network states and only acts on their invariant maps. It is in principle possible to choose a basis in the set of invariant maps such that the volume operator is diagonal, however not analytically, as this would require to diagonalise matrices of arbitrary dimension. This is in particular problematic since the volume operator enters the dynamics of loop quantum gravity in the construction of the Hamiltonian constraint. 
So far, it is only known that the spectrum of the volume operator is discrete.

\subsection{Quantum geometry}

\begin{figure}[h]
	\centering\includegraphics[trim = 0mm 200mm 100mm 0mm, clip, scale=0.61]{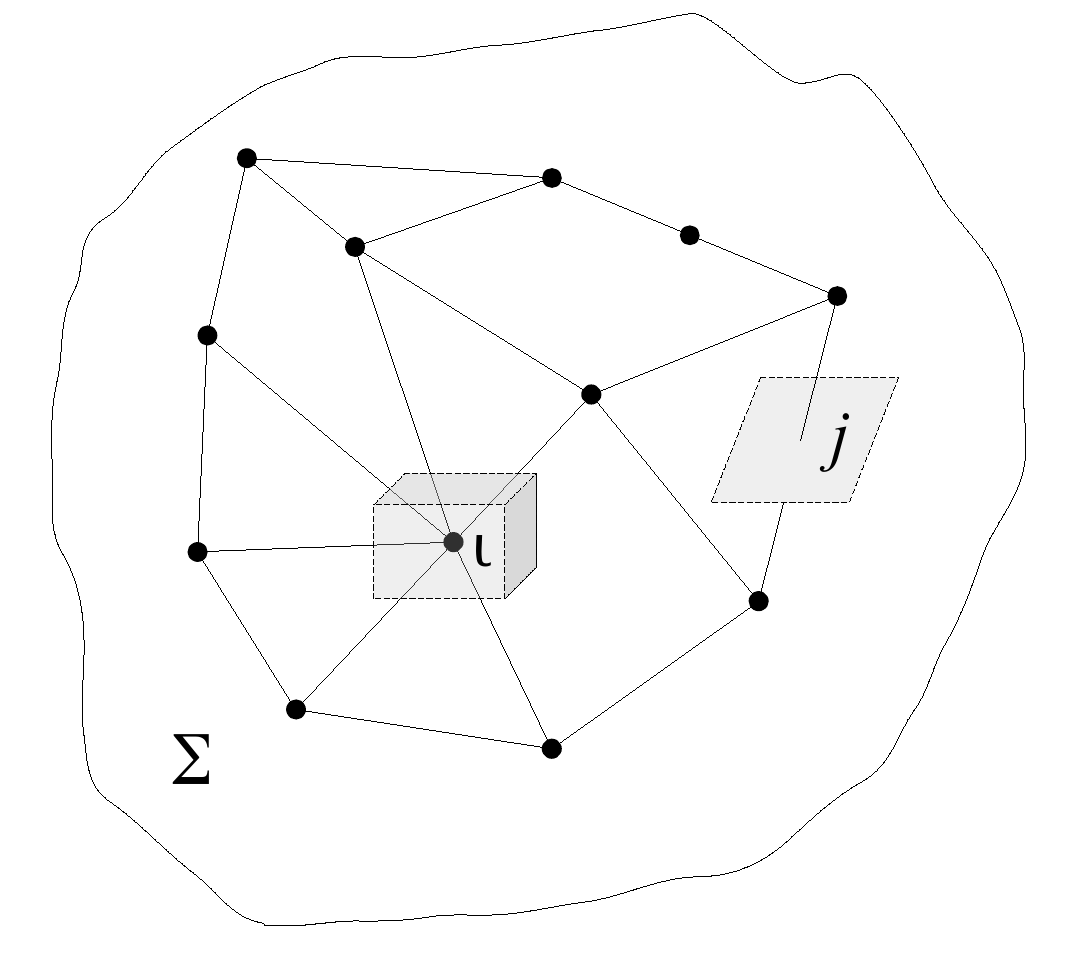} 
		\caption{A spin network is embedded into the spatial slice $\Sigma$. At vertices of the graph underlying the spin network, the volume operator can have a non-vanishing action determined by the invariant map $\iota$. The area operator is non-trivial when the underlying area is intersected by edges of the graph labelled with $j \neq 0$. The emerging picture is that of a discrete geometry where quanta of volume are connected by the graph edges, which, at the same time, determine the area quanta separating two volume quanta. }
		\label{fig:QuantumGeometry}
	\end{figure}

We are now in a position to give an interpretation to spin network states as quantum geometries. For this, we recall that a spin network function was defined as a coloured graph embedded in the spatial slice $\Sigma$. From the discussion of the geometric operators, we recall that the area operator was non-vanishing only if it intersects a holonomy. The volume operator on the other hand was non-vanishing only when acting on a vertex. A spin network thus corresponds to a quantum state where the geometry is excited in such a way that there are quanta of volume at the vertices of the graph, as well as quanta of area on surfaces intersected by it. The edges of the graph thus define a certain notion of connectedness for two neighbouring quanta of volume, associated with the magnitude of a surface separating them. The situation is depicted in figure \ref{fig:QuantumGeometry}.

It is important to note that the picture of quantum geometry drawn here is very qualitative, but already enough to understand how to couple matter to the theory. More detailed investigations of the quantum geometry underlying loop quantum gravity can be found in \cite{FreidelTwistedGeometriesA, BianchiPolyhedraInLoop}.

The geometric operators are a priori defined only on the kinematical Hilbert space, where spatial diffeomorphism invariance has not been implemented yet. In order to construct spatially diffeomorphism invariant geometric operators, we need to localise the surfaces and regions which we want to measure w.r.t. other entities in the theory, such as additional matter fields. This can for example be done at the classical level using deparametrisation, see for example \cite{GieselAQG4}. Another possibility is to consider spatially diffeomorphism invariant quantities such as the total volume of the universe $V(\Sigma)$ or the area of a boundary of $\Sigma$, such as a spatial section of a black hole horizon. 

Another issue of spin networks states is that they are sharply peaked on the intrinsic geometry characterised by the fluxes, whereas holonomies are maximally uncertain. To overcome this, one can consider superpositions of spin networks defined on the same graph to construct coherent states peaked on both fluxes and holonomies with minimal uncertainty. This task was achieved in \cite{ThiemannGCS1, ThiemannGCS2, ThiemannGCS3} and applied to compute the expectation value of the Master constraint (a sum of the squared constraints) in \cite{GieselAQG2, GieselAQG3}, showing consistency with lattice regularised general relativity.

\subsection{Matter}

So far, we only considered pure general relativity without any matter couplings. In order to couple matter, we add suitable matter contributions already at the classical level and include them in the quantisation procedure. The details of doing this can be motivated by looking at lattice gauge theory, where one needs to express continuum actions on a discretised spacetime. 
A spatial slice of the underlying lattice then corresponds to the quantum geometry encoded by a spin network state. 
We will again be brief in this section and discuss only the most important conceptual steps in coupling matter. 

The incorporation of other gauge fields with compact gauge group is straight forward. We coordinatise the their phase spaces again by holonomies and fluxes. The cylindrical functions that we considered in the gravitational case now are also allowed to depend on a finite number of holonomies of other gauge fields. The additional gauge constraints lead us to consider generalised spin networks, where the label set includes representation labels and invariant maps for all gauge groups. The scalar product is generalised by adding integrations w.r.t. the Haar measure of the other gauge groups.

Next, we consider fermions. We recall that fermions carry spinor indices which transform under the Lorentz group, as well as colour indices (e.g. gluons), which transform under the internal gauge groups. The spinor indices corresponding to the Lorentz group turn out to transform under the gravitational gauge group SU$(2)$ once a properly gauge fixed Hamiltonian formulation of general relativity coupled to fermions is established \cite{KibbleCanonicalVariablesFor, AshtekarInclusionOfMatter, ThiemannQSD5}. 
In order to construct gauge invariant states, one should therefore consider fermions at vertices of the spin networks. There, the free indices of the fermions can be contracted with invariant maps, whose other free indices are contracted with holonomies. The details of the precise derivation of these statements are somewhat technical, especially at the classical level. In particular, it turns out that the proper density weight of a fermion should be $1/2$ in order to have simple Poisson brackets \cite{ThiemannQSD5}.

Let us now consider scalar fields. Due to its geometric properties, a scalar field $\phi$ should be evaluated at a point with no further smearing. In order to construct a background independent representation for the classical Poisson algebra of the scalar field and its momentum, we will consider so called point holonomies $h_x^\lambda(\phi) = e^{i \lambda \phi(x)}$ of the scalar field. (We will specify the range of $\lambda$ later.) Cylindrical functions $\Psi$ are again defined as complex valued functions depending on a finite number of point holonomies. The momentum $\pi_\phi$ conjugate to a scalar field is geometrically a density, so we integrate it over a 3-dimensional region $R$. $\pi_\phi(R)$ then corresponds to a flux since we have
\be
	\left\{ h_x^\lambda (\phi), \pi_\phi(R) \right\} = \epsilon(x,R) \, i \,  \lambda \, h_x^\lambda (\phi), ~~~~~~ \epsilon(x,R) = \begin{cases} 1  &\mbox{if } x \in R \backslash \partial R \\ 
	\frac{1}{2} &\mbox{if } x \in  \partial R \\ 
	0 &\mbox{if } x \notin R \cup \partial R \text{,}
	 \end{cases}  
\ee
and it acts on a cylindrical function as
\be
	\hat \pi_\phi (R) \ket{\Psi \left(h_{x_1}^{\lambda_1}, \ldots, h_{x_n}^{\lambda_n}\right)} = \sum_{i=1}^n \epsilon(x_i, R) \lambda_i  \ket{\Psi \left(h_{x_1}^{\lambda_1}, \ldots, h_{x_n}^{\lambda_n}\right)} \text{.}
\ee

The scalar product between two cylindrical functions is again defined by expressing them on a common refinement and integrating against the Haar measure. Here, we need to make a distinction between $\lambda \in \mathbb Z$, which corresponds to using U$(1)$, and $\lambda \in \mathbb R$, which corresponds to using the Bohr compactification $\mathbb R_\text{Bohr}$ of the real line\footnote{The Bohr compactification $\mathbb R_\text{Bohr}$ of the real line is a compact Abelian group with representations labelled by $\lambda \in \mathbb R$. In particular, a normalised Haar measure exists, which allows us to use this group in the context of loop quantum gravity. $\mathbb R_\text{Bohr}$ is used mainly in the context of loop quantum cosmology, where a Hilbert space based on this group \cite{AshtekarMathematicalStructureOf} is necessary to define viable dynamics \cite{AshtekarQuantumNatureOf} for choices of variables different than in section \ref{sec:LQCBouncebv}.}. For concreteness, let us write in the U$(1)$ case
\be
	\braket{\Psi}{\Phi} = \int_{U(1)^n} dg^n \,\overline {\Psi\left(g_1, \ldots, g_n \right)} \Phi\left(g_1, \ldots, g_n \right) \text{.}
\ee
While we can in principle have a non-trivial dependency on the scalar field at arbitrary points in $\Sigma$, it makes most sense to put scalar field excitations at spin network vertices. Otherwise, the contribution of this excitation to the Hamiltonian constraint vanishes due to the coupling to geometry, as we will see in more detail in the next section.

The intuitive picture of matter coupling is thus the same as in lattice gauge theory. Scalar fields and fermions are sitting at vertices, where quanta of volume are located, whereas (1-form) gauge fields have support on edges connecting vertices. The main difference is that our theory is in addition subject to the spatial diffeomorphism and Hamiltonian constraints, which act on both the gravitational and matter sector. The spatial diffeomorphism constraint enforces that the location of vertices and edges on $\Sigma$ is washed out, while only their relations among each other, e.g. their connectedness, remain. The physics of the quantised Hamiltonian constraint, encoding the dynamics, is so far not well understood. We will consider various approaches to defining the dynamics in the next section. More exotic and so far not observed particles such as higher form fields or spin $3/2$ particles appearing in (higher-dimensional) supergravity theories can also be treated with similar methods \cite{AriasSecondQuantizationOf, Armand-UgonTowardsALoop, LingSupersymmetricSpinNetworks, BTTVI, BTTVII}, but we will not detail this here, save for one exercise below.

\subsection{Exercises}

\begin{enumerate}

	\item {\bf Induced metric density} \\ 
	Show \eqref{eq:Induced2dAreaApprox}. Use $ds_a = \epsilon_{abc} dx^b \wedge dx^c = \frac{1}{2} \epsilon_{abc} \epsilon^{bc} d^2x$, where $\epsilon^{bc}$ is the totally antisymmetric tensor in $2$ dimensions with $\epsilon^{12} = 1$. 
	
	\item {\bf State counting and black hole entropy} \\ 
	Consider a surface $S$ of total area $A$, e.g. a spatial slice of the horizon of a black hole. The goal of this exercise is to determine the number of microscopic quantum states corresponding to a total area $A$ of $S$. We make the assumption that $S$ is intersected by holonomies only transversally and that there are no spin network vertices on $S$, such that we can use the simple form \eqref{eq:AreaOpSingleHol} for the area operator, and that two intersections with the same $j$ can be distinguished. The dimension of a representation space of SU$(2)$ of spin $j$, i.e. the dimension of the Hilbert space associated to a single holonomy, is $2j+1$. Parts of the spin networks which are not holonomies intersecting $A$ are traced over, i.e. not taken into account in the counting. We thus only consider ``surface states'', i.e. do not imply a strong notion of holography that would count also the states inside the black hole. \\
	Show that the number of states corresponding to a configuration $s_j$, where $s_j$ denotes the number of intersections with spin $j$ is given by 
	\be
		N = \frac{\left( \sum_j s_j\right)!} {\prod_j \left(s_j ! \right)} \prod_j (2j+1)^{s_j}
	\ee
	Approximate $\log N$ by using Stirling's formula. Extremise $\log N$ w.r.t. $s_j$ under the total area constraint $8 \pi l_p^2 \beta \sum_j s_j \sqrt{j(j+1)} = A$ by using the Lagrange multiplier method, i.e. solve $\delta (\log N + \lambda (8 \pi l_p^2 \beta \sum_j s_j \sqrt{j(j+1)} - A)) = 0$, where $\delta$ varies only $s_j$. Derive an implicit equation for $\lambda$ as a function of $\beta$. Derive an explicit expression for the function $s_j$. What is the value of $\log N$? \\
		An approach to computing black hole entropy is to identify the logarithm $\log N$ of the number of states $N$ leading to a total area $A$ as the black hole entropy. One would expect that this quantity is maximised as in this computation.
		What is the value of $\beta$ so that the Bekenstein-Hawking entropy $2 \pi A$  in units $8 \pi G = 1 = \hbar$ is obtained? Is this a sound method to derive black hole entropy?
		This computation underlying this exercise has been given in \cite{GhoshAnImprovedEstimate}.

	\item {\bf Charge quantisation in electrodynamics}
	
	Following \cite{CorichiAmbiguitiesInLoop}, consider Maxwell theory in a Hamiltonian formulation, that is with the Poisson bracket $\{A_a(x), E^b(y)\} = \delta_a^b \delta^{(3)}(x,y)$ and the Gau{\ss} law $\partial_a E^a \approx 0$. $E^b$ here denotes the electric field, which is geometrically a densitised vector field, as $E^a_i$ in the main text. $A_a$ is the vector potential of the magnetic field.\\
	Quantise this theory using the methods developed for the gravitational field by using the group U$(1)$ instead of SU$(2)$. Show that the electric charge is quantised by analysing the operator corresponding to the electric flux through a closed surface. \\
	What happens if you substitute U$(1)$ by $\mathbb R_{\text{Bohr}}$, where the integers labelling the representations of U$(1)$ are substituted by arbitrary real numbers?

	\item {\bf Higher $p$-form fields}
	
	Consider an Abelian 2-form field, that is a density 0, rank 2 covariant tensor field $A_{ab} = - A_{ba}$, with Poisson bracket $\{A_{ab}(x), E^{cd}(y)\} = \delta_{[a}^c \delta_{b]}^d \delta^{(3)}(x,y)$, subject to the Gau{\ss} law $\partial_a E^{ab} \approx 0$. Generalise the quantisation techniques from the main text to this case. What is the gauge invariant field strength? How does the holonomy-flux algebra look like? What is the Hilbert space? How do we solve the Gau{\ss} law?
	What is the generalisation to $p$-forms, $p \in \mathbb N$, in arbitrary spatial dimensions? These questions were discussed in \cite{AriasSecondQuantizationOf, BTTVII}.

	\item {\bf Alternative connection variables and non-degenerate vacua}\\
	Show that $E^{a}_i \rightarrow E^{a}_i + \theta \epsilon^{abc} F_{bc}^i(A)$, $\theta \in \mathbb R$ is a canonical transformation (up to a boundary term to be neglected here). Use the same quantisation techniques as for standard Ashtekar-Barbero variables. What are the implications for the quantum geometry in the vacuum $\Psi(A) = 1$? The answer and further references can be found in \cite{BodendorferSomeNotesOn}.

\end{enumerate}

\section{Quantum dynamics and outlook} \label{sec:DynamicsAndOutlook}

\subsection{Dynamics}

\subsubsection{Canonical definitions}

\paragraph{Interpretation} \mbox{}\\
\indent In the Hamiltonian formulation, the dynamics of the theory is implemented by the Hamiltonian constraint $\mathcal H$. At the classical level, physical observables $\mathcal O$ are defined as functions Poisson commuting with the Hamiltonian constraint and evaluated on the constraint surface where $\mathcal H = 0$. An example of such an observable was given in section \ref{sec:LQC} in the context of a homogeneous and isotropic spacetime. The key idea was that the change in the physical system has to be seen w.r.t. a clock inherent in the system, e.g. the value of a scalar field. Evolution w.r.t. coordinate time is not physically meaningful by itself due to the diffeomorphism symmetry. 

At the quantum level, observables have the same interpretation, as already mentioned in section \ref{sec:LQC}, but are defined as operators $\hat{\mathcal O}$ which commute with the Hamiltonian constraint operator $\hat {\mathcal H}$. Physical states of the system are annihilated by the Hamiltonian constraint. In order to define the quantum dynamics, it is therefore left to properly define an operator $\hat {\mathcal H}$ and study its properties, in particular its kernel. This form of solving the constraints (at the quantum level) is referred to as Dirac quantisation. There exists however also an alternative approach:

It is also possible to solve the constraint already at the classical level, proceed to the reduced (``physical'') phase space, and quantise a complete set of observables. While this is often believed to be too complicated to achieve, several examples are known where this is in fact possible. The key point here is to choose suitable clock and rod fields at the classical level, which serve as physical coordinates. Depending on the choice of such fields, the resulting true Hamiltonian can be very complicated and non-local. However, for certain choices of matter fields, it turns out to be simple enough. For original literature, see for example \cite{BrownDustAsStandard}, as well as \cite{GieselScalarMaterialReference} for a recent review. Let us just mention two examples here:

The most straight forward example is to consider a massless scalar field as done in \cite{RovelliThePhysicalHamiltonian, DomagalaGravityQuantizedLoop}, and also in section \ref{sec:LQC}. The Hamiltonian constraint in this case reads
\be
	\mathcal H[N]= \int d^3x \, N \left( \frac{2}{\sqrt{q}} \left( P^{ab} P_{ab} - \frac{1}{2} P^2 \right) - \frac{\sqrt q}{2} \stackrel{(3)} R + \frac{P_\phi^2}{2 \sqrt{q}}\right) \label{eq:HamScalarConstr}
\ee
Deparametrisation (i.e. solving the constraints) amounts to rewriting\footnote{See \cite{DomagalaGravityQuantizedLoop} for a discussion of the different choices of signs.} $\mathcal H[N] = 0$ as 
\be
	P_\phi = \pm \sqrt{q \stackrel{(3)} R-4 \left( P^{ab} P_{ab} - \frac{1}{2} P^2 \right) } =: H_{\text{true}} \label{eq:HamScalarDepara}
\ee
Then, we can quantise \eqref{eq:HamScalarDepara} and treat it as a Schr{\"o}dinger equation in the physical time $\phi$. Quantum states then only depend on the metric variables and their time evolution is generated by the ``true'' (physical) Hamiltonian $H_{\text{true}}$. Since we choose the scalar field to be massless (in fact we choose the whole potential to vanish), the resulting Hamiltonian is time-independent. This is however not true in general, in particular for non-vanishing potential. 

An inconvenient feature of \eqref{eq:HamScalarDepara} is that it involves a square root, which makes its definition at the quantum level cumbersome. The square root can be avoided if one uses a suitable choice of dust as a means of clocks and rods \cite{BrownDustAsStandard, HusainTimeAndA, SwiezewskiOnTheProperties, GieselScalarMaterialReference}. The respective clock field is given by the proper time passing for an irrotational dust field. In this case, the true Hamiltonian is simply given by the gravitational part of the Hamiltonian constraint (plus additional matter fields that are coupled). From the point of view of recovering quantum field theory on curved spacetimes, this seems to be an optimal choice of clock fields, as it is algebraically as simple as possible. On the other hand, the fundamental physical status of dust fields may be questioned. 

In all cases, we have to regularise the Hamiltonian (constraint) in terms of holonomies and fluxes. This regularisation procedure is graph-adapted, i.e. defined separately for every possible graph underlying a quantum state separately. Cylindrical consistency is demanded in the case of trivial refinements of the graph. Such a procedure can be either graph-preserving or graph-changing, as we will discuss below.

\paragraph{Holonomies and fluxes} \mbox{}\\
\indent As a first step, we have to regularise the Hamiltonian constraint \eqref{eq:HamConstrAE} in terms of holonomies and fluxes. Let us first consider the so called ``Euclidean part''  of  \eqref{eq:HamConstrAE} 
\be
	\mathcal{H}_E[N] = \int_\Sigma d^3x \,N \beta^2 \frac{E^{ai} E^{bj}}{2 \sqrt{q}} \epsilon^{ijk} F_{ab}^k \text{.}
\ee
The first problem we see is the inverse power of $\sqrt{q}$. While an operator corresponding to a positive power of $\sqrt{q}$ can be defined via the spectral theorem based on the volume operator, this is not possible for inverse powers as the volume operator has zero in its spectrum. In \cite{ThiemannQSD1}, Thiemann has proposed a set of classical Poisson bracket identities which can circumvent this problem. We first note that
\be
	 \beta^2 \frac{E^{ai} E^{bj}}{2 \sqrt{q}} \epsilon^{ijk} (x) = \frac{1}{2} \epsilon^{abc} e_c^k (x)= \frac{1}{\beta}\epsilon^{abc} \frac{\delta V(R)}{\delta E^c_k(x)} = \frac{1}{\beta}\epsilon^{abc} \left\{A_c^k(x) , V(R)\right\} \text{,} \label{eq:RewritingEEe}
\ee
where $R$ is a three-dimensional region containing $x$ in its interior. The resulting expression contains the volume of a region $R$, which exists as a well defined operator, as well as a connection, which we can approximate by a holonomy using
\be
	h^{j}_{s}(A)= 1+\eta\, \dot{s}^a(0)\, A^i_a (s(0)) \, \tau_i^{(j)}+{\cal O}(\eta^2) \text{,}
\ee
where $s$ is a (short) segment of an edge with $s(0)=x$ and $\eta$ is a measure of the coordinate length of $s$. Therefore, an expression such as \eqref{eq:RewritingEEe} can be quantised by substituting the Poisson bracket by a commutator after all connections have been expressed via holonomies. 

\begin{figure}[h]
	\centering\includegraphics[trim = 60mm 70mm 60mm 70mm, clip, scale=0.5]{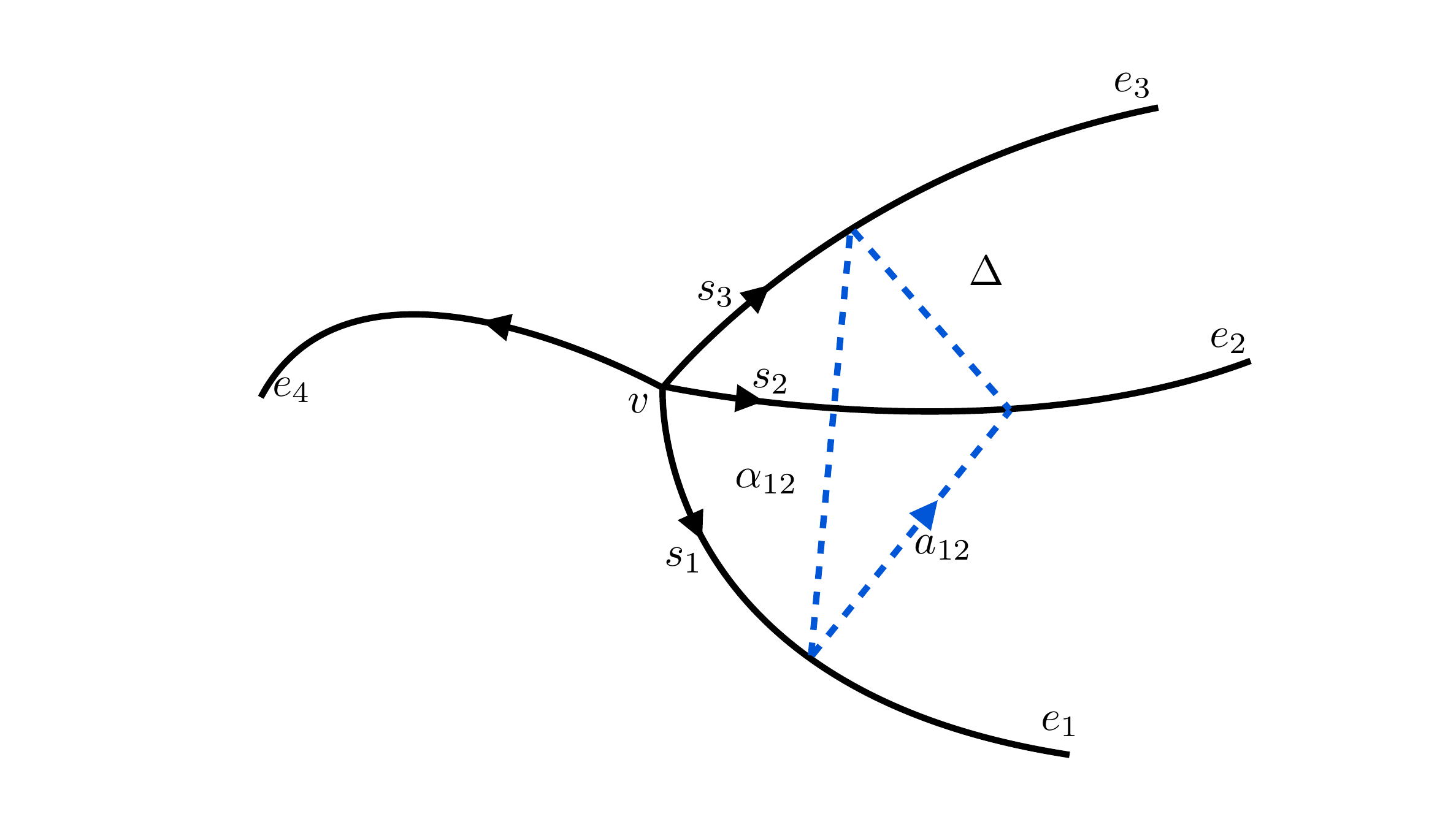} 
		\caption{The construction of a triangulation used in the regularisation of the Hamiltonian constraint acting on a vertex $v$ is depicted. Three edges $e_1$, $e_2$, and $e_3$ are chosen. The edges are oriented to be outgoing at $v$. Their starting segments are denotes by $s_1$, $s_2$, and $s_3$. The endpoints of these segments are connected as in the figure, giving rise to the arcs $a_{12}$, $a_{23}$, and $a_{31}$. Segments and arcs are connected to loops $\alpha_{ij} = s_i \circ a_{ij} \circ s_j^{-1}$. The loop to regularise the field strength in the Hamiltonian constraint can now for example be taken as $\alpha_{12}$, whereas the remaining holonomy is taken along $s_3$. The three segments and three arcs together form a tetrahedron $\Delta$, which is included in the triangulation of $\Sigma$.
	In the construction, one averages over all such choices, as well as over all triplets of edges incident at $v$.}
	\label{fig:HamRegGraphChange}
	\end{figure}

In order to approximate the field strength via holonomies, we use the classical identity	
\be
	h^{(j)}_{\alpha_{12}(\Delta)}(A)=1+\frac{\eta^2}{2} \,\dot{s}_1^a \dot{s}_2^b\, F^i_{ab}(v(\Delta)) \, \tau_i^{(j)} +{\cal O}(\eta^3) \text{,}
\ee
where the involved quantities are explained in figure \ref{fig:HamRegGraphChange}. We note that if we would have regularised the curvature along a rectangle, we would have gotten an additional factor of $2$ in the $\eta^2$ term.

The regularisation of the remaining terms in the Hamiltonian constraint works similarly, although the resulting expression is more cumbersome. We will not detail this here and refer to the exercises for further useful Poisson bracket identities. 

What remains to be defined is which holonomies and volumes we actually mean in our regularisation when we act on a given spin network state. As discussed up to now, a rewriting in terms of holonomies and fluxes always involves a certain approximation, labelled by a regulator given by the length of holonomies and areas over which the fluxes are defined. Ideally, one would like to remove this regulator after the regularised action of the operator has been defined. This then corresponds to taking a continuum limit\footnote{However, this has to be contrasted with the continuum limit discussed in section \eqref{sec:CriticismOfLoops}, which is a continuum limit for the states on which the constraint acts.} in the sense that the lattice on which the Hamiltonian constraint was defined is shrunk to zero edge length. Precisely this has been achieved by Thiemann in his original definition \cite{ThiemannQSD1} of the Hamiltonian constraint. The resulting operator turns out to change the underlying graph by adding a certain class of ``extraordinary'' loops. 
Another possibility is to consider a regularisation which preserves the underlying graph. In this approach, the regulator cannot be completely removed and the holonomies and fluxes are chosen as in a lattice approximation of $\mathcal H$ based on the graph underlying the spin network acted upon.
We will now give some further details on the two possibilities. Earlier work in the same direction includes \cite{RovelliAshtekarFormulationOf, HusainIntersectingLoopSolutions, BrugmannIntersectingNLoop, GambiniLoopSpaceRepresentation, BrugmannJonesPolynomialsFor}.

\paragraph{Graph-changing} \mbox{}\\
\indent Thiemann's original construction of the Hamiltonian was graph-changing \cite{ThiemannQSD1}. This is more satisfactory from a fundamental point of view as otherwise an infinite number of conserved quantities (the graph) emerges at the quantum level. We will only give a rough overview of the construction here.

\begin{figure}[h]
	\centering\includegraphics[trim = 1mm 60mm 1mm 30mm, clip, scale=0.36]{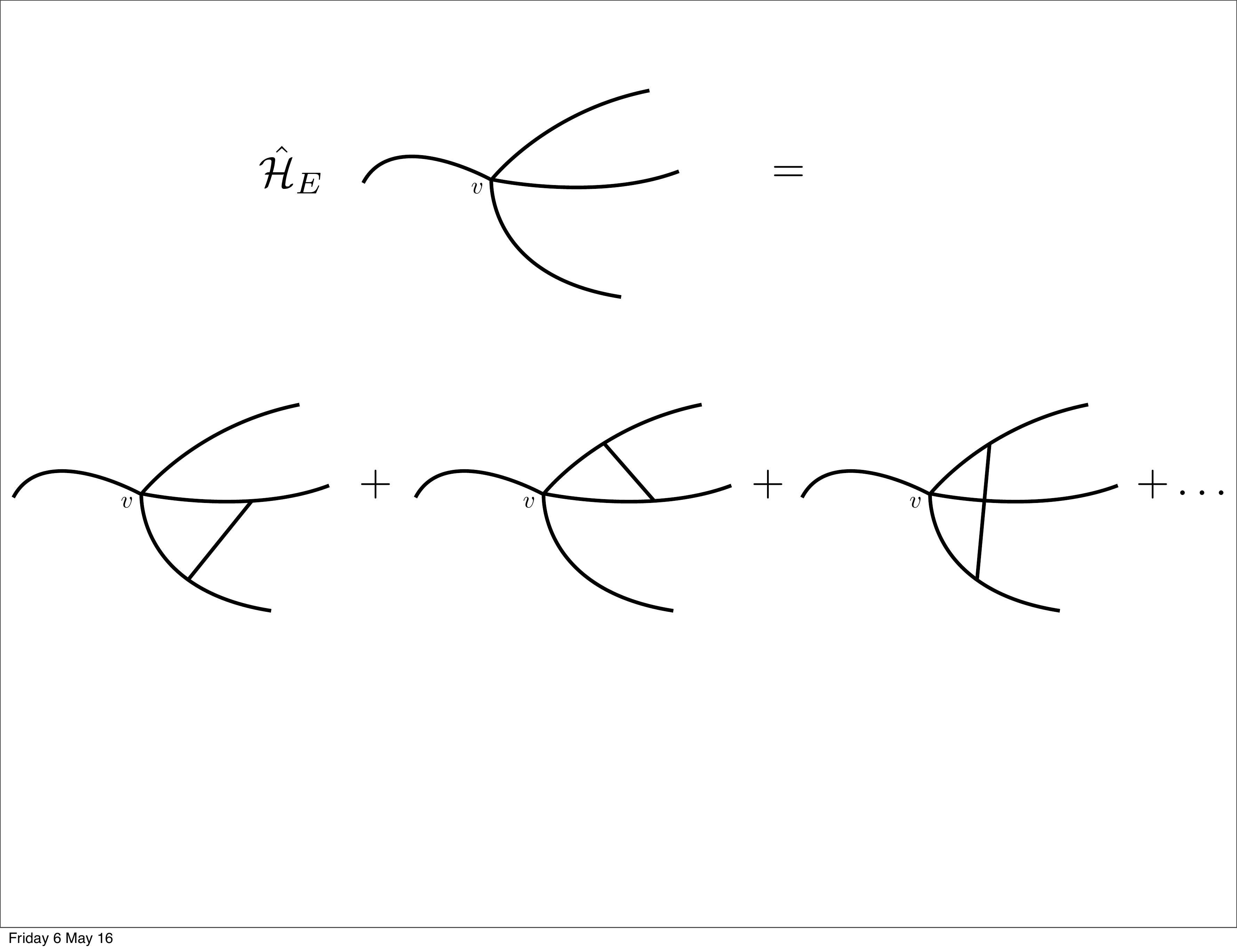} 
		\caption{The action of the Euclidean part of the Hamiltonian constraint at a vertex $v$ is visualised. The spins labelling the edges overlapping with the loop $\alpha$ are determined by SU$(2)$ recouping theory. In addition to the terms shown, one obtains further terms coming from an averaging over all possible triplets of edges used for the regularisation shown in figure \ref{fig:HamRegGraphChange}.}
		\label{fig:HamActionGraphChange}
	\end{figure}

In order to prescribe the edges on which the holonomies in the operator are defined, one first constructs a triangulation of $\Sigma$ adapted to the graph $\gamma$ underlying the spin network acted upon. The triangulation has to satisfy that all edges of $\gamma$ can be build as unions of edges of the triangulation. The ordering in the Hamiltonian is chosen in such a way that the final operator acts only on spin network vertices (by ordering the commutator $[\hat h, \hat V]$ to the right). Then, around a vertex, the triangulation is used to prescribe a segment $s$ of an edge as well as a loop $\alpha$, as shown in figure \ref{fig:HamRegGraphChange}. One averages over all possible such prescriptions. The size (fineness) of the triangulation functions so far as a finite regulator in this definition. However, when one evaluates the result on a diffeomorphism-invariant state, this regulator can be removed, i.e. the triangulation infinitely refined, since two arcs of different size are related by a diffeomorphism. The action of the constraint is visualised in figure \ref{fig:HamActionGraphChange}. Various modifications of Thiemann's original prescription have been proposed in \cite{LewandowskiLoopConstraintsA, LewandowskiSymmetricScalarConstraint, AlesciHamiltonianOperatorFor, AssanioussiNewScalarConstraint}. 

The precise choices made in the regularisation \cite{ThiemannQSD1} lead to a certain notion of on-shell anomaly-freedom. The precise statement is that the commutator of two Hamiltonian constraints acting on a state in the kinematical (non-diff-invariant) Hilbert space is non-vanishing, but vanishes when evaluated on a diffeomorphism invariant state. The reason is that it contains differences of spin networks whose graphs are related by a diffeomorphism. This notion of anomaly-freedom has been criticised for example in \cite{NicolaiLoopQuantumGravity}. Progress in deriving an off-shell notion of anomaly freedom properly implementing the classical Dirac algebra has been recently made in simplified models \cite{LaddhaTheDiffeomorphismConstraint, TomlinTowardsAnAnomaly, HendersonConstraintAlgebraInI, LaddhaHamiltonianConstraintIn}. 

\paragraph{Non-graph-changing} \mbox{}\\
\indent An alternative approach to defining a Hamiltonian constraint is to prescribe a graph-preserving regularisation \cite{GieselAQG1, GieselAQG2, GieselAQG3, GieselAQG4}. In doing this, one fixes once and for all an underlying graph on which quantum states can have support. The regularisation then strongly resembles that of lattice gauge theory, with the difference that the underlying metric is an operator. A strong reason to consider a fixed graph is the current lack of good coherent states for graph-changing operators \cite{ThiemannGCS1, ThiemannGCS2, ThiemannGCS3}, which only approximate a certain subset of degrees of freedom supported on the graph on which they are defined. Technically, such a fixed graph would be considered as an algebraic graph, which does not retain any embedding information such as knotting \cite{GieselAQG1}. 

Taking this approach, it has been verified that the expectation value of the Master constraint (a sum of squares of the constraints, see below) takes its classical (discretised) value \cite{GieselAQG2, GieselAQG3}. Non-graph-changing definitions of Hamiltonians are currently also used in the context of deriving symmetry reduced sectors of loop quantum gravity from the full theory \cite{AlesciANewPerspective, AlesciLoopQuantumCosmology, BIII, BZI, BVI} as well as via a midisuperspace quantisation \cite{BojowaldSphericallySymmetricQuantumGeometryHamiltonian, AlvarezLocalHamiltonianFor, GambiniLoopQuantizationOf}, as they are more tractable in practical calculations due to their restriction of the degrees of freedom to those of the underlying graph. 

From a fundamental point of view however, one would prefer a graph-changing operator that creates new generic\footnote{The vertices created using Thiemann's prescription \cite{ThiemannQSD1} are called ``extraordinary'' and have a vanishing volume as well as vanishing action of the Hamiltonian constraint on them. This property is essential in the current proof of anomaly freedom \cite{ThiemannQSD1}.} vertices,  in particular to be able to describe an expanding universe only in terms of low spin. Currently, it seems that the group field theory formalism (see below) comes closest to achieve this goal \cite{GielenEmergenceOfA}.

\paragraph{Master constraint} \mbox{}\\
\indent The master constraint approach \cite{ThiemannThePhoenixProject, ThiemannQSD8, HanMasterConstraintOperators} to defining the dynamics has been developed as an alternative to Dirac quantisation where the constraint algebra is trivialised. Instead of quantising (some of) the constraints $C_\alpha(x)$ of the theory individually, one quantises a single master constraint
\be
	\mathbb M = \int_\Sigma d^3x \, C_\alpha(x) K^{\alpha \beta}(x) C_\beta(x)
\ee
where $K^{\alpha \beta}(x)$ is an invertible phase space function. Hence, imposing $\mathbb M = 0$ is classically equivalent to imposing the vanishing of all the constraints. The selection of observables proceeds somewhat differently as
\be
	\left\{\mathcal{O}, \left\{ \mathcal O, \mathbb M \right\} \right\} \approx 0 \text{,}
\ee
but is equivalent to the standard method on the constraint surface. 
The operator corresponding to $\mathbb M$ is quantised using the same methods that were developed for the Hamiltonian constraint. The main motivation for the master constraint is of technical nature, e.g. the simplifications arising due to a trivialised constraint algebra and a rigorous proof of the existence of the physical Hilbert space. The expectation value of the master has been computed using coherent states and shown to coincide with a suitable discretisation of the classical theory \cite{GieselAQG2}. 

The master constraint approach has been tested in a number of systems \cite{DittrichTestingTheMasterI, DittrichTestingTheMasterII, DittrichTestingTheMasterIII, DittrichTestingTheMasterIV, DittrichTestingTheMasterV} and has been employed for imposing the simplicity constraints in the spinfoam models \cite{EngleLQGVertexWith} as well as in imposing a symmetry reduction \cite{AlesciANewPerspective}.

\subsubsection{Spin foams}

Spinfoam models (see \cite{RovelliBook2} for a textbook and \cite{BaezAnIntroductionTo, RovelliLecturesOnLoop, PerezSpinFoamModels} for reviews) are a covariant path integral approach to defining the dynamics of loop quantum gravity. They grew out of state sum models \cite{BarrettStateSumModels, EngleLQGVertexWith} and their development \cite{ReisenbergerSumOverSurfaces} was influenced by the dynamics defined by the Hamiltonian constraint as well as the quantum kinematics. The first important model was the Barrett-Crane model \cite{BarrettRelativisticSpinNetworks}, followed by the improved ERPL / FK model \cite{FreidelANewSpin}, which cured problems with the graviton propagator \cite{AlesciCompleteLQGPropagatorI, AlesciCompleteLQGPropagatorII, BianchiLorentzianSpinfoamPropagator}. 

\begin{figure}[h]
	\centering\includegraphics[trim = 70mm 60mm 1mm 90mm, clip, scale=0.42]{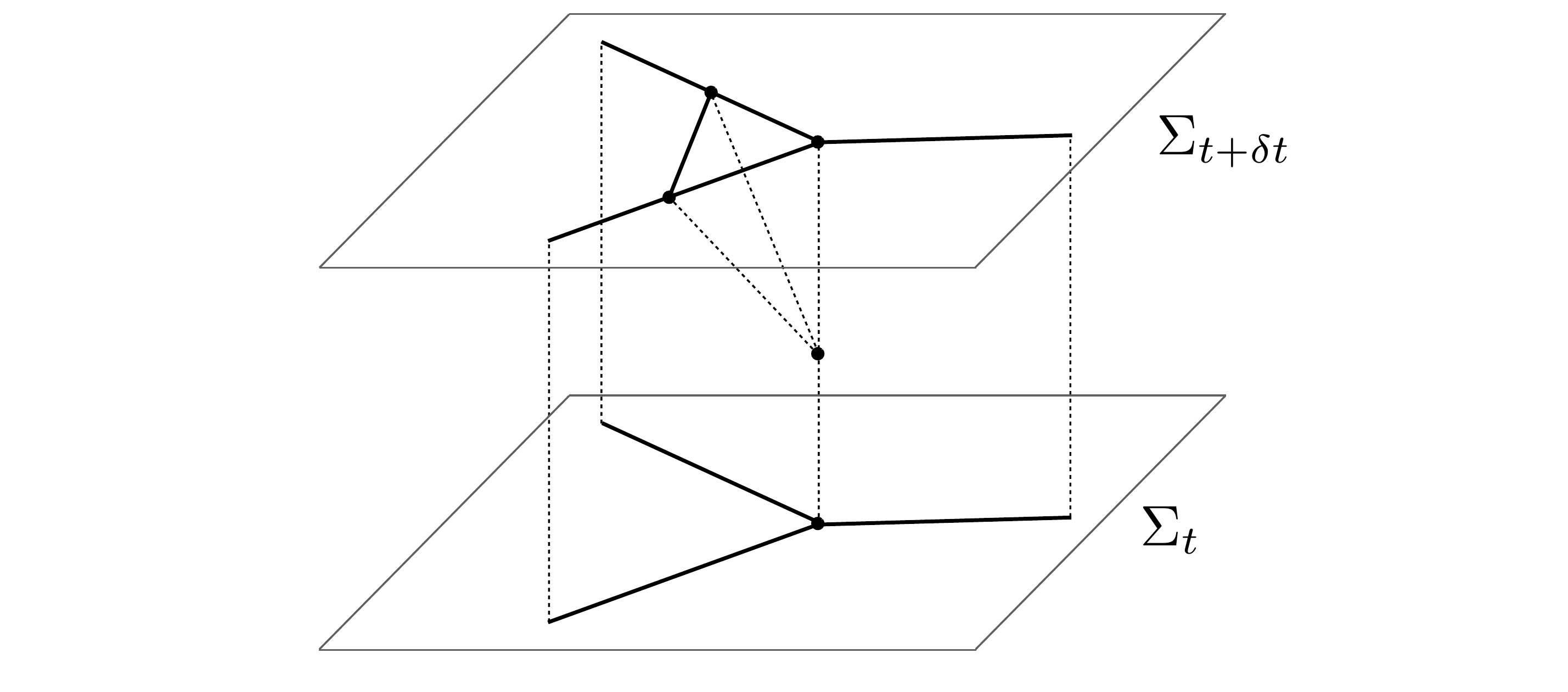} 
		\caption{A a spin network containing a three-valent vertex evolves under the path integral into a new spin network, containing an additional arc, reminiscent of the action of the Hamiltonian constraint in figure \ref{fig:HamActionGraphChange}. The spinfoam depicted here is the simplest contribution to the depicted process between two equal time hypersurfaces $\Sigma_t$ and $\Sigma_{t+\delta t}$, originating from a single action of the Hamiltonian constraint, i.e. the linear term in equation \eqref{eq:SpinFoamRiggingMap}. }
		\label{fig:SpinfoamH1}
	\end{figure}

There are two basic strategies to arrive at the currently known spinfoam models. First, one formally tries to define a projector on the physical Hilbert space by giving sense to the expression 
\be
	\ket {\Psi_\text{phys}} := \delta (\hat {\mathcal H}) \ket{\Psi} := \int [D N] \exp \left(i \int_\Sigma d^3x \, N(x) \hat{\mathcal H}(x) \right) \ket{\Psi} \text{.}
\ee
In practise, one then computes a path integral between two kinematical (or diff-invariant) boundary states whose value defines the physical scalar product as
\be
	\braket{\Psi_\text{phys}} {\Psi'_\text{phys}} := \braopket{\Psi_\text{diff}} {\delta (\hat{\mathcal H})}{\Psi'_\text{diff}}_{\text{diff}} := \sum_{n=0}^{\infty} \frac{i^n}{n!} \int [DN] \braopket{\Psi_\text{diff}}{\hat{\mathcal H}[N]^n}{\Psi'_\text{diff}}_\text{diff} \text{.} \label{eq:SpinFoamRiggingMap}
\ee
Different terms in this sum than can be interpreted as Feynman graphs, with the simplest example shown in figure \ref{fig:SpinfoamH1}.

Second, one can start with the formulation of general relativity as a constrained BF-theory going back to Plebanski \cite{PlebanskiOnTheSeparation}. A BF-theory \cite{HorowitzExactlySolubleDiffeomorphism} is a topological theory with the action 
\be
	S_{\text{BF}} = \int_{\mathcal M} d^4x \, \text{Tr} \left[B \wedge F(A) \right] \text{,} \label{eq:BFTheoryAction}
\ee
where $F$ is the curvature of a connection $A$ and $B$ is a Lie-algebra valued two-form. The equations of motion of this action tell us that $F(A) = 0$, so that the theory is topological. Moreover, $D(A) B = 0$ corresponds to the Gau{\ss} law.
Using the gauge group SO$(1,3)$, the condition $B_{IJ} = \epsilon_{IJKL} e^K \wedge e^L$ ensures equivalence to general relativity, where $e^I = e^I_\mu dx^{\mu}$ is the co-vierbein. The reason is that $D(A) B = 0$ turns into the torsion-freeness condition w.r.t. the vierbein, and the action reduces to the Einstein-Hilbert action (see exercises). The condition $B_{IJ} = \epsilon_{IJKL} e^K \wedge e^L$ can be conveniently expressed in terms of a quadratic expression $\epsilon_{IJKL} B^{IJ}_{\mu \nu}  B^{KL}_{\rho \sigma} \propto\epsilon_{\mu \nu \rho \sigma}$ (up to a topological sector), known as simplicity constraints. The main idea is now to quantise \eqref{eq:BFTheoryAction} as a topological quantum field theory, which is a well established subject, and to impose the simplicity constraints at the quantum level. Much care has to be taken here since the simplicity constraints turn out to be non-commuting and imposing all of them strongly seems to restrict the physical degrees of freedom too much, as observed in the Barrett-Crane model \cite{AlesciCompleteLQGPropagatorI}.

The situation concerning the current understanding of the dynamics of spinfoam models is comparable to that in the canonical theory. Much is known in the context of large spins, where a relation to discretised general relativity has been established \cite{BarrettLorentzianSpinFoam, HanAsymptoticsOfSpinfoam} (see however \cite{HellmannGeometricAsymptoticsFor}). On the other hand, as in the canonical theory, little is known in the context of many small spins. While the spinfoam approach is clearly motivated by the canonical framework and with a (roughly) identical kinematics, it is now known whether the dynamics are the same, see. e.g. \cite{AlexandrovSpinFoamsAnd, HanCommutingSimplicityAnd, AlesciLinkingCovariantAnd, ThiemannLinkingCovariantAndII}. 
More recent work focusses mainly on the issue of renormalisation in spin foam models, see for example \cite{LivineCouplingOfSpacetime, RielloSelfEnergyOf, BonzomBubbleDivergencesAnd, DittrichCoarseGrainingOf, BahrOnBackgroundIndependent}.

\subsubsection{Group field theory} 

Group field theories (GFTs) \cite{OritiTheGroupField, BaratinTenQuestionsOn, OritiGroupFieldTheoryAndLoop} are a class of quantum field theories that have grown out of tensor models \cite{GurauColoredTensorModels}, enriched with additional group theoretic data. 
They can be interpreted as a second quantised version of loop quantum gravity \cite{OritiGroupFieldTheoryAsTheSecond} and in particular provide an organisational principle to systematically derive spinfoam amplitudes, see \cite{OritiGroupFieldTheoryAndLoop} for a review. 

GFTs are defined as quantum field theories on group manifolds and should be interpreted as quantum field theories {\it of} spacetime, instead of quantum field theories {\it on} spacetime. Spacetime itself only obtains a meaning once it is constructed from its atoms, which are quanta that can be excited over the ``no-space'' Fock vacuum of GFT. These quanta of space correspond to the quanta that we already saw in the context of loop quantum gravity for an appropriate choice of dimension, gauge group, and constraints in the GFT. 

\begin{figure}[h]
	\centering\includegraphics[trim = 10mm 60mm 1mm 50mm, clip, scale=0.48]{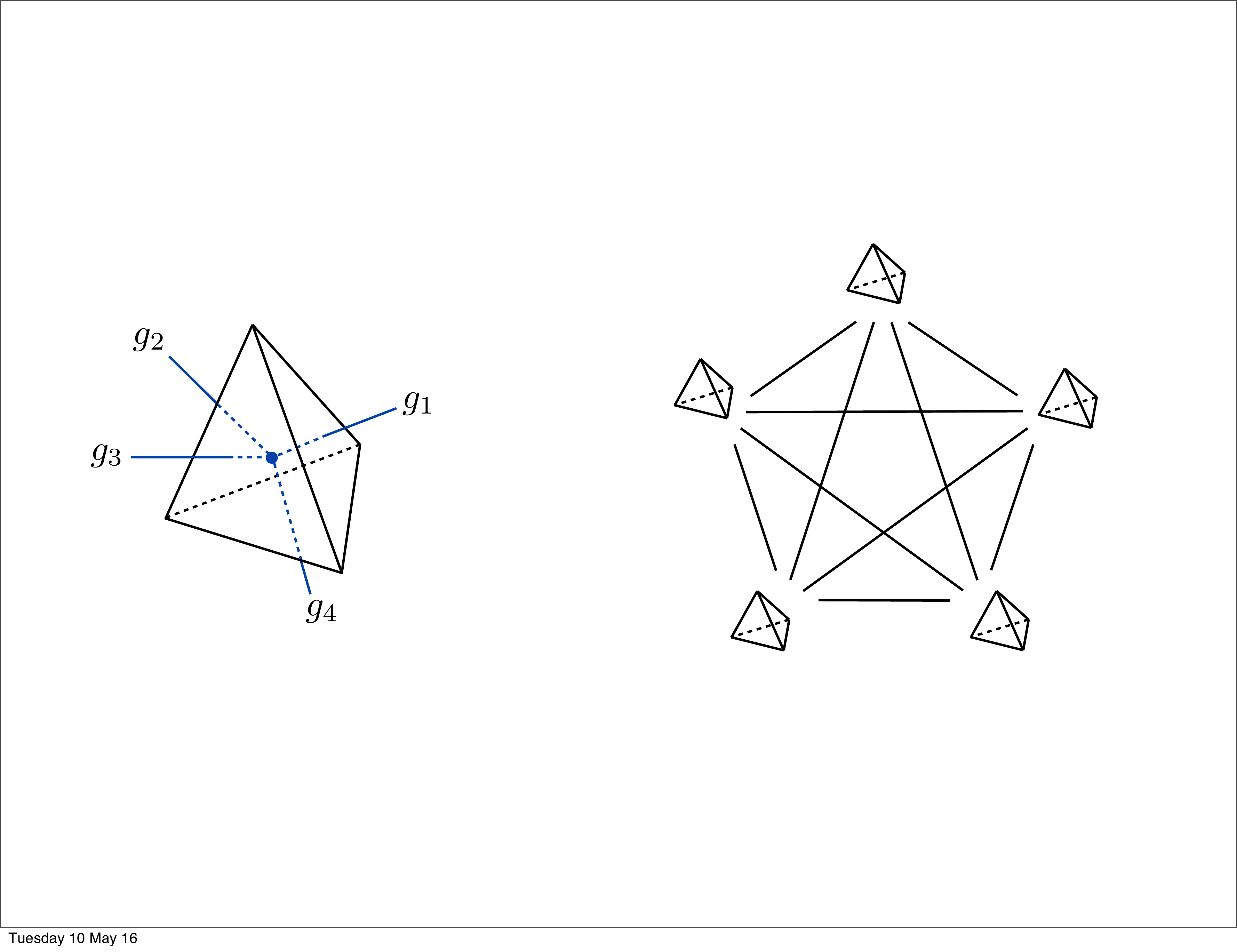} 
		\caption{A quantum state of a tetrahedron is specified by a wave function of four group elements $g_1, g_2, g_3, g_4$. The group field theory potential can be chosen such that it glues five tetrahedra into a four-simplex. The kinetic term in the action glues two tetrahedra of neighbouring four-simplexes.  
		The non-locality (on the group manifold) in the interaction vertex is restricted within a single four-simplex.}
		\label{fig:GFTInt}
	\end{figure}

The classical starting point for defining a (single field) GFT (in d spacetime dimensions) is the action principle (e.g., \cite{OritiGroupFieldTheoryAndLoop})
\begin{align}
	S &= \int [dg_I] \, [dg'_J]  \, \bar{\phi}(g_I) \mathcal K(g_I, g_J) \phi(g'_J) \\
		& ~~ + \sum_i \frac{\lambda_i}{D_i!} \int [dg_{1 I}] \ldots [dg_{{D_i}I}] \phi^*(g_{1I}) \ldots \mathcal V_i(g_{1I}, \ldots, g_{D_i I}) \ldots \phi({g_{{D_i}I}})
\end{align}
where $\phi : G^d \rightarrow \mathbb C$ is a complex field on $d$ copies of a group $G$, $\mathcal K$ is an interaction kernel for the kinetic term, and $\mathcal V_i$ are interaction kernels for the potential, and $I,J = 1, \ldots, d$ . This action is quantised by standard quantum field theory techniques and its Fock vacuum corresponds to a state without any space. 
A single quantum of the field $\phi$ corresponds to an atom of space. Its four arguments correspond to the holonomies orthogonal to the four sides of a tetrahedron. An additional gauge invariance $\phi(g_1, \ldots, g_d) = \phi(h g_1, \ldots, h g_d) , ~ \forall h \in G$ restricts us to contract the holonomies with invariant tensors, as for spin networks. 
The interaction term appropriate for an interaction corresponding to a gluing of five tetrahedra to a four-simplex is visualised in figure \ref{fig:GFTInt}. 
The dynamics of the theory is then defined in the standard way as an expansion into Feynman amplitudes. 

The basic philosophy of defining dynamics in group field theory is opposite\footnote{In principle, one can also try to implement the dynamics of a specific quantisation of the Hamiltonian constraint in a group field theory, however this approach has not been followed so far \cite{OritiGroupFieldTheoryAndLoop}.} of that in canonical loop quantum gravity. Instead of starting with general relativity at the classical level and quantising its dynamics, one starts with a fundamental dynamics that is determined by symmetries and simplicity. The dynamics of general relativity only have to emerge in the continuum limit after a suitable coarse graining procedure. Still, the kinematics underlying group field theory is strongly influence by the quantisation of diffeomorphism-invariant\footnote{The kinematical structures used for loop quantum gravity type quantisations of more general diffeomorphism-invariant theories turns out to agree with those used in standard general relativity, see e.g. \cite{MaExtensionOfLoop, BNII}.} field theories via loop quantum gravity methods.

The main current trends in group field theory research is the extraction of cosmological dynamics, including corrections to the classical FRW scenario \cite{GielenCosmologyFromGroup, GielenHomogeneousCosmologiesAs, OritiBouncingCosmologiesFrom, OritiEmergentFriedmannDynamics}, as well as the study of renormalisation, see e.g. \cite{CarrozzaRenormalizationOfASU, LahocheRenormalizationOfAn, CarrozzaTensorialMethodsAnd}. Especially in the context of cosmology, much progress could be made using condensate states and the loop quantum cosmology dynamics could be extracted up to a further quantum correction \cite{OritiBouncingCosmologiesFrom, OritiEmergentFriedmannDynamics}. Also, a low spin phase seems to be favoured by the dynamics in this context \cite{GielenEmergenceOfA}, which is very interesting from the point of view of the large spin limit, as such configurations (with no coarse graining implied) seem to be dynamically unstable.

\subsection{Open questions and future directions}

\subsubsection{Quantising classically symmetry reduced models}

Two of the main areas of interest of quantum gravity theories, cosmology and black holes, are symmetric sectors of the classical theory. Progress in understanding quantum gravity in these situations can thus be made by quantising a symmetry reduced phase space, known as mini- or midisuperspace quantisation. Most progress along this route has been made in loop quantum cosmology, as outlined in section \ref{sec:LQC}. The main open question here is how to go beyond a homogeneous and isotropic spacetime, in particular to properly understand how perturbations propagate on the quantum spacetime. This is of paramount importance for extracting cosmological predictions from the theory and no consensus has been reached so far \cite{BollietConceptualIssuesIn}.

The quantisation of spherically symmetric spacetimes has also advanced recently \cite{AlvarezLocalHamiltonianFor, GambiniLoopQuantizationOf}, but it is by far not as advanced as the homogeneous and isotropic sector of the theory, which can be solved exactly in a suitable ordering and for which a complete set of Dirac observables are known. 
Also, dynamical issues have mostly been ignored so far  (see however \cite{BojowaldLemaitreTolmanBondi}), in particular spherical collapse and a dynamical resolution thereof along the lines of the big bounce found in loop quantum cosmology.
Understanding this properly in a quantum gravity scenario is especially important for the resolution of the black hole information paradox, which might simply be absent once the central singularity is resolved and substituted by a regular evolution of a suitable quantum state, as e.g. sketched in \cite{AshtekarBlackHoleEvaporation}.

\subsubsection{Deriving loop quantum cosmology from full loop quantum gravity}

In recent years, progress has been made in deriving loop quantum cosmology from full loop quantum gravity on different fronts. Since the issue of the continuum limit and proper continuum states, as well as the issue of coarse grained large spin descriptions have not been settled so far, it is currently not possible to judge any one approach to be fully satisfactory. In particular, one should distinguish between two a priori different questions:
 \begin{enumerate}
	\item What is the cosmological sector of loop quantum gravity?
	\item Is there a theory of quantum gravity that reduces to loop quantum cosmology in a suitable truncation?
\end{enumerate} 
The second question turns out to be easier, since one can taylor the full theory in such a way that the relevant cosmological observables are, in a suitable sense, orthogonal on the remaining inhomogeneous degrees of freedom \cite{BIII, BVI}. Moreover, one can use the same gauge fixings that are employed in loop quantum cosmology already at the classical level, resulting in Abelian gauge theories rather than non-Abelian ones. In this context, it turns out that a truncation of the full theory to maximally coarse graphs matches the dynamics of loop quantum cosmology for similar choices of factor ordering. Since the involved gauge theories have $\mathbb R_{\text{Bohr}}$ as a (gauge) group, Thiemann's original regularisation prescription for the Hamiltonian constraint can be refined to incorporate the improved loop quantum cosmology dynamics (such a refined regularisation prescription is the natural one for graph-preserving Hamiltonians \cite{BIII, BVI}). 

Different approaches have been taken to answer the first question. On the one hand, a gauge fixing and truncation has employed at the quantum level in \cite{AlesciANewPerspective, AlesciLoopQuantumCosmology} to simplify the Hamiltonian constraint. The resulting quantum states still capture sufficient information, as one can show that the expectation value of the Hamiltonian constraint agrees with the loop quantum cosmology effective Hamiltonian \cite{AlesciQuantumReducedLoopGravitySemiclassicalLimit, AlesciLoopQuantumCosmology}. A suitable sum over graphs then also leads to the improved LQC dynamics \cite{AlesciImprovedRegularizationFrom}, however still only in the context of expectation values of the Hamiltonian constraint operator. 
On the other hand, a condition for a homogeneous and isotropic spacetime was derived \cite{BeetleDiffeomorphismInvariantCosmological}, with the aim of suppressing inhomogeneous fluctuations in both the connection and its conjugate momentum. This approach is different from the one in \cite{AlesciANewPerspective} in that one quantises a homogeneity condition instead of trying to construct homogeneous spin networks, i.e. with homogeneously distributed spins, at the quantum level. 
Other, more mathematically minded studies have focussed on the embeddability of the LQC state space into that of standard loop quantum gravity, see e.g. \cite{EngleRelatingLoopQuantum, EngleEmbeddingLoopQuantum, BrunnemannSymmetryReductionOf, HanuschInvariantConnectionsIn, FleischhackKinematicalFoundationsOf}. 

Within group field theory, condensate states have been considered as suitable candidates for continuum states encoding sufficient information to extract cosmology \cite{GielenCosmologyFromGroup, GielenHomogeneousCosmologiesAs}. Here, the full quantum dynamics are computed in a hydrodynamic limit using the Gross-Pitaevskii equation. It can be shown that one obtains the improved loop quantum cosmology dynamics for a suitable choice of parameters in the GFT action, including a new quantum correction \cite{OritiEmergentFriedmannDynamics, OritiBouncingCosmologiesFrom}. In particular, the dynamics exponentially suppress large spins and thus probe a genuine low spin regime \cite{GielenEmergenceOfA}.    

Concerning observations, it is of great importance to study how cosmological perturbations propagate on the homogeneous and isotropic quantum spacetimes described in any of these approaches, and to compare it to the standard loop quantum cosmology picture \cite{AshtekarLoopQuantumCosmologyFrom}, see for example \cite{GielenIdentifyingCosmologicalPerturbations} for some seminal work. Also, some work has been done on defining a notion of spherical symmetry at the quantum level \cite{BojowaldSphericallySymmetricQuantum, BLSI, OritiGeneralizedQuantumGravity, BZI, BVI}. 

Irrespective of their final relevance for the cosmological sector of LQG, all of these works are valuable in that they shed some light on the hard issue of the continuum limit and effective dynamics. In any case, they provide simplified models in which coarse graining can be studied and new techniques useful for full LQG can be developed.

\subsubsection{Black hole entropy}

The first computations of black hole entropy within loop quantum gravity \cite{SmolinLinkingTopologicalQuantum, KrasnovGeometricalEntropyFrom, RovelliBlackHoleEntropy} were based on counting the number of ways in which a total area can be subdivided into the quanta of area suggested by the LQG area operator. The result is that the entropy 
\be
	S = \text{const}(\beta)  \times A \label{eq:Entropy}
\ee
is proportional to the area $A$, with the proportionality constant depending on the Barbero-Immirzi parameter $\beta$. 
Based on the seminal work \cite{SmolinLinkingTopologicalQuantum}, these computations were later incorporated into the framework of isolated horizons \cite{AshtekarIsolatedAndDynamical}, where the black hole horizon is modelled as a boundary of spacetime \cite{AshtekarQuantumGeometryAndBlackHoleEntropy, AshtekarIsolatedHorizonsThe, AshtekarQuantumGeometryOf, EngleBlackHoleEntropyFrom, BTTXII}, see \cite{Diaz-PoloIsolatedHorizonsAnd} for a review. While isolated horizons provide a satisfactory local notion of a horizon at the classical level, the conditions which are actually imposed in the quantum theory turn out to be insufficient to characterise a quantum horizon \cite{BNII, BodendorferSomeNotesOn}. In fact, it turns out that all the ingredients necessary to perform the computations according to \cite{AshtekarQuantumGeometryOf} are available for general boundaries \cite{BII} and one computes the entanglement entropy of the gravitational field in a maximally mixed state \cite{HusainApparentHorizonsBlack, DonnellyEntanglementEntropyIn, BII}. This in particular agrees with recent results from other approaches to quantum gravity \cite{BianchiOnTheArchitecture}. 
The two main open questions are thus the following: 
\begin{enumerate}
	\item Are (black hole) horizons somehow special in the quantum theory?
	\item How should the prefactor in front of the area be interpreted?
\end{enumerate} 

\noindent First, as noted in \cite{HusainApparentHorizonsBlack}, it would be appropriate to quantise the marginally outer trapped surface condition 
\be
	(q^{ab} -s^a s^b) (K_{ab} + \nabla_a s_b) = 0
\ee
where $s^a$ is an outward pointing normal of the surface under question, and to investigate its kernel. While this seems rather cumbersome without any gauge fixing, the condition greatly simplifies is one considers it in the radial gauge \cite{DuchObservablesForGeneral, BLSII} (with $\Lambda^2 = q_{rr} = q_{rr}(r)$ \cite{BZI}), where it simply reads
\be
	P_\Lambda \Lambda +  \partial_r  \sqrt{ \det q_{AB}}=0
\ee
for a surface embedded in a sphere of constant geodesic distance from the origin. 
At first sight, nothing noteworthy seems to happen, as the solution to this equation simply restricts the quantum numbers in a certain way. This seems also consistent with the analysis of \cite{DasguptaSemiclassicalQuantizationOf}, where the apparent horizon condition introduces some correlations across the horizon, but it is unclear that only those correlations would lead to an area-proportional entropy. 
More generally, it would be interesting to quantise a sufficient set of conditions to characterise a spacetime boundary as an isolated horizon. 
From the point of view of the value that the entropy takes, the answer to the first question crucially depends on the following:

Second, the computation of black hole entropy following \cite{AshtekarQuantumGeometryOf} takes place in the (continuum) limit of many low spins, where we have little knowledge of the dynamics. In particular, the prefactor in \eqref{eq:Entropy} contains the bare Newton constant, whereas the Bekenstein-Hawking entropy should contain the renormalised one. Ideally, one would like to understand a possible renormalisation group flow of the Newton constant and check whether \eqref{eq:Entropy} expressed in terms of renormalised quantities gives the Bekenstein-Hawking formula. If one were able to do this, one should also check how the prefactor behaves if the area of which one computes the entropy is not a slice of a horizon, hopefully providing insights to the first question. 

Some interesting progress on these questions has been made recently in the context of the large spin regime. Here, it was first noted that one can analytically continue\footnote{While the classical theory simplifies from $\beta = \pm i$, no quantum theory for complex $\beta$ is known due to the complicated reality conditions. The only way to derive results for complex $\beta$ so far is thus to compute with real $\beta$ and analytically continue the result.} the formula for the dimension of the horizon Hilbert space from real to complex $\beta$. In particular, one obtains the Bekenstein-Hawking formula for $\beta = \pm i$ \cite{FroddenBlackHoleEntropy}. At the same time, consistency of the large spin four-simplex amplitude with the classical on-shell action including the boundary term also demands $\beta = \pm i$, so that a match between the effective action and the entropy computation is achieved \cite{BNI}. However, these computations have been in the realm of (fundamental) large spins and do not apply to the continuum, at least naively. It is noteworthy that these computations do not require the boundary to be a horizon slice, supporting \cite{BianchiOnTheArchitecture}. If one furthermore assumes an energy-area relation appropriate for black holes \cite{GhoshBlackHoleEntropy}, additional results specific to horizons can be derived \cite{PranzettiBlackHoleEntropy, GeillerNearHorizonRadiation}.

\subsubsection{Gauge / gravity}

The gauge / gravity duality is a very interesting recent development in theoretical physics, which relates quantum gravity theories defined on a suitable class of spacetimes to gauge theories on the boundary of these spacetimes. Since Maldacena's original proposal \cite{MaldacenaTheLargeN} for a duality between type IIB string theory and $\mathcal N = 4$ super Yang-Mills theory, the field has evolved considerably and many new dualities have emerged, see e.g. \cite{NatsuumeAdSCFTDuality, AmmonGaugeGravityDuality}.  

The practical importance of the gauge / gravity duality is that it allows to study strongly coupled gauge theories, where perturbative methods are not applicable, by doing computations in gravitational theories. In particular, the most commonly used technique is to approximate the string theory side by a classical gravitational theory, which translates into an infinite 't Hooft coupling and infinite number of colours in the Yang-Mills theory. If one wants to go beyond this limit, string theory corrections to the classical gravitational action have to be taken into account. The two relevant parameters on the string theory side are $\alpha'$ (higher-curvature corrections) as well as $g_s$ (quantum corrections). 

Concerning loop quantum gravity, two questions now have to be asked:
\begin{enumerate}
	\item Is there a limit of the existing dualities in which loop quantum gravity techniques can be helpful to compute corrections to the classical gravity approximation?
	\item Are there other dualities which apply specifically to loop quantum gravity?
\end{enumerate} 

The first point was briefly discussed in \cite{BIV}, where it was argued that the large spin expansion could be related to the $1/N$ expansion in a dual gauge theory while neglecting $\alpha'$ corrections. Here, one has to be careful about the limits that can be taken on the loop quantum gravity side: the large spin limit refers to a limit on a fixed underlying graph, so that there is an associated notion of discreteness. One would therefore expect the dual gauge theory to be a lattice gauge theory, where the lattice is specified by the intersection of the graph with the boundary. Such fixed graph dualities would thus be discretised versions of a gauge / gravity duality that would only recover the usual dualities upon taking a continuum limit. A recent calculation proving the potential usefulness of loop quantum gravity techniques for gauge / gravity dualities has been given in \cite{Bonzom3DHolography}.

For the second point, one could in principle start from scratch and try to prove or motivate a new duality. However, one should keep in mind that it is so far unclear whether quantum gravity \`a la loop quantum gravity should even possess a holographic dual like string theory is conjectured to have\footnote{So far, there does not seem to be any LQG-specific evidence to support the existence of a holographic dual. In particular, the black hole entropy computations in LQG refer to the number of states accessible at the horizon, instead of the number of states contained within the horizon.}. A recent success along this road has been to proof a duality of the Ponzano-Regge model, which can be considered as Euclidean 3d LQG, to the 2d Ising model \cite{BonzomDualityBetweenSpin}, see also \cite{DittrichIsingModelFrom}. Here, the lattice sides for the Ising spins are vertices of the boundary spin network state, giving an example of the discrete duality mentioned above. Of course, 3d gravity is very special, i.e. topological, and it is therefore unclear if higher-dimensional generalisations can be rigorously established. Still, \cite{BonzomDualityBetweenSpin} should provide us with valuable hints on how potential dualities can be formulated (see also \cite{FreidelReconstructingAdSCFT}).

\subsubsection{Anomaly freedom}

If one is interested in defining the quantum dynamics from a purely canonical point of view, anomaly freedom of the constraint algebra is the key topic that needs to be solved in a satisfactory way. Properly implementing a quantum version of the Dirac algebra so far seems to greatly restrict the possible regularisation choices even before any semi-classical consistency with general relativity is demanded. Thiemann's original proposal \cite{ThiemannQSD1} discussed above already implements a certain on-shell notion of anomaly freedom, however it would be more satisfactory to see a (suitably defined) generator of spatial diffeomorphisms emerge from the commutator of two Hamiltonian constraints. 

Based on insights coming from parametrised field theory \cite{LaddhaHamiltonianConstraintInPolymer}, recent progress was made in simplified toy models, where one for example works in $2+1$ dimensions, the Euclidean setting, or using the gauge group $U(1)^3$ \cite{TomlinTowardsAnAnomaly, TomlinTowardsAnAnomalyII, HendersonConstraintAlgebraInI, HendersonConstraintAlgebraInII, LaddhaHamiltonianConstraintIn}. Of these works, \cite{LaddhaHamiltonianConstraintIn} (four-dimensional, gauge group SU$(2)$, but Euclidean) is most similar to full LQG. However, in order to achieve a non-trivial representation of the constraint algebra, self-dual variables were used (corresponding to $\beta = \pm i$ in the Lorentzian case), in which the action of the Hamiltonian constraint can be written in terms of certain Lie derivatives. These allow for a geometric interpretation of the flow generated by the Hamiltonian constraint and thus greatly simplify the task of finding an anomaly free representation. 
Within full loop quantum gravity, a new regularisation of the spatial diffeomorphism constraint using similar methods than for the Hamiltonian constraint seems especially satisfying \cite{LaddhaTheDiffeomorphismConstraint}. It should be mentioned however that the constraint operators defined in the works cited here are not defined on the usual kinematical Hilbert space discussed in section \ref{sec:Quantisation}, but on a suitable generalisation of the so-called Lewandowski-Marolf habitat \cite{LewandowskiLoopConstraintsA}. In particular, the constraints feature non-unit density weights. 

The restrictions imposed by finding a quantum representation of the Dirac algebra can be further strengthened if one consider supergravity quantised via the methods of loop quantum gravity \cite{Armand-UgonTowardsALoop, LingSupersymmetricSpinNetworks, BTTVI, BTTVII}. The constraint algebra then obtains additional relations involving the supersymmetry generator $S$ which have to be satisfied, among them 
\begin{align}
	\{ S[\ldots] , S[\ldots] \} = \mathcal H[\ldots] +  \mathcal H_a[\ldots] + \ldots
\end{align}
This is in particular interesting to constrain different ways to couple matter fields present in the supersymmetry multiplets. 

Furthermore, the issue of anomaly freedom is currently being investigated in the context of effective constraints for cosmological and spherically symmetric scenarios, see \cite{CailleteauAnomalyFreeScalar, CailleteauAnomalyFreePerturbations, BojowaldDiscretenessCorrectionsAnd, BarrauAnomalyFreeCosmological, BojowaldSomeImplicationsOf}, where interesting physical effects such as signature change at high energy density are found.

\subsubsection{Coarse graining}

Coarse graining here refers to an effective description of a quantum geometry on large scales by a reduced number of degrees of freedom. A prototypical example would be to consider the Ising model and ingrate out a certain subset of spins, leading to an effective Hamiltonian for the remaining ones. In practise, integrating out some degrees of freedom turns out to be very complicated already in simple systems, in particular the qualitative form of the effective dynamics might be different from the fundamental one. For example, already in the 2-dimensional Ising model, new interaction terms arise after integrating out some of the spins. 

The complexity of the full loop quantum gravity dynamics as encoded in different proposals for the dynamics has so far prohibited to study coarse graining in detail. Still, some results are available in the context of simplified models, see e.g. \cite{DittrichDecoratedTensorNetwork, DittrichQuantumGroupSpin, DittrichCoarseGrainingOf, DittrichTimeEvolutionAs}. For a more general discussion, see \cite{DittrichTheContinuumLimit}. 
In studying coarse graining, a particularly interesting question is whether there exists a good coarse grained description in terms of large spins, and how different the effective coarse grained dynamics is from the fundamental dynamics acting on states with large spins. 
This is especially relevant to determine the range of validity of existing semiclassical techniques in the context of large spins while using the fundamental dynamics.

\subsection{Exercises}

\begin{enumerate}

	\item {\bf Further Poisson bracket identities} \\
	For the remaining ``Lorentzian'' part of the Hamiltonian constraint, we need to construct an operator corresponding to components of the extrinsic curvature. Show that 
	\be
		\sqrt{q} K =  \sqrt{q} K_{ab} q^{ab}(x) = \beta K_{ai} E^{ai}(x) = \frac{1}{\beta^2} \left\{\mathcal H_E(x)|_{N=1}, V(\Sigma) \right\} \text{.}
	\ee
	Next, show that $\beta K_a^i(x) = \left\{A_{a}^i(x), \int_R d^3x \, \sqrt{q} K\right\}$. To simplify the calculation, it is helpful to view $\int_R d^3x \, \sqrt{q} K$ as the generator of global conformal transformations on the $K_a^i$, $E^b_j$ variables. How does $\Gamma_a^i$ transform under such transformations? Using your results, write down an expression for the full Hamiltonian constraint in terms of connections $A_a^i$, curvatures $F_{ab}^i$, as well as volume operators. 
	
		\item {\bf An obstruction for self-adjoint constraint operators?}\\
	Show that there is an obstruction to quantise the Hamiltonian and spatial diffeomorphism constraints as self-adjoint operators such that the constraints appearing in their commutators are ordered to the right. Is there a logical necessity for constraints to be quantised as self-adjoint operators? Can this obstruction be avoided in a Dirac quantisation (such that $\hat C_i \ket{\Psi_\text{phys}} = 0$ for all constraints $\hat C_i$) while insisting on self-adjointness?

\item {\bf General relativity form BF-theory} \\
	Upon imposing the simplicity constraints, we get $B_{\mu \nu}^{IJ} = \epsilon^{IJKL} e_{\mu K} e_{\nu L}$ (see \cite{FreidelBFDescriptionOf} for a proof). Show that the equation $D_{[\mu}(A) B_{\nu \rho]}^{IJ} = 0$ translate to the no-torsion condition $D_{[\mu}(A) e^{I}_{\nu]} = 0$. Solve this equation for $A$. Hint: contract the tensor indices $\mu, \nu$ with vierbeins $e^\mu_K e^{\nu}_L$ to obtain an expression with free indices $IKL$, then take suitable linear combinations of cyclic permutations in these indices. The resulting connection is known as the (spacetime) spin connection $\Gamma_\mu^{IJ}$. 
	Show that $F_{\mu \nu IJ}(\Gamma) e^I_\rho e^J_\sigma$ coincides with the four-dimensional Riemann tensor $R_{\mu \nu \rho \sigma}$. Insert back into the action and show that we obtain the Einstein-Hilbert action. 
	
	\item {\bf Quantum polyhedra \cite{BianchiPolyhedraInLoop}} \\
	A theorem due to Minkowski states that given $F$ (pairwise non-parallel) unit normals $\vec n_\alpha$ and $F$ areas $A_i$ satisfying $\sum_{\alpha=1}^F \vec n_\alpha A_\alpha = 0$ determine a unique (up to translations) convex polyhedron with face areas $A_\alpha$ and outward pointing face normals $\vec n_\alpha$. In 3 dimensions, the space of polyhedra up to translations and rotations can be turned into the Kapovich-Millson phase space by considering the $F$ (non-unit) vectors $\vec A_\alpha$ and equipping them with the Poisson bracket $\{A^i_\alpha, A^j_\beta\} = \epsilon^{ijk} A_\alpha^k \delta_{\alpha \beta}$ and the first class constraint $\vec C = \sum_{\alpha=1}^F \vec A_\alpha = 0$. \\
	Show that $\vec C$ generates rotations of the polyhedra. Quantise this phase space and impose $\vec C = 0$ at the quantum level. Derive a quantisation condition on the allowed face areas. How many degrees of freedom (independent quantum numbers) does a quantum tetrahedron have? Compare to the classical case. What about polyhedra with more than four faces? What is the relation to the LQG Hilbert space? Hint: all invariant tensors in SU$(2)$ can be constructed by contracting 3-valent ones, which are unique for 3 given representations to be intertwined.

	\item {\bf Group field theory} \\
	Visualise the gluing of tetrahedra to a four-simplex shown in figure \ref{fig:GFTInt} in lower dimensions. 
	
\end{enumerate}

\section*{Acknowledgements}

These lectures were prepared for an introductory course given at the University of Regensburg in June 2016 in response to the receipt of a ``Dozentur Professor Bernhard He{\ss}'' prize. The author thanks Mehdi Assanioussi, Martin Bojowald, Wojciech Kami{\'n}ski, and Antonia Zipfel for helpful discussions. Support by the Polish National Science Centre grant No. 2012/05/E/ST2/03308 and an FNP START fellowship is gratefully acknowledged.


\begin{thebibliography}{300}\setlength{\itemsep}{2.75mm}

\bibitem{LiberatiTestsOfLorentz}
S.~Liberati, ``{Tests of Lorentz invariance: a 2013 update},'' {\em Classical
  and Quantum Gravity} {\bf 30} (2013) 133001, {\tt arXiv:1304.5795 [gr-qc]}.

\bibitem{HaggardBlackHoleFireworks}
H.~M. Haggard and C.~Rovelli, ``{Quantum-gravity effects outside the horizon
  spark black to white hole tunneling},'' {\em Physical Review D} {\bf 92}
  (2015) 104020, {\tt arXiv:1407.0989 [gr-qc]}.

\bibitem{RovelliPlanckStars}
C.~Rovelli and F.~Vidotto, ``{Planck stars},'' {\em International Journal of
  Modern Physics D} {\bf 23} (2014) 1442026, {\tt arXiv:1401.6562 [gr-qc]}.

\bibitem{BarrauFastRadioBursts}
A.~Barrau, C.~Rovelli, and F.~Vidotto, ``{Fast radio bursts and white hole
  signals},'' {\em Physical Review D} {\bf 90} (2014) 127503, {\tt
  arXiv:1409.4031 [gr-qc]}.

\bibitem{DimopoulosBlackHolesAt}
S.~Dimopoulos and G.~Landsberg, ``{Black holes at the Large Hadron Collider},''
  {\em Physical Review Letters} {\bf 87} (2001) 161602, {\tt
  arXiv:hep-ph/0106295}.

\bibitem{Arkani-HamedTheHierarchyProblem}
N.~Arkani-Hamed, S.~Dimopoulos, and G.~Dvali, ``{The hierarchy problem and new
  dimensions at a millimeter},'' {\em Physics Letters B} {\bf 429} (1998)
  263--272, {\tt arXiv:hep-ph/9803315}.

\bibitem{AshtekarQuantumGravityExtension}
I.~Agullo, A.~Ashtekar, and W.~Nelson, ``{Quantum Gravity Extension of the
  Inflationary Scenario},'' {\em Physical Review Letters} {\bf 109} (2012)
  251301, {\tt arXiv:1209.1609 [gr-qc]}.

\bibitem{AshtekarLoopQuantumCosmologyFrom}
A.~Ashtekar and A.~Barrau, ``{Loop quantum cosmology: from pre-inflationary
  dynamics to observations},'' {\em Classical and Quantum Gravity} {\bf 32}
  (2015) 234001, {\tt arXiv:1504.07559 [gr-qc]}.

\bibitem{LuestTheLHCString}
D.~L\"{u}st, S.~Stieberger, and T.~R. Taylor, ``{The LHC string Hunter's
  companion},'' {\em Nuclear Physics B} {\bf 808} (2009) 1--52, {\tt
  arXiv:0807.3333 [hep-th]}.

\bibitem{MaharanaModelsOfParticle}
A.~Maharana and E.~Palti, ``{Models of Particle Physics from Type IIB String
  Theory and F-theory: A Review},'' {\em International Journal of Modern
  Physics A} {\bf 28} (2013) 1330005, {\tt arXiv:1212.0555 [hep-th]}.

\bibitem{DenefQuantumOscillationsAnd}
F.~Denef, S.~A. Hartnoll, and S.~Sachdev, ``{Quantum oscillations and black
  hole ringing},'' {\em Physical Review D} {\bf 80} (2009) 126016, {\tt
  arXiv:0908.1788 [hep-th]}.

\bibitem{DenefBlackHoleDeterminants}
F.~Denef, S.~A. Hartnoll, and S.~Sachdev, ``{Black hole determinants and
  quasinormal modes},'' {\em Classical and Quantum Gravity} {\bf 27} (2010)
  125001, {\tt arXiv:0908.2657 [hep-th]}.

\bibitem{Caron-HuotHydrodynamicLongTime}
S.~Caron-Huot and O.~Saremi, ``{Hydrodynamic long-time tails from Anti de
  Sitter space},'' {\em Journal of High Energy Physics} {\bf 2010} (2010) 13,
  {\tt arXiv:0909.4525 [hep-th]}.

\bibitem{OritiApproachesToQuantum}
D.~Oriti, {\em {Approaches to Quantum Gravity: Toward a New Understanding of
  Space, Time and Matter}}.
\newblock Cambridge University Press, 2009.

\bibitem{RovelliNotesForA}
C.~Rovelli, ``{Notes for a brief history of quantum gravity},'' {\tt
  arXiv:gr-qc/0006061}.

\bibitem{FlanaganDoesBackReaction}
E.~E. Flanagan and R.~M. Wald, ``{Does back reaction enforce the averaged null
  energy condition in semiclassical gravity?},'' {\em Physical Review D} {\bf
  54} (1996) 6233--6283, {\tt arXiv:gr-qc/9602052}.

\bibitem{DappiaggiStableCosmologicalModels}
C.~Dappiaggi, K.~Fredenhagen, and N.~Pinamonti, ``{Stable cosmological models
  driven by a free quantum scalar field},'' {\em Physical Review D} {\bf 77}
  (2008) 104015, {\tt arXiv:0801.2850 [gr-qc]}.

\bibitem{GoroffQuantumGravityAt}
M.~H. Goroff and A.~Sagnotti, ``{Quantum gravity at two loops},'' {\em Physics
  Letters B} {\bf 160} (1985) 81--86.

\bibitem{DonoghueTheEffectiveField}
J.~F. Donoghue, ``{The effective field theory treatment of quantum gravity},''
  in {\em AIP Conf.Proc. 1483}, pp.~73--94, 2012.
\newblock {\tt arXiv:1209.3511 [gr-qc]}.

\bibitem{DeserNonrenormalizabilityOfLast}
S.~Deser, ``{Nonrenormalizability of (last hope) D= 11 supergravity, with a
  terse survey of divergences in quantum gravities},'' {\tt
  arXiv:hep-th/9905017}.

\bibitem{BernUltravioletBehaviorOf}
Z.~Bern, J.~Carrasco, L.~Dixon, H.~Johansson, and R.~Roiban, ``{Ultraviolet
  Behavior of N=8 Supergravity at Four Loops},'' {\em Physical Review Letters}
  {\bf 103} (2009) 081301, {\tt arXiv:0905.2326 [hep-th]}.

\bibitem{ReuterQuantumEinsteinGravity}
M.~Reuter and F.~Saueressig, ``{Quantum Einstein gravity},'' {\em New Journal
  of Physics} {\bf 14} (2012) 055022, {\tt arXiv:1202.2274 [hep-th]}.

\bibitem{DonaMatterMattersIn}
P.~Dona, A.~Eichhorn, and R.~Percacci, ``{Matter matters in asymptotically safe
  quantum gravity},'' {\em Physical Review D} {\bf 89} (2014) 084035, {\tt
  arXiv:1311.2898 [hep-th]}.

\bibitem{LaihoEvidenceForAsymptotic}
J.~Laiho and D.~Coumbe, ``{Evidence for asymptotic safety from lattice quantum
  gravity.},'' {\em Physical review letters} {\bf 107} (2011) 161301, {\tt
  arXiv:1104.5505 [hep-lat]}.

\bibitem{ShaposhnikovAsymptoticSafetyOf}
M.~Shaposhnikov and C.~Wetterich, ``{Asymptotic safety of gravity and the Higgs
  boson mass},'' {\em Physics Letters B} {\bf 683} (2010) 196--200, {\tt
  arXiv:0912.0208 [hep-th]}.

\bibitem{WheelerGeometrodynamics}
J.~A. Wheeler, {\em {Geometrodynamics}}.
\newblock Academic Press, New York, 1962.

\bibitem{DeWittQuantumTheoryOfI}
B.~DeWitt, ``{Quantum Theory of Gravity. I. The Canonical Theory},'' {\em
  Physical Review} {\bf 160} (1967) 1113--1148.

\bibitem{ArnowittTheDynamicsOf}
R.~Arnowitt, S.~Deser, and C.~W. Misner, ``{Republication of: The dynamics of
  general relativity},'' {\em General Relativity and Gravitation} {\bf 40}
  (2008) 1997--2027, {\tt arXiv:gr-qc/0405109}.

\bibitem{IshamCanonicalQuantumGravity}
C.~J. Isham, ``{Canonical Quantum Gravity and the Problem of Time},'' {\tt
  arXiv:gr-qc/9210011}.

\bibitem{GieselScalarMaterialReference}
K.~Giesel and T.~Thiemann, ``{Scalar material reference systems and loop
  quantum gravity},'' {\em Classical and Quantum Gravity} {\bf 32} (2015)
  135015, {\tt arXiv:1206.3807 [gr-qc]}.

\bibitem{HamberQuantumGravitation}
H.~Hamber, {\em {Quantum Gravitation}}.
\newblock Springer, Berlin, Heidelberg, 2008.

\bibitem{AmbjornQuantumGravityVia}
J.~Ambjorn, A.~Goerlich, J.~Jurkiewicz, and R.~Loll, ``{Quantum Gravity via
  Causal Dynamical Triangulations},'' {\tt arXiv:1302.2173 [hep-th]}.

\bibitem{LaihoLatticeQuantumGravity}
J.~Laiho, S.~Bassler, D.~Coumbe, D.~Du, and J.~T. Neelakanta, ``{Lattice
  Quantum Gravity and Asymptotic Safety},'' {\tt arXiv:1604.02745 [hep-th]}.

\bibitem{PolchinskiBook1}
J.~Polchinski, {\em {String Theory, Vol. 1: An Introduction to the bosonic
  string}}.
\newblock Cambridge University Press, Cambridge, 1998.

\bibitem{PolchinskiBook2}
J.~Polchinski, {\em {String Theory, Vol. 2: Superstring theory and beyond}}.
\newblock Cambridge University Press, Cambridge, 1998.

\bibitem{MukhiStringTheoryThe}
S.~Mukhi, ``{String theory: the first 25 years},'' {\em Classical and Quantum
  Gravity} {\bf 28} (2011) 153001, {\tt arXiv:1110.2569 [physics.pop-ph]}.

\bibitem{AmmonGaugeGravityDuality}
M.~Ammon and J.~Erdmenger, {\em {Gauge/Gravity Duality: Foundations and
  Applications}}.
\newblock Cambridge University Press, 2015.

\bibitem{MaldacenaTheLargeN}
J.~Maldacena, ``{The Large N Limit of Superconformal Field Theories and
  Supergravity},'' {\em Advances in Theoretical and Mathematical Physics} {\bf
  2} (1998) 231--252, {\tt arXiv:hep-th/9711200}.

\bibitem{GubserGaugeTheoryCorrelators}
S.~Gubser, I.~Klebanov, and A.~Polyakov, ``{Gauge theory correlators from
  non-critical string theory},'' {\em Physics Letters B} {\bf 428} (1998)
  105--114, {\tt arXiv:hep-th/9802109}.

\bibitem{WittenAntiDeSitter}
E.~Witten, ``{Anti De Sitter Space And Holography},'' {\em Advances in
  Theoretical and Mathematical Physics} {\bf 2} (1998) 253--291, {\tt
  arXiv:hep-th/9802150}.

\bibitem{RovelliQuantumGravity}
C.~Rovelli, {\em {Quantum Gravity}}.
\newblock Cambridge University Press, Cambridge, 2004.

\bibitem{ThiemannModernCanonicalQuantum}
T.~Thiemann, {\em {Modern Canonical Quantum General Relativity}}.
\newblock Cambridge University Press, Cambridge, 2007.

\bibitem{GambiniAFirstCourse}
J.~Pullin and R.~Gambini, {\em {A First Course in Loop Quantum Gravity}}.
\newblock Oxford University Press, USA, 2011.

\bibitem{AshtekarNewVariablesFor}
A.~Ashtekar, ``{New Variables for Classical and Quantum Gravity},'' {\em
  Physical Review Letters} {\bf 57} (1986) 2244--2247.

\bibitem{BarberoRealAshtekarVariables}
J.~Barbero, ``{Real Ashtekar variables for Lorentzian signature space-times},''
  {\em Physical Review D} {\bf 51} (1995) 5507--5510, {\tt
  arXiv:gr-qc/9410014}.

\bibitem{AshtekarLoopQuantumCosmology}
A.~Ashtekar and P.~Singh, ``{Loop quantum cosmology: a status report},'' {\em
  Classical and Quantum Gravity} {\bf 28} (2011) 213001, {\tt arXiv:1108.0893
  [gr-qc]}.

\bibitem{BollietConceptualIssuesIn}
B.~Bolliet and A.~Barrau, ``{Conceptual issues in loop quantum cosmology},''
  {\tt arXiv:1602.04452 [gr-qc]}.

\bibitem{PerezTheSpinFoam}
A.~Perez, ``{The Spin-Foam Approach to Quantum Gravity},'' {\em Living Reviews
  in Relativity} {\bf 16} (2013) {\tt arXiv:1205.2019 [gr-qc]}.

\bibitem{BaratinTenQuestionsOn}
A.~Baratin and D.~Oriti, ``{Ten questions on Group Field Theory (and their
  tentative answers)},'' {\em Journal of Physics: Conference Series} {\bf 360}
  (2012) 012002, {\tt arXiv:1112.3270 [gr-qc]}.

\bibitem{GiuliniOnTheGenerality}
D.~Giulini and D.~Marolf, ``{On the generality of refined algebraic
  quantization},'' {\em Classical and Quantum Gravity} {\bf 16} (1999)
  2479--2488, {\tt arXiv:gr-qc/9812024}.

\bibitem{GiuliniGroupAveragingAnd}
D.~Giulini, ``{Group averaging and refined algebraic quantization},'' {\em
  Nuclear Physics B - Proceedings Supplements} {\bf 88} (2000) 385--388, {\tt
  arXiv:gr-qc/0003040}.

\bibitem{ThiemannQSD5}
T.~Thiemann, ``{Quantum spin dynamics (QSD) V: Quantum Gravity as the Natural
  Regulator of Matter Quantum Field Theories},'' {\em Classical and Quantum
  Gravity} {\bf 15} (1998) 1281--1314, {\tt arXiv:gr-qc/9705019}.

\bibitem{HusainApparentHorizonsBlack}
V.~Husain, ``{Apparent horizons, black hole entropy, and loop quantum
  gravity},'' {\em Physical Review D} {\bf 59} (1999) 084019, {\tt
  arXiv:gr-qc/9806115}.

\bibitem{DonnellyEntanglementEntropyIn}
W.~Donnelly, ``{Entanglement entropy in loop quantum gravity},'' {\em Physical
  Review D} {\bf 77} (2008) 104006, {\tt arXiv:0802.0880 [gr-qc]}.

\bibitem{BII}
N.~Bodendorfer, ``{A note on entanglement entropy and quantum geometry},'' {\em
  Classical and Quantum Gravity} {\bf 31} (2014) 214004, {\tt arXiv:1402.1038
  [gr-qc]}.

\bibitem{BLSI}
N.~Bodendorfer, J.~Lewandowski, and J.~Swiezewski, ``{A quantum reduction to
  spherical symmetry in loop quantum gravity},'' {\em Physics Letters B} {\bf
  747} (2015) 18--21, {\tt arXiv:1410.5609 [gr-qc]}.

\bibitem{BIII}
N.~Bodendorfer, ``{Quantum reduction to Bianchi I models in loop quantum
  gravity},'' {\em Physical Review D} {\bf 91} (2015) 081502(R), {\tt
  arXiv:1410.5608 [gr-qc]}.

\bibitem{BZI}
N.~Bodendorfer and A.~Zipfel, ``{On the relation between reduced quantisation
  and quantum reduction for spherical symmetry in loop quantum gravity},'' {\em
  Classical and Quantum Gravity} {\bf 33} (2016) 155014, {\tt arXiv:1512.00221
  [gr-qc]}.

\bibitem{BVI}
N.~Bodendorfer, ``{An embedding of loop quantum cosmology in (b,v) variables
  into a full theory context},'' {\em Classical and Quantum Gravity} {\bf 33}
  (2016) 125014, {\tt arXiv:1512.00713 [gr-qc]}.

\bibitem{BSTI}
N.~Bodendorfer, A.~Stottmeister, and A.~Thurn, ``{Loop quantum gravity without
  the Hamiltonian contraint},'' {\em Classical and Quantum Gravity} {\bf 30}
  (2013) 082001, {\tt arXiv:1203.6525 [gr-qc]}.

\bibitem{BLSII}
N.~Bodendorfer, J.~Lewandowski, and J.~Swiezewski, ``{General relativity in the
  radial gauge: Reduced phase space and canonical structure},'' {\em Physical
  Review D} {\bf 92} (2015) 084041, {\tt arXiv:1506.09164 [gr-qc]}.

\bibitem{GieselAQG2}
K.~Giesel and T.~Thiemann, ``{Algebraic quantum gravity (AQG): II.
  Semiclassical analysis},'' {\em Classical and Quantum Gravity} {\bf 24}
  (2007) 2499--2564, {\tt arXiv:gr-qc/0607100}.

\bibitem{GieselAQG3}
K.~Giesel and T.~Thiemann, ``{Algebraic quantum gravity (AQG): III.
  Semiclassical perturbation theory},'' {\em Classical and Quantum Gravity}
  {\bf 24} (2007) 2565--2588, {\tt arXiv:gr-qc/0607101}.

\bibitem{BarrettLorentzianSpinFoam}
J.~W. Barrett, R.~J. Dowdall, W.~J. Fairbairn, F.~Hellmann, and R.~Pereira,
  ``{Lorentzian spin foam amplitudes: graphical calculus and asymptotics},''
  {\em Classical and Quantum Gravity} {\bf 27} (2010) 165009, {\tt
  arXiv:0907.2440 [gr-qc]}.

\bibitem{HanAsymptoticsOfSpinfoam}
M.~Han and M.~Zhang, ``{Asymptotics of spinfoam amplitude on simplicial
  manifold: Lorentzian theory},'' {\em Classical and Quantum Gravity} {\bf 30}
  (2013) 165012, {\tt arXiv:1109.0499 [gr-qc]}.

\bibitem{DittrichTheContinuumLimit}
B.~Dittrich, ``{The continuum limit of loop quantum gravity - a framework for
  solving the theory},'' {\tt arXiv:1409.1450 [gr-qc]}.

\bibitem{BahrOnBackgroundIndependent}
B.~Bahr, ``{On background-independent renormalization of spin foam models},''
  {\tt arXiv:1407.7746 [gr-qc]}.

\bibitem{BahrNumericalEvidenceFor}
B.~Bahr and S.~Steinhaus, ``{Numerical evidence for a phase transition in 4d
  spin foam quantum gravity},'' {\tt arXiv:1605.07649 [gr-qc]}.

\bibitem{GielenEmergenceOfA}
S.~Gielen, ``{Emergence of a Low Spin Phase in Group Field Theory
  Condensates},'' {\tt arXiv:1604.06023 [gr-qc]}.

\bibitem{BianchiSqueezedVacuaIn}
E.~Bianchi, J.~Guglielmon, L.~Hackl, and N.~Yokomizo, ``{Squeezed vacua in loop
  quantum gravity},'' {\tt arXiv:1605.05356 [gr-qc]}.

\bibitem{SahlmannTowardsTheQFT}
H.~Sahlmann and T.~Thiemann, ``{Towards the QFT on curved spacetime limit of
  QGR: I. A general scheme},'' {\em Classical and Quantum Gravity} {\bf 23}
  (2006) 867--908, {\tt arXiv:gr-qc/0207030}.

\bibitem{SahlmannTowardsTheQFT2}
H.~Sahlmann and T.~Thiemann, ``{Towards the QFT on curved spacetime limit of
  QGR: II. A concrete implementation},'' {\em Classical and Quantum Gravity}
  {\bf 23} (2006) 909--954, {\tt arXiv:gr-qc/0207031}.

\bibitem{StottmeisterCoherentStatesQuantumI}
A.~Stottmeister and T.~Thiemann, ``{Coherent states, quantum gravity, and the
  Born-Oppenheimer approximation. I. General considerations},'' {\em Journal of
  Mathematical Physics} {\bf 57} (2016) 063509, {\tt arXiv:1504.02169
  [math-ph]}.

\bibitem{StottmeisterCoherentStatesQuantumII}
A.~Stottmeister and T.~Thiemann, ``{Coherent states, quantum gravity and the
  Born-Oppenheimer approximation, II: Compact Lie Groups},'' {\tt
  arXiv:1504.02170 [math-ph]}.

\bibitem{StottmeisterCoherentStatesQuantumIII}
A.~Stottmeister and T.~Thiemann, ``{Coherent states, quantum gravity and the
  Born-Oppenheimer approximation, III: Applications to loop quantum gravity},''
  {\tt arXiv:1504.02171 [math-ph]}.

\bibitem{RovelliReconcilePlanckScale}
C.~Rovelli and S.~Speziale, ``{Reconcile Planck-scale discreteness and the
  Lorentz-Fitzgerald contraction},'' {\em Physical Review D} {\bf 67} (2003)
  064019, {\tt arXiv:gr-qc/0205108}.

\bibitem{LivineAboutLorentzInvariance}
E.~R. Livine and D.~Oriti, ``{About Lorentz invariance in a discrete quantum
  setting},'' {\em Journal of High Energy Physics} {\bf 2004} (2004) 050--050,
  {\tt arXiv:gr-qc/0405085}.

\bibitem{SnyderQuantizedSpaceTime}
H.~S. Snyder, ``{Quantized Space-Time},'' {\em Physical Review} {\bf 71} (1947)
  38--41.

\bibitem{DowkerQuantumGravityPhenomenology}
F.~Dowker, J.~Henson, and R.~D. Sorkin, ``{Quantum Gravity Phenomenology,
  Lorentz Invariance and Discreteness},'' {\em Modern Physics Letters A} {\bf
  19} (2004) 1829--1840, {\tt arXiv:gr-qc/0311055}.

\bibitem{RovelliLorentzCovarianceOf}
C.~Rovelli and S.~Speziale, ``{Lorentz covariance of loop quantum gravity},''
  {\em Physical Review D} {\bf 83} (2011) 104029, {\tt arXiv:1012.1739
  [gr-qc]}.

\bibitem{EngleLQGVertexWith}
J.~Engle, E.~R. Livine, R.~Pereira, and C.~Rovelli, ``{LQG vertex with finite
  Immirzi parameter},'' {\em Nuclear Physics B} {\bf 799} (2008) 136--149, {\tt
  arXiv:0711.0146 [gr-qc]}.

\bibitem{AlexandrovHilbertSpaceStructure}
S.~Alexandrov, ``{Hilbert space structure of covariant loop quantum gravity},''
  {\em Physical Review D} {\bf 66} (2002) 024028, {\tt arXiv:gr-qc/0201087}.

\bibitem{CianfraniTowardsLoopQuantum}
F.~Cianfrani and G.~Montani, ``{Towards loop quantum gravity without the time
  gauge.},'' {\em Physical review letters} {\bf 102} (2009) 091301, {\tt
  arXiv:0811.1916 [gr-qc]}.

\bibitem{GeillerALorentzCovariant}
M.~Geiller, M.~Lachieze-Rey, K.~Noui, and F.~Sardelli, ``{A Lorentz-Covariant
  Connection for Canonical Gravity},'' {\em SIGMA} {\bf 7} (2011) 083, {\tt
  arXiv:1103.4057 [gr-qc]}.

\bibitem{BTTI}
N.~Bodendorfer, T.~Thiemann, and A.~Thurn, ``{New variables for classical and
  quantum gravity in all dimensions: I. Hamiltonian analysis},'' {\em Classical
  and Quantum Gravity} {\bf 30} (2013) 045001, {\tt arXiv:1105.3703 [gr-qc]}.

\bibitem{CailleteauAnomalyFreeScalar}
T.~Cailleteau, J.~Mielczarek, A.~Barrau, and J.~Grain, ``{Anomaly-free scalar
  perturbations with holonomy corrections in loop quantum cosmology},'' {\em
  Classical and Quantum Gravity} {\bf 29} (2012) 095010, {\tt arXiv:1111.3535
  [gr-qc]}.

\bibitem{CailleteauAnomalyFreePerturbations}
T.~Cailleteau, L.~Linsefors, and A.~Barrau, ``{Anomaly-free perturbations with
  inverse-volume and holonomy corrections in loop quantum cosmology},'' {\em
  Classical and Quantum Gravity} {\bf 31} (2014) 125011, {\tt arXiv:1307.5238
  [gr-qc]}.

\bibitem{BojowaldDiscretenessCorrectionsAnd}
M.~Bojowald, G.~M. Paily, and J.~D. Reyes, ``{Discreteness corrections and
  higher spatial derivatives in effective canonical quantum gravity},'' {\em
  Physical Review D} {\bf 90} (2014) 025025, {\tt arXiv:1402.5130 [gr-qc]}.

\bibitem{BarrauAnomalyFreeCosmological}
A.~Barrau, M.~Bojowald, G.~Calcagni, J.~Grain, and M.~Kagan, ``{Anomaly-free
  cosmological perturbations in effective canonical quantum gravity},'' {\em
  Journal of Cosmology and Astroparticle Physics} {\bf 2015} (2015) 051--051,
  {\tt arXiv:1404.1018 [gr-qc]}.

\bibitem{BojowaldSomeImplicationsOf}
M.~Bojowald and J.~Mielczarek, ``{Some implications of signature-change in
  cosmological models of loop quantum gravity},'' {\em Journal of Cosmology and
  Astroparticle Physics} {\bf 2015} (2015) 052--052, {\tt arXiv:1503.09154
  [gr-qc]}.

\bibitem{CollinsLorentzInvarianceAnd}
J.~Collins, A.~Perez, D.~Sudarsky, L.~Urrutia, and H.~Vucetich, ``{Lorentz
  invariance and quantum gravity: an additional fine-tuning problem?},'' {\em
  Physical review letters} {\bf 93} (2004) 191301, {\tt arXiv:gr-qc/0403053}.

\bibitem{BelenchiaLorentzViolationNaturalness}
A.~Belenchia, A.~Gambassi, and S.~Liberati, ``{Lorentz violation naturalness
  revisited},'' {\em Journal of High Energy Physics} {\bf 2016} (2016) 49, {\tt
  arXiv:1601.06700 [hep-th]}.

\bibitem{AgulloThePreInflationary}
I.~Agullo, A.~Ashtekar, and W.~Nelson, ``{The pre-inflationary dynamics of loop
  quantum cosmology: confronting quantum gravity with observations},'' {\em
  Classical and Quantum Gravity} {\bf 30} (2013) 085014, {\tt arXiv:1302.0254
  [gr-qc]}.

\bibitem{BollietObservationalExclusionOf}
B.~Bolliet, A.~Barrau, J.~Grain, and S.~Schander, ``{Observational exclusion of
  a consistent loop quantum cosmology scenario},'' {\em Physical Review D} {\bf
  93} (2016) 124011, {\tt arXiv:1510.08766 [gr-qc]}.

\bibitem{ImmirziQuantumGravityAnd}
G.~Immirzi, ``{Quantum gravity and Regge calculus},'' {\em Nuclear Physics B -
  Proceedings Supplements} {\bf 57} (1997) 65--72, {\tt arXiv:gr-qc/9701052}.

\bibitem{AshtekarQuantumGeometryOf}
A.~Ashtekar, J.~Baez, and K.~Krasnov, ``{Quantum Geometry of Isolated Horizons
  and Black Hole Entropy},'' {\em Advances in Theoretical and Mathematical
  Physics} {\bf 4} (2000) 1--94, {\tt arXiv:gr-qc/0005126}.

\bibitem{FroddenBlackHoleEntropy}
E.~Frodden, M.~Geiller, K.~Noui, and A.~Perez, ``{Black-hole entropy from
  complex Ashtekar variables},'' {\em Europhysics Letters} {\bf 107} (2014)
  10005, {\tt arXiv:1212.4060 [gr-qc]}.

\bibitem{BNI}
N.~Bodendorfer and Y.~Neiman, ``{Imaginary action, spinfoam asymptotics and the
  'transplanckian' regime of loop quantum gravity},'' {\em Classical and
  Quantum Gravity} {\bf 30} (2013) 195018, {\tt arXiv:1303.4752 [gr-qc]}.

\bibitem{ThiemannQSD1}
T.~Thiemann, ``{Quantum spin dynamics (QSD)},'' {\em Classical and Quantum
  Gravity} {\bf 15} (1998) 839--873, {\tt arXiv:gr-qc/9606089}.

\bibitem{NicolaiLoopQuantumGravity}
H.~Nicolai, K.~Peeters, and M.~Zamaklar, ``{Loop quantum gravity: an outside
  view},'' {\em Classical and Quantum Gravity} {\bf 22} (2005) R193--R247, {\tt
  arXiv:hep-th/0501114}.

\bibitem{LaddhaTheDiffeomorphismConstraint}
A.~Laddha and M.~Varadarajan, ``{The diffeomorphism constraint operator in loop
  quantum gravity},'' {\em Classical and Quantum Gravity} {\bf 28} (2011)
  195010, {\tt arXiv:1105.0636 [gr-qc]}.

\bibitem{TomlinTowardsAnAnomaly}
C.~Tomlin and M.~Varadarajan, ``{Towards an anomaly-free quantum dynamics for a
  weak coupling limit of Euclidean gravity},'' {\em Physical Review D} {\bf 87}
  (2013) 044039, {\tt arXiv:1210.6869 [gr-qc]}.

\bibitem{HendersonConstraintAlgebraInI}
A.~Henderson, A.~Laddha, and C.~Tomlin, ``{Constraint algebra in LQG reloaded :
  Toy model of a U(1)\^{}\{3\} Gauge Theory I},'' {\tt arXiv:1204.0211
  [gr-qc]}.

\bibitem{LaddhaHamiltonianConstraintIn}
A.~Laddha, ``{Hamiltonian constraint in Euclidean LQG revisited: First hints of
  off-shell Closure},'' {\tt arXiv:1401.0931 [gr-qc]}.

\bibitem{JacobsonThermodynamicsofSpacetime}
T.~Jacobson, ``{Thermodynamics of Spacetime: The Einstein Equation of State},''
  {\em Physical Review Letters} {\bf 75} (1995) 1260--1263, {\tt
  arXiv:gr-qc/9504004}.

\bibitem{ChirocoSpacetimeThermodynamicsWithout}
G.~Chirco, H.~M. Haggard, A.~Riello, and C.~Rovelli, ``{Spacetime
  thermodynamics without hidden degrees of freedom},'' {\em Physical Review D}
  {\bf 90} (2014) 044044, {\tt arXiv:1401.5262 [gr-qc]}.

\bibitem{AshtekarRobustnessOfKey}
A.~Ashtekar, A.~Corichi, and P.~Singh, ``{Robustness of key features of loop
  quantum cosmology},'' {\em Physical Review D} {\bf 77} (2008) 024046, {\tt
  arXiv:0710.3565 [gr-qc]}.

\bibitem{AshtekarAnIntroductionTo}
A.~Ashtekar, ``{An Introduction to Loop Quantum Gravity Through Cosmology},''
  {\em Nuovo Cim.} {\bf 122B} (2007) {\tt arXiv:gr-qc/0702030}.

\bibitem{KaminksiQuantumConstraintsDirac}
W.~Kaminski, J.~Lewandowski, and T.~Pawlowski, ``{Quantum constraints, Dirac
  observables and evolution: group averaging versus the Schr\"{o}dinger picture
  in LQC},'' {\em Classical and Quantum Gravity} {\bf 26} (2009) 245016, {\tt
  arXiv:0907.4322 [gr-qc]}.

\bibitem{CraigDynamicalEigenfunctionsAnd}
D.~A. Craig, ``{Dynamical eigenfunctions and critical density in loop quantum
  cosmology},'' {\em Classical and Quantum Gravity} {\bf 30} (2013) 035010,
  {\tt arXiv:1207.5601 [gr-qc]}.

\bibitem{BojowaldLargeScaleEffective}
M.~Bojowald, ``{Large scale effective theory for cosmological bounces},'' {\em
  Physical Review D} {\bf 75} (2007) 081301, {\tt arXiv:gr-qc/0608100}.

\bibitem{MielczarekExactSolutionsFor}
J.~Mielczarek, T.~Stachowiak, and M.~Szydlowski, ``{Exact solutions for a big
  bounce in loop quantum cosmology},'' {\em Physical Review D} {\bf 77} (2008)
  123506, {\tt arXiv:0801.0502 [gr-qc]}.

\bibitem{BojowaldAbsenceOfSingularity}
M.~Bojowald, ``{Absence of a Singularity in Loop Quantum Cosmology},'' {\em
  Physical Review Letters} {\bf 86} (2001) 5227--5230, {\tt
  arXiv:gr-qc/0102069}.

\bibitem{AshtekarMathematicalStructureOf}
A.~Ashtekar, M.~Bojowald, and J.~Lewandowski, ``{Mathematical structure of loop
  quantum cosmology},'' {\em Adv.Theor.Math.Phys.} {\bf 7} (2003) 233--268,
  {\tt arXiv:gr-qc/0304074}.

\bibitem{AshtekarQuantumNatureOf}
A.~Ashtekar, T.~Pawlowski, and P.~Singh, ``{Quantum nature of the big bang:
  Improved dynamics},'' {\em Physical Review D} {\bf 74} (2006) 084003, {\tt
  arXiv:gr-qc/0607039}.

\bibitem{DiracLecturesOnQuantum}
P.~A.~M. Dirac, {\em {Lectures on Quantum Mechanics}}.
\newblock Belfer Graduate School of Science, Yeshiva University Press, New
  York, 1964.

\bibitem{SenovillaSingularityTheoremsIn}
J.~M. Senovilla, ``{Singularity Theorems in General Relativity: Achievements
  and Open Questions},'' {\tt arXiv:physics/0605007 [physics.hist-ph]}.

\bibitem{HiguchiQuantumLinearizationInstabilities}
A.~Higuchi, ``{Quantum linearization instabilities of de Sitter spacetime.
  II},'' {\em Classical and Quantum Gravity} {\bf 8} (1991) 1983--2004.

\bibitem{CraigConsistentProbabilitiesInWheeler}
D.~A. Craig and P.~Singh, ``{Consistent probabilities in Wheeler-DeWitt quantum
  cosmology},'' {\em Physical Review D} {\bf 82} (2010) 123526, {\tt
  arXiv:1006.3837 [gr-qc]}.

\bibitem{CraigConsistentProbabilitiesInLoop}
D.~A. Craig and P.~Singh, ``{Consistent probabilities in loop quantum
  cosmology},'' {\em Classical and Quantum Gravity} {\bf 30} (2013) 205008,
  {\tt arXiv:1306.6142 [gr-qc]}.

\bibitem{FalcianoWheelerDeWittQuantization}
F.~T. Falciano, N.~Pinto-Neto, and W.~Struyve, ``{Wheeler-DeWitt quantization
  and singularities},'' {\em Physical Review D} {\bf 91} (2015) 043524, {\tt
  arXiv:1501.04181 [gr-qc]}.

\bibitem{CraigTheConsistentHistories}
D.~A. Craig, ``{The consistent histories approach to loop quantum cosmology},''
  {\em International Journal of Modern Physics D} {\bf 25} (2016) 1642009, {\tt
  arXiv:1604.01385 [gr-qc]}.

\bibitem{ThiemannLecturesOnLoop}
T.~Thiemann, ``{Lectures on loop quantum gravity},'' {\em Lecture Notes in
  Physics} {\bf 631} (2002) 41--135, {\tt arXiv:gr-qc/0210094}.

\bibitem{PerezIntroductionToLoop}
A.~Perez, ``{Introduction to loop quantum gravity and spin foams},'' {\tt
  arXiv:gr-qc/0409061}.

\bibitem{MercuriIntroductionToLoop}
S.~Mercuri, ``{Introduction to Loop Quantum Gravity},'' {\tt arXiv:1001.1330
  [gr-qc]}.

\bibitem{GieselFromClassicalTo}
K.~Giesel and H.~Sahlmann, ``{From Classical To Quantum Gravity: Introduction
  to Loop Quantum Gravity},'' {\tt arXiv:1203.2733 [gr-qc]}.

\bibitem{BilsonLQGForThe}
S.~Bilson-Thompson and D.~Vaid, ``{LQG for the Bewildered},'' {\tt
  arXiv:1402.3586 [gr-qc]}.

\bibitem{RovelliLoopSpaceRepresentation}
C.~Rovelli and L.~Smolin, ``{Loop space representation of quantum general
  relativity},'' {\em Nuclear Physics B} {\bf 331} (1990) 80--152.

\bibitem{AshtekarRepresentationsOfThe}
A.~Ashtekar and C.~J. Isham, ``{Representations of the holonomy algebras of
  gravity and non-Abelian gauge theories},'' {\em Classical and Quantum
  Gravity} {\bf 9} (1992) 1433--1468, {\tt arXiv:hep-th/9202053}.

\bibitem{AshtekarRepresentationTheoryOf}
A.~Ashtekar and J.~Lewandowski, ``{Representation Theory of Analytic Holonomy
  C* Algebras},'' in {\em Knots and Quantum Gravity} (J.~Baez, ed.), (Oxford),
  Oxford University Press1994.
\newblock {\tt arXiv:gr-qc/9311010}.

\bibitem{AshtekarDifferentialGeometryOn}
A.~Ashtekar and J.~Lewandowski, ``{Differential geometry on the space of
  connections via graphs and projective limits},'' {\em Journal of Geometry and
  Physics} {\bf 17} (1995) 191--230, {\tt arXiv:hep-th/9412073}.

\bibitem{AshtekarProjectiveTechniquesAnd}
A.~Ashtekar and J.~Lewandowski, ``{Projective techniques and functional
  integration for gauge theories},'' {\em Journal of Mathematical Physics} {\bf
  36} (1995) 2170--2191, {\tt arXiv:gr-qc/9411046}.

\bibitem{MarolfOnTheSupport}
D.~Marolf and J.~M. Mour\~{a}o, ``{On the support of the Ashtekar-Lewandowski
  measure},'' {\em Communications in Mathematical Physics} {\bf 170} (1995)
  583--605, {\tt arXiv:hep-th/9403112}.

\bibitem{AshtekarQuantizationOfDiffeomorphism}
A.~Ashtekar, J.~Lewandowski, D.~Marolf, J.~M. Mour\~{a}o, and T.~Thiemann,
  ``{Quantization of diffeomorphism invariant theories of connections with
  local degrees of freedom},'' {\em Journal of Mathematical Physics} {\bf 36}
  (1995) 6456--6493, {\tt arXiv:gr-qc/9504018}.

\bibitem{AlexandrovSU(2)LoopQuantum}
S.~Alexandrov and E.~Livine, ``{SU(2) loop quantum gravity seen from covariant
  theory},'' {\em Physical Review D} {\bf 67} (2003) 044009, {\tt
  arXiv:gr-qc/0209105}.

\bibitem{BTTII}
N.~Bodendorfer, T.~Thiemann, and A.~Thurn, ``{New variables for classical and
  quantum gravity in all dimensions: II. Lagrangian analysis},'' {\em Classical
  and Quantum Gravity} {\bf 30} (2013) 045002, {\tt arXiv:1105.3704 [gr-qc]}.

\bibitem{LewandowskiUniquenessOfDiffeomorphism}
J.~Lewandowski, A.~Okol\'{o}w, H.~Sahlmann, and T.~Thiemann, ``{Uniqueness of
  Diffeomorphism Invariant States on Holonomy-Flux Algebras},'' {\em
  Communications in Mathematical Physics} {\bf 267} (2006) 703--733, {\tt
  arXiv:gr-qc/0504147}.

\bibitem{SmolinRecentDevelopmentsIn}
L.~Smolin, ``{Recent developments in non-perturbative quantum gravity},'' in
  {\em Proceedings of the XXIIth GIFT International Seminar on Theoretical
  Physics}, pp.~3--84, World Scientific, 1992.
\newblock {\tt arXiv:hep-th/9202022}.

\bibitem{AshtekarQuantumTheoryOf1}
A.~Ashtekar and J.~Lewandowski, ``{Quantum theory of gravity I: Area
  operators},'' {\em Classical and Quantum Gravity} {\bf 14} (1997) A55--A81,
  {\tt arXiv:gr-qc/9602046}.

\bibitem{RovelliDiscretenessOfArea}
C.~Rovelli and L.~Smolin, ``{Discreteness of area and volume in quantum
  gravity},'' {\em Nuclear Physics B} {\bf 442} (1995) 593--619, {\tt
  arXiv:gr-qc/9411005}.

\bibitem{AshtekarQuantumTheoryOf2}
A.~Ashtekar and J.~Lewandowski, ``{Quantum theory of geometry. 2. Volume
  operators},'' {\em Adv.Theor.Math.Phys.} {\bf 1} (1998) 388--429, {\tt
  arXiv:gr-qc/9711031}.

\bibitem{ThiemannKinematicalHilbertSpaces}
T.~Thiemann, ``{Kinematical Hilbert spaces for Fermionic and Higgs quantum
  field theories},'' {\em Classical and Quantum Gravity} {\bf 15} (1998)
  1487--1512, {\tt arXiv:gr-qc/9705021}.

\bibitem{AshtekarPolymerAndFock}
A.~Ashtekar, J.~Lewandowski, and H.~Sahlmann, ``{Polymer and Fock
  representations for a scalar field},'' {\em Classical and Quantum Gravity}
  {\bf 20} (2003) L11--L21, {\tt arXiv:gr-qc/0211012}.

\bibitem{AshtekarGravitonsAndLoops}
A.~Ashtekar, C.~Rovelli, and L.~Smolin, ``{Gravitons and loops},'' {\em
  Physical Review D} {\bf 44} (1991) 1740--1755, {\tt arXiv:hep-th/9202054}.

\bibitem{VaradarajanPhotonsFromQuantized}
M.~Varadarajan, ``{Photons from quantized electric flux representations},''
  {\em Physical Review D} {\bf 64} (2001) 104003, {\tt arXiv:gr-qc/0104051}.

\bibitem{VaradarajanGravitonsFromA}
M.~Varadarajan, ``{Gravitons from a loop representation of linearized
  gravity},'' {\em Physical Review D} {\bf 66} (2002) 024017, {\tt
  arXiv:gr-qc/0204067}.

\bibitem{FreidelTwistedGeometriesA}
L.~Freidel and S.~Speziale, ``{Twisted geometries: A geometric parametrization
  of SU(2) phase space},'' {\em Physical Review D} {\bf 82} (2010) 084040, {\tt
  arXiv:1001.2748 [gr-qc]}.

\bibitem{BianchiPolyhedraInLoop}
E.~Bianchi, P.~Dona, and S.~Speziale, ``{Polyhedra in loop quantum gravity},''
  {\em Physical Review D} {\bf 83} (2011) 044035, {\tt arXiv:1009.3402
  [gr-qc]}.

\bibitem{GieselAQG4}
K.~Giesel and T.~Thiemann, ``{Algebraic quantum gravity (AQG): IV. Reduced
  phase space quantization of loop quantum gravity},'' {\em Classical and
  Quantum Gravity} {\bf 27} (2010) 175009, {\tt arXiv:0711.0119 [gr-qc]}.

\bibitem{ThiemannGCS1}
T.~Thiemann, ``{Gauge field theory coherent states (GCS): I. General
  properties},'' {\em Classical and Quantum Gravity} {\bf 18} (2001)
  2025--2064, {\tt arXiv:hep-th/0005233}.

\bibitem{ThiemannGCS2}
T.~Thiemann and O.~Winkler, ``{Gauge field theory coherent states (GCS): II.
  Peakedness properties},'' {\em Classical and Quantum Gravity} {\bf 18} (2001)
  2561--2636, {\tt arXiv:hep-th/0005237}.

\bibitem{ThiemannGCS3}
T.~Thiemann and O.~Winkler, ``{Gauge field theory coherent states (GCS): III.
  Ehrenfest theorems},'' {\em Classical and Quantum Gravity} {\bf 18} (2001)
  4629--4681, {\tt arXiv:hep-th/0005234}.

\bibitem{KibbleCanonicalVariablesFor}
T.~W.~B. Kibble, ``{Canonical Variables for the Interacting Gravitational and
  Dirac Fields},'' {\em Journal of Mathematical Physics} {\bf 4} (1963), no.~11
  1433--1437.

\bibitem{AshtekarInclusionOfMatter}
A.~Ashtekar, J.~Romano, and R.~Tate, ``{New variables for gravity: Inclusion of
  matter},'' {\em Physical Review D} {\bf 40} (1989) 2572--2587.

\bibitem{AriasSecondQuantizationOf}
P.~J. Arias, C.~di~Bartolo, X.~Fustero, R.~Gambini, and A.~Trias, ``{Second
  Quantization of the Antisymmetric Potential in the Space of Abelian
  Surfaces},'' {\em International Journal of Modern Physics A} {\bf 7} (1992)
  737--753.

\bibitem{Armand-UgonTowardsALoop}
D.~Armand-Ugon, R.~Gambini, O.~Obr\'{e}gon, and J.~Pullin, ``{Towards a loop
  representation for quantum canonical supergravity},'' {\em Nuclear Physics B}
  {\bf 460} (1996) 615--631, {\tt arXiv:hep-th/9508036}.

\bibitem{LingSupersymmetricSpinNetworks}
Y.~Ling and L.~Smolin, ``{Supersymmetric spin networks and quantum
  supergravity},'' {\em Physical Review D} {\bf 61} (2000) 044008, {\tt
  arXiv:hep-th/9904016}.

\bibitem{BTTVI}
N.~Bodendorfer, T.~Thiemann, and A.~Thurn, ``{Towards loop quantum supergravity
  (LQSG): I. Rarita-Schwinger sector},'' {\em Classical and Quantum Gravity}
  {\bf 30} (2013) 045006, {\tt arXiv:1105.3709 [gr-qc]}.

\bibitem{BTTVII}
N.~Bodendorfer, T.~Thiemann, and A.~Thurn, ``{Towards loop quantum supergravity
  (LQSG): II. p -form sector},'' {\em Classical and Quantum Gravity} {\bf 30}
  (2013) 045007, {\tt arXiv:1105.3710 [gr-qc]}.

\bibitem{GhoshAnImprovedEstimate}
A.~Ghosh and P.~Mitra, ``{An improved estimate of black hole entropy in the
  quantum geometry approach},'' {\em Physics Letters B} {\bf 616} (2005)
  114--117, {\tt arXiv:gr-qc/0411035}.

\bibitem{CorichiAmbiguitiesInLoop}
A.~Corichi and K.~Krasnov, ``{Ambiguities in loop quantization: Area versus
  electric charge},'' {\em Modern Physics Letters A} {\bf 13} (1998)
  1339--1346, {\tt arXiv:hep-th/9703177}.

\bibitem{BodendorferSomeNotesOn}
N.~Bodendorfer, ``{Some notes on the Kodama state, maximal symmetry, and the
  isolated horizon boundary condition},'' {\em Physical Review D} {\bf 93}
  (2016) 124042, {\tt arXiv:1602.05499 [gr-qc]}.

\bibitem{BrownDustAsStandard}
J.~Brown and K.~Kuchar, ``{Dust as a standard of space and time in canonical
  quantum gravity},'' {\em Physical Review D} {\bf 51} (1995) 5600--5629, {\tt
  arXiv:gr-qc/9409001}.

\bibitem{RovelliThePhysicalHamiltonian}
C.~Rovelli and L.~Smolin, ``{The physical Hamiltonian in nonperturbative
  quantum gravity},'' {\em Physical Review Letters} {\bf 72} (1994) 446--449,
  {\tt arXiv:gr-qc/9308002}.

\bibitem{DomagalaGravityQuantizedLoop}
M.~Domagala, K.~Giesel, W.~Kaminski, and J.~Lewandowski, ``{Gravity quantized:
  Loop quantum gravity with a scalar field},'' {\em Physical Review D} {\bf 82}
  (2010) 104038, {\tt arXiv:1009.2445 [gr-qc]}.

\bibitem{HusainTimeAndA}
V.~Husain and T.~Pawlowski, ``{Time and a physical Hamiltonian for quantum
  gravity},'' {\em Physical Review Letters} {\bf 108} (2012) 141301, {\tt
  arXiv:1108.1145 [gr-qc]}.

\bibitem{SwiezewskiOnTheProperties}
J.~Swiezewski, ``{On the properties of the irrotational dust model},'' {\em
  Classical and Quantum Gravity} {\bf 30} (2013) 237001, {\tt arXiv:1307.4687
  [gr-qc]}.

\bibitem{RovelliAshtekarFormulationOf}
C.~Rovelli, ``{Ashtekar formulation of general relativity and loop-space
  nonperturbative quantum gravity: a report},'' {\em Classical and Quantum
  Gravity} {\bf 8} (1991) 1613--1676.

\bibitem{HusainIntersectingLoopSolutions}
V.~Husain, ``{Intersecting loop solutions of the hamiltonian constraint of
  quantum general relativity},'' {\em Nuclear Physics B} {\bf 313} (1989)
  711--724.

\bibitem{BrugmannIntersectingNLoop}
B.~Br\"{u}gmann and J.~Pullin, ``{Intersecting N loop solutions of the
  hamiltonian constraint of quantum gravity},'' {\em Nuclear Physics B} {\bf
  363} (1991) 221--244.

\bibitem{GambiniLoopSpaceRepresentation}
R.~Gambini, ``{Loop space representation of quantum general relativity and the
  group of loops},'' {\em Physics Letters B} {\bf 255} (1991) 180--188.

\bibitem{BrugmannJonesPolynomialsFor}
B.~Br\"{u}gmann, R.~Gambini, and J.~Pullin, ``{Jones polynomials for
  intersecting knots as physical states of quantum gravity},'' {\em Nuclear
  Physics B} {\bf 385} (1992) 587--603.

\bibitem{LewandowskiLoopConstraintsA}
J.~Lewandowski and D.~Marolf, ``{Loop constraints: A habitat and their
  algebra},'' {\em International Journal of Modern Physics D} {\bf 07} (1998)
  299--330, {\tt arXiv:gr-qc/9710016}.

\bibitem{LewandowskiSymmetricScalarConstraint}
J.~Lewandowski and H.~Sahlmann, ``{Symmetric scalar constraint for loop quantum
  gravity},'' {\em Physical Review D} {\bf 91} (2015) 044022, {\tt
  arXiv:1410.5276 [gr-qc]}.

\bibitem{AlesciHamiltonianOperatorFor}
E.~Alesci, M.~Assanioussi, J.~Lewandowski, and I.~M\"{a}kinen, ``{Hamiltonian
  operator for loop quantum gravity coupled to a scalar field},'' {\em Physical
  Review D} {\bf 91} (2015) 124067, {\tt arXiv:1504.02068 [gr-qc]}.

\bibitem{AssanioussiNewScalarConstraint}
M.~Assanioussi, J.~Lewandowski, and I.~M\"{a}kinen, ``{New scalar constraint
  operator for loop quantum gravity},'' {\em Physical Review D} {\bf 92} (2015)
  044042, {\tt arXiv:1506.00299 [gr-qc]}.

\bibitem{GieselAQG1}
K.~Giesel and T.~Thiemann, ``{Algebraic quantum gravity (AQG): I. Conceptual
  setup},'' {\em Classical and Quantum Gravity} {\bf 24} (2007) 2465--2497,
  {\tt arXiv:gr-qc/0607099}.

\bibitem{AlesciANewPerspective}
E.~Alesci and F.~Cianfrani, ``{A new perspective on cosmology in Loop Quantum
  Gravity},'' {\em Europhysics Letters} {\bf 104} (2013) 10001, {\tt
  arXiv:1210.4504 [gr-qc]}.

\bibitem{AlesciLoopQuantumCosmology}
E.~Alesci and F.~Cianfrani, ``{Loop quantum cosmology from quantum reduced loop
  gravity},'' {\em Europhysics Letters} {\bf 111} (2015) 40002, {\tt
  arXiv:1410.4788 [gr-qc]}.

\bibitem{BojowaldSphericallySymmetricQuantumGeometryHamiltonian}
M.~Bojowald and R.~Swiderski, ``{Spherically symmetric quantum geometry:
  Hamiltonian constraint},'' {\em Classical and Quantum Gravity} {\bf 23}
  (2006) 2129--2154, {\tt arXiv:gr-qc/0511108}.

\bibitem{AlvarezLocalHamiltonianFor}
N.~Alvarez, R.~Gambini, and J.~Pullin, ``{Local Hamiltonian for Spherically
  Symmetric Gravity Coupled to a Scalar Field},'' {\em Physical Review Letters}
  {\bf 108} (2012) 051301, {\tt arXiv:1111.4962 [gr-qc]}.

\bibitem{GambiniLoopQuantizationOf}
R.~Gambini and J.~Pullin, ``{Loop Quantization of the Schwarzschild Black
  Hole},'' {\em Physical Review Letters} {\bf 110} (2013) 211301, {\tt
  arXiv:1302.5265 [gr-qc]}.

\bibitem{ThiemannThePhoenixProject}
T.~Thiemann, ``{The Phoenix Project: master constraint programme for loop
  quantum gravity},'' {\em Classical and Quantum Gravity} {\bf 23} (2006)
  2211--2247, {\tt arXiv:gr-qc/0305080}.

\bibitem{ThiemannQSD8}
T.~Thiemann, ``{Quantum spin dynamics: VIII. The master constraint},'' {\em
  Classical and Quantum Gravity} {\bf 23} (2006) 2249--2265, {\tt
  arXiv:gr-qc/0510011}.

\bibitem{HanMasterConstraintOperators}
M.~Han and Y.~Ma, ``{Master constraint operators in loop quantum gravity},''
  {\em Physics Letters B} {\bf 635} (2006) 225--231, {\tt arXiv:gr-qc/0510014}.

\bibitem{DittrichTestingTheMasterI}
B.~Dittrich and T.~Thiemann, ``{Testing the master constraint programme for
  loop quantum gravity: I. General framework},'' {\em Classical and Quantum
  Gravity} {\bf 23} (2006) 1025--1065, {\tt arXiv:gr-qc/0411138}.

\bibitem{DittrichTestingTheMasterII}
B.~Dittrich and T.~Thiemann, ``{Testing the master constraint programme for
  loop quantum gravity: II. Finite-dimensional systems},'' {\em Classical and
  Quantum Gravity} {\bf 23} (2006) 1067--1088, {\tt arXiv:gr-qc/0411139}.

\bibitem{DittrichTestingTheMasterIII}
B.~Dittrich and T.~Thiemann, ``{Testing the master constraint programme for
  loop quantum gravity: III. models},'' {\em Classical and Quantum Gravity}
  {\bf 23} (2006) 1089--1120, {\tt arXiv:gr-qc/0411140}.

\bibitem{DittrichTestingTheMasterIV}
B.~Dittrich and T.~Thiemann, ``{Testing the master constraint programme for
  loop quantum gravity: IV. Free field theories},'' {\em Classical and Quantum
  Gravity} {\bf 23} (2006) 1121--1142, {\tt arXiv:gr-qc/0411141}.

\bibitem{DittrichTestingTheMasterV}
B.~Dittrich and T.~Thiemann, ``{Testing the master constraint programme for
  loop quantum gravity: V. Interacting field theories},'' {\em Classical and
  Quantum Gravity} {\bf 23} (2006) 1143--1162, {\tt arXiv:gr-qc/0411142}.

\bibitem{RovelliBook2}
C.~Rovelli and F.~Vidotto, {\em {Covariant Loop Quantum Gravity: An Elementary
  Introduction to Quantum Gravity and Spinfoam Theory}}.
\newblock Cambridge University Press, 2014.

\bibitem{BaezAnIntroductionTo}
J.~C. Baez, ``{An Introduction to Spin Foam Models of Quantum Gravity and BF
  Theory},'' {\em Lecture Notes in Physics} {\bf 543} (2000) 25--94, {\tt
  arXiv:gr-qc/9905087}.

\bibitem{RovelliLecturesOnLoop}
C.~Rovelli, ``{Zakopane lectures on loop gravity},'' {\tt arXiv:1102.3660
  [gr-qc]}.

\bibitem{PerezSpinFoamModels}
A.~Perez, ``{Spin foam models for quantum gravity},'' {\em Classical and
  Quantum Gravity} {\bf 20} (2003) R43--R104, {\tt arXiv:gr-qc/0301113}.

\bibitem{BarrettStateSumModels}
J.~W. Barrett, ``{State sum models for quantum gravity},'' {\tt
  arXiv:gr-qc/0010050}.

\bibitem{ReisenbergerSumOverSurfaces}
M.~P. Reisenberger and C.~Rovelli, ``{'Sum over surfaces' form of loop quantum
  gravity},'' {\em Physical Review D} {\bf 56} (1997) 3490--3508, {\tt
  arXiv:gr-qc/9612035}.

\bibitem{BarrettRelativisticSpinNetworks}
J.~W. Barrett and L.~Crane, ``{Relativistic spin networks and quantum
  gravity},'' {\em Journal of Mathematical Physics} {\bf 39} (1998) 3296--3302,
  {\tt arXiv:gr-qc/9709028}.

\bibitem{FreidelANewSpin}
L.~Freidel and K.~Krasnov, ``{A new spin foam model for 4D gravity},'' {\em
  Classical and Quantum Gravity} {\bf 25} (2008) 125018, {\tt arXiv:0708.1595
  [gr-qc]}.

\bibitem{AlesciCompleteLQGPropagatorI}
E.~Alesci and C.~Rovelli, ``{Complete LQG propagator: Difficulties with the
  Barrett-Crane vertex},'' {\em Physical Review D} {\bf 76} (2007) 104012, {\tt
  arXiv:0708.0883 [gr-qc]}.

\bibitem{AlesciCompleteLQGPropagatorII}
E.~Alesci and C.~Rovelli, ``{Complete LQG propagator. II. Asymptotic behavior
  of the vertex},'' {\em Physical Review D} {\bf 77} (2008) 044024, {\tt
  arXiv:0711.1284 [gr-qc]}.

\bibitem{BianchiLorentzianSpinfoamPropagator}
E.~Bianchi and Y.~Ding, ``{Lorentzian spinfoam propagator},'' {\em Physical
  Review D} {\bf 86} (2012) 104040, {\tt arXiv:1109.6538 [gr-qc]}.

\bibitem{PlebanskiOnTheSeparation}
J.~F. Plebanski, ``{On the separation of Einsteinian substructures},'' {\em
  Journal of Mathematical Physics} {\bf 18} (1977) 2511--2520.

\bibitem{HorowitzExactlySolubleDiffeomorphism}
G.~T. Horowitz, ``{Exactly soluble diffeomorphism invariant theories},'' {\em
  Communications in Mathematical Physics} {\bf 125} (1989) 417--437.

\bibitem{HellmannGeometricAsymptoticsFor}
F.~Hellmann and W.~Kaminski, ``{Geometric asymptotics for spin foam lattice
  gauge gravity on arbitrary triangulations},'' {\tt arXiv:1210.5276 [gr-qc]}.

\bibitem{AlexandrovSpinFoamsAnd}
S.~Alexandrov, M.~Geiller, and K.~Noui, ``{Spin Foams and Canonical
  Quantization},'' {\em Symmetry, Integrability and Geometry: Methods and
  Applications} {\bf 8} (2012) 55, {\tt arXiv:1112.1961 [gr-qc]}.

\bibitem{HanCommutingSimplicityAnd}
M.~Han and T.~Thiemann, ``{Commuting simplicity and closure constraints for 4D
  spin-foam models},'' {\em Classical and Quantum Gravity} {\bf 30} (2013)
  235024, {\tt arXiv:1010.5444 [gr-qc]}.

\bibitem{AlesciLinkingCovariantAnd}
E.~Alesci, T.~Thiemann, and A.~Zipfel, ``{Linking covariant and canonical loop
  quantum gravity: New solutions to the Euclidean scalar constraint},'' {\em
  Physical Review D} {\bf 86} (2012) 024017, {\tt arXiv:1109.1290 [gr-qc]}.

\bibitem{ThiemannLinkingCovariantAndII}
T.~Thiemann and A.~Zipfel, ``{Linking covariant and canonical LQG II: spin foam
  projector},'' {\em Classical and Quantum Gravity} {\bf 31} (2014) 125008,
  {\tt arXiv:1307.5885 [gr-qc]}.

\bibitem{LivineCouplingOfSpacetime}
E.~R. Livine and D.~Oriti, ``{Coupling of spacetime atoms in 4D spin foam
  models from group field theory},'' {\em Journal of High Energy Physics} {\bf
  2007} (2007) 092--092, {\tt arXiv:gr-qc/0512002}.

\bibitem{RielloSelfEnergyOf}
A.~Riello, ``{Self-energy of the Lorentzian Engle-Pereira-Rovelli-Livine and
  Freidel-Krasnov model of quantum gravity},'' {\em Physical Review D} {\bf 88}
  (2013) 024011, {\tt arXiv:1302.1781 [gr-qc]}.

\bibitem{BonzomBubbleDivergencesAnd}
V.~Bonzom and B.~Dittrich, ``{Bubble divergences and gauge symmetries in spin
  foams},'' {\em Physical Review D} {\bf 88} (2013) 124021, {\tt
  arXiv:1304.6632 [gr-qc]}.

\bibitem{DittrichCoarseGrainingOf}
B.~Dittrich, M.~Mart\'{\i}n-Benito, and E.~Schnetter, ``{Coarse graining of
  spin net models: dynamics of intertwiners},'' {\em New Journal of Physics}
  {\bf 15} (2013) 103004, {\tt arXiv:1306.2987 [gr-qc]}.

\bibitem{OritiTheGroupField}
D.~Oriti, ``{The Group field theory approach to quantum gravity},'' in {\em
  Approaches to Quantum Gravity - toward a new understanding of space, time,
  and matter} (D.~Oriti, ed.).
\newblock Cambridge University Press,  2006.
\newblock {\tt arXiv:gr-qc/0607032}.

\bibitem{OritiGroupFieldTheoryAndLoop}
D.~Oriti, ``{Group Field Theory and Loop Quantum Gravity},'' {\tt
  arXiv:1408.7112 [gr-qc]}.

\bibitem{GurauColoredTensorModels}
R.~Gurau, ``{Colored Tensor Models - a Review},'' {\em Symmetry, Integrability
  and Geometry: Methods and Applications} (2012) {\tt arXiv:1109.4812
  [hep-th]}.

\bibitem{OritiGroupFieldTheoryAsTheSecond}
D.~Oriti, ``{Group field theory as the second quantization of loop quantum
  gravity},'' {\em Classical and Quantum Gravity} {\bf 33} (2016) 085005, {\tt
  arXiv:1310.7786 [gr-qc]}.

\bibitem{MaExtensionOfLoop}
Y.~Ma, ``{Extension of Loop Quantum Gravity to Metric Theories beyond General
  Relativity},'' {\em Journal of Physics: Conference Series} {\bf 360} (2012)
  012006, {\tt arXiv:1112.2085 [gr-qc]}.

\bibitem{BNII}
N.~Bodendorfer and Y.~Neiman, ``{Wald entropy formula and loop quantum
  gravity},'' {\em Physical Review D} {\bf 90} (2014) 084054, {\tt
  arXiv:1304.3025 [gr-qc]}.

\bibitem{GielenCosmologyFromGroup}
S.~Gielen, D.~Oriti, and L.~Sindoni, ``{Cosmology from Group Field Theory
  Formalism for Quantum Gravity},'' {\em Physical Review Letters} {\bf 111}
  (2013) 031301, {\tt arXiv:1303.3576 [gr-qc]}.

\bibitem{GielenHomogeneousCosmologiesAs}
S.~Gielen, D.~Oriti, and L.~Sindoni, ``{Homogeneous cosmologies as group field
  theory condensates},'' {\em Journal of High Energy Physics} {\bf 2014} (2014)
  13, {\tt arXiv:1311.1238 [gr-qc]}.

\bibitem{OritiBouncingCosmologiesFrom}
D.~Oriti, L.~Sindoni, and E.~Wilson-Ewing, ``{Bouncing cosmologies from quantum
  gravity condensates},'' {\tt arXiv:1602.08271 [gr-qc]}.

\bibitem{OritiEmergentFriedmannDynamics}
D.~Oriti, L.~Sindoni, and E.~Wilson-Ewing, ``{Emergent Friedmann dynamics with
  a quantum bounce from quantum gravity condensates},'' {\tt arXiv:1602.05881
  [gr-qc]}.

\bibitem{CarrozzaRenormalizationOfASU}
S.~Carrozza, D.~Oriti, and V.~Rivasseau, ``{Renormalization of a SU(2)
  Tensorial Group Field Theory in Three Dimensions},'' {\em Communications in
  Mathematical Physics} {\bf 330} (2014) 581--637, {\tt arXiv:1303.6772
  [hep-th]}.

\bibitem{LahocheRenormalizationOfAn}
V.~Lahoche, D.~Oriti, and V.~Rivasseau, ``{Renormalization of an Abelian tensor
  group field theory: solution at leading order},'' {\em Journal of High Energy
  Physics} {\bf 2015} (2015) 95, {\tt arXiv:1501.02086 [hep-th]}.

\bibitem{CarrozzaTensorialMethodsAnd}
S.~Carrozza, {\em {Tensorial Methods and Renormalization in Group Field
  Theories}}.
\newblock PhD thesis,
\newblock 2014.
\newblock {\tt arXiv:1310.3736 [hep-th]}.

\bibitem{BojowaldLemaitreTolmanBondi}
M.~Bojowald, T.~Harada, and R.~Tibrewala, ``{Lemaitre-Tolman-Bondi collapse
  from the perspective of loop quantum gravity},'' {\em Physical Review D} {\bf
  78} (2008) 064057, {\tt arXiv:0806.2593 [gr-qc]}.

\bibitem{AshtekarBlackHoleEvaporation}
A.~Ashtekar and M.~Bojowald, ``{Black hole evaporation: a paradigm},'' {\em
  Classical and Quantum Gravity} {\bf 22} (2005) 3349--3362, {\tt
  arXiv:gr-qc/0504029}.

\bibitem{AlesciQuantumReducedLoopGravitySemiclassicalLimit}
E.~Alesci and F.~Cianfrani, ``{Quantum reduced loop gravity: Semiclassical
  limit},'' {\em Physical Review D} {\bf 90} (2014) 024006, {\tt
  arXiv:1402.3155 [gr-qc]}.

\bibitem{AlesciImprovedRegularizationFrom}
E.~Alesci and F.~Cianfrani, ``{Improved regularization from Quantum Reduced
  Loop Gravity},'' {\tt arXiv:1604.02375 [gr-qc]}.

\bibitem{BeetleDiffeomorphismInvariantCosmological}
C.~Beetle, J.~S. Engle, M.~E. Hogan, and P.~Mendonca, ``{Diffeomorphism
  invariant cosmological symmetry in full quantum gravity},'' {\tt
  arXiv:1603.01128 [gr-qc]}.

\bibitem{EngleRelatingLoopQuantum}
J.~Engle, ``{Relating loop quantum cosmology to loop quantum gravity: symmetric
  sectors and embeddings},'' {\em Classical and Quantum Gravity} {\bf 24}
  (2007) 5777--5802, {\tt arXiv:gr-qc/0701132}.

\bibitem{EngleEmbeddingLoopQuantum}
J.~Engle, ``{Embedding loop quantum cosmology without piecewise linearity},''
  {\em Classical and Quantum Gravity} {\bf 30} (2013) 085001, {\tt
  arXiv:1301.6210 [gr-qc]}.

\bibitem{BrunnemannSymmetryReductionOf}
J.~Brunnemann and T.~A. Koslowski, ``{Symmetry reduction of loop quantum
  gravity},'' {\em Classical and Quantum Gravity} {\bf 28} (2011) 245014, {\tt
  arXiv:1012.0053 [gr-qc]}.

\bibitem{HanuschInvariantConnectionsIn}
M.~Hanusch, ``{Invariant Connections in Loop Quantum Gravity},'' {\em
  Communications in Mathematical Physics} {\bf 343} (2016) 1--38, {\tt
  arXiv:1307.5303 [math-ph]}.

\bibitem{FleischhackKinematicalFoundationsOf}
C.~Fleischhack, ``{Kinematical Foundations of Loop Quantum Cosmology},'' {\tt
  arXiv:1505.04400 [math-ph]}.

\bibitem{GielenIdentifyingCosmologicalPerturbations}
S.~Gielen, ``{Identifying cosmological perturbations in group field theory
  condensates},'' {\em Journal of High Energy Physics} {\bf 2015} (2015) 10,
  {\tt arXiv:1505.07479 [gr-qc]}.

\bibitem{BojowaldSphericallySymmetricQuantum}
M.~Bojowald, ``{Spherically symmetric quantum geometry: states and basic
  operators},'' {\em Classical and Quantum Gravity} {\bf 21} (2004) 3733--3753,
  {\tt arXiv:gr-qc/0407017}.

\bibitem{OritiGeneralizedQuantumGravity}
D.~Oriti, D.~Pranzetti, J.~P. Ryan, and L.~Sindoni, ``{Generalized quantum
  gravity condensates for homogeneous geometries and cosmology},'' {\em
  Classical and Quantum Gravity} {\bf 32} (2015) 235016, {\tt arXiv:1501.00936
  [gr-qc]}.

\bibitem{SmolinLinkingTopologicalQuantum}
L.~Smolin, ``{Linking Topological Quantum Field Theory and Nonperturbative
  Quantum Gravity},'' {\em Journal of Mathematical Physics} {\bf 36} (1995)
  6417--6455, {\tt arXiv:gr-qc/9505028}.

\bibitem{KrasnovGeometricalEntropyFrom}
K.~Krasnov, ``{Geometrical entropy from loop quantum gravity},'' {\em Physical
  Review D} {\bf 55} (1997) 3505 -- 3513, {\tt arXiv:gr-qc/9603025}.

\bibitem{RovelliBlackHoleEntropy}
C.~Rovelli, ``{Black Hole Entropy from Loop Quantum Gravity},'' {\em Physical
  Review Letters} {\bf 77} (1996) 3288--3291, {\tt arXiv:gr-qc/9603063}.

\bibitem{AshtekarIsolatedAndDynamical}
A.~Ashtekar and B.~Krishnan, ``{Isolated and Dynamical Horizons and Their
  Applications},'' {\em Living Reviews in Relativity} {\bf 7} (2004) {\tt
  arXiv:gr-qc/0407042}.

\bibitem{AshtekarQuantumGeometryAndBlackHoleEntropy}
A.~Ashtekar, J.~Baez, A.~Corichi, and K.~Krasnov, ``{Quantum Geometry and Black
  Hole Entropy},'' {\em Physical Review Letters} {\bf 80} (1998) 904--907, {\tt
  arXiv:gr-qc/9710007}.

\bibitem{AshtekarIsolatedHorizonsThe}
A.~Ashtekar, A.~Corichi, and K.~Krasnov, ``{Isolated Horizons: the Classical
  Phase Space},'' {\em Advances in Theoretical and Mathematical Physics} {\bf
  3} (1999) 419--478, {\tt arXiv:gr-qc/9905089}.

\bibitem{EngleBlackHoleEntropyFrom}
J.~Engle, K.~Noui, A.~Perez, and D.~Pranzetti, ``{Black hole entropy from an
  SU(2)-invariant formulation of Type I isolated horizons},'' {\em Physical
  Review D} {\bf 82} (2010) 044050, {\tt arXiv:1006.0634 [gr-qc]}.

\bibitem{BTTXII}
N.~Bodendorfer, T.~Thiemann, and A.~Thurn, ``{New variables for classical and
  quantum gravity in all dimensions: V. Isolated horizon boundary degrees of
  freedom},'' {\em Classical and Quantum Gravity} {\bf 31} (2014) 055002, {\tt
  arXiv:1304.2679 [gr-qc]}.

\bibitem{Diaz-PoloIsolatedHorizonsAnd}
J.~Diaz-Polo and D.~Pranzetti, ``{Isolated Horizons and Black Hole Entropy in
  Loop Quantum Gravity},'' {\em SIGMA} {\bf 8} (2012) 048, {\tt arXiv:1112.0291
  [gr-qc]}.

\bibitem{BianchiOnTheArchitecture}
E.~Bianchi and R.~C. Myers, ``{On the architecture of spacetime geometry},''
  {\em Classical and Quantum Gravity} {\bf 31} (2014) 214002, {\tt
  arXiv:1212.5183 [hep-th]}.

\bibitem{DuchObservablesForGeneral}
P.~Duch, W.~Kaminski, J.~Lewandowski, and J.~Swiezewski, ``{Observables for
  general relativity related to geometry},'' {\em Journal of High Energy
  Physics} {\bf 2014} (2014) 77, {\tt arXiv:1403.8062 [gr-qc]}.

\bibitem{DasguptaSemiclassicalQuantizationOf}
A.~Dasgupta, ``{Semi-classical quantization of spacetimes with apparent
  horizons},'' {\em Classical and Quantum Gravity} {\bf 23} (2006) 635--671,
  {\tt arXiv:gr-qc/0505017}.

\bibitem{GhoshBlackHoleEntropy}
A.~Ghosh and A.~Perez, ``{Black Hole Entropy and Isolated Horizons
  Thermodynamics},'' {\em Physical Review Letters} {\bf 107} (2011) 241301,
  {\tt arXiv:1107.1320 [gr-qc]}.

\bibitem{PranzettiBlackHoleEntropy}
D.~Pranzetti, ``{Geometric temperature and entropy of quantum isolated
  horizons},'' {\em Physical Review D} {\bf 89} (2014) 104046, {\tt
  arXiv:1305.6714 [gr-qc]}.

\bibitem{GeillerNearHorizonRadiation}
M.~Geiller and K.~Noui, ``{Near-horizon radiation and self-dual loop quantum
  gravity},'' {\em Europhysics Letters} {\bf 105} (2014) 60001, {\tt
  arXiv:1402.4138 [gr-qc]}.

\bibitem{NatsuumeAdSCFTDuality}
M.~Natsuume, ``{AdS/CFT Duality User Guide},'' {\em Lecture Notes in Physics}
  {\bf 903} (2015) {\tt arXiv:1409.3575 [hep-th]}.

\bibitem{BIV}
N.~Bodendorfer, ``{A note on quantum supergravity and AdS/CFT},'' {\tt
  arXiv:1509.02036 [hep-th]}.

\bibitem{Bonzom3DHolography}
V.~Bonzom and B.~Dittrich, ``{3D holography: from discretum to continuum},''
  {\em Journal of High Energy Physics} {\bf 2016} (2016) 208, {\tt
  arXiv:1511.05441 [hep-th]}.

\bibitem{BonzomDualityBetweenSpin}
V.~Bonzom, F.~Costantino, and E.~R. Livine, ``{Duality Between Spin Networks
  and the 2D Ising Model},'' {\em Communications in Mathematical Physics}
  (2016) {\tt arXiv:1504.02822 [math-ph]}.

\bibitem{DittrichIsingModelFrom}
B.~Dittrich and J.~Hnybida, ``{Ising Model from Intertwiners},'' {\tt
  arXiv:1312.5646 [gr-qc]}.

\bibitem{FreidelReconstructingAdSCFT}
L.~Freidel, ``{Reconstructing AdS/CFT},'' {\tt arXiv:0804.0632v1 [hep-th]}.

\bibitem{LaddhaHamiltonianConstraintInPolymer}
A.~Laddha and M.~Varadarajan, ``{Hamiltonian constraint in polymer parametrized
  field theory},'' {\em Physical Review D} {\bf 83} (2011) 025019, {\tt
  arXiv:1011.2463 [gr-qc]}.

\bibitem{TomlinTowardsAnAnomalyII}
M.~Varadarajan, ``{Towards an anomaly-free quantum dynamics for a weak coupling
  limit of Euclidean gravity: Diffeomorphism covariance},'' {\em Physical
  Review D} {\bf 87} (2013) 044040, {\tt arXiv:1210.6877 [gr-qc]}.

\bibitem{HendersonConstraintAlgebraInII}
A.~Henderson, A.~Laddha, and C.~Tomlin, ``{Constraint algebra in loop quantum
  gravity reloaded. II. Toy model of an Abelian gauge theory: Spatial
  diffeomorphisms},'' {\em Physical Review D} {\bf 88} (2013) 044029, {\tt
  arXiv:1210.3960 [gr-qc]}.

\bibitem{DittrichDecoratedTensorNetwork}
B.~Dittrich, S.~Mizera, and S.~Steinhaus, ``{Decorated tensor network
  renormalization for lattice gauge theories and spin foam models},'' {\em New
  Journal of Physics} {\bf 18} (2016) 053009, {\tt arXiv:1409.2407 [gr-qc]}.

\bibitem{DittrichQuantumGroupSpin}
B.~Dittrich, M.~Martin-Benito, and S.~Steinhaus, ``{Quantum group spin nets:
  Refinement limit and relation to spin foams},'' {\em Physical Review D} {\bf
  90} (2014) 024058, {\tt arXiv:1312.0905 [gr-qc]}.

\bibitem{DittrichTimeEvolutionAs}
B.~Dittrich and S.~Steinhaus, ``{Time evolution as refining, coarse graining
  and entangling},'' {\em New Journal of Physics} {\bf 16} (2014) 123041, {\tt
  arXiv:1311.7565 [gr-qc]}.

\bibitem{FreidelBFDescriptionOf}
L.~Freidel, K.~Krasnov, and R.~Puzio, ``{BF description of higher-dimensional
  gravity theories},'' {\em Advances in Theoretical and Mathematical Physics}
  {\bf 3} (1999) 1289--1324, {\tt arXiv:hep-th/9901069}.

\end{thebibliography}

\end{document}